\documentclass[acmsmall,screen]{acmart}

\AtBeginDocument{%
  \providecommand\BibTeX{{%
    \normalfont B\kern-0.5em{\scshape i\kern-0.25em b}\kern-0.8em\TeX}}}

\acmJournal{JACM}

\usepackage{balance}
\usepackage{graphicx}
\usepackage{tabularx}
\usepackage{enumitem}
\usepackage{siunitx} 
\usepackage{hyperref}
\usepackage{pifont}
 \usepackage{booktabs}
\usepackage{indentfirst}
\usepackage{array} %
\newcolumntype{P}[1]{>{\centering\arraybackslash}p{#1}} %
\usepackage{subcaption} 
\usepackage{comment}
\usepackage{xspace}
\usepackage{amsmath}
\usepackage[nameinlink]{cleveref}

\usepackage{multirow}
\usepackage{makecell}
\usepackage{threeparttable} 

\usepackage{tikz}

\usepackage{todonotes} 
\usepackage{xcolor}   

\hypersetup{
  colorlinks=true,
  linkcolor=transparent,
  anchorcolor=transparent,
  citecolor=black,
  urlcolor=black
}

\definecolor{myred}{RGB}{200, 50, 50}
\definecolor{darkpastelred}{rgb}{0.76, 0.23, 0.13}
\definecolor{ao(english)}{rgb}{0.0, 0.5, 0.0}
\definecolor{darkpastelred}{rgb}{0.76, 0.23, 0.13}
\definecolor{ao(english)}{rgb}{0.0, 0.5, 0.0}
\definecolor{yellow}{RGB}{255,255,153}
\definecolor{grey}{RGB}{224,224,224}

\definecolor{DarkOrange}{rgb}{0.8,0.3,0.0}
\definecolor{DarkCyan}{rgb}{0.0, 0.55, 0.55}
\definecolor{DarkCyel}{rgb}{1.0, 0.49, 0.0}
\definecolor{yellow-green}{rgb}{0.6, 0.8, 0.2}

\newcolumntype{?}{!{\vrule width 1pt}}

\usepackage[most]{tcolorbox}
\usepackage{listings}

\usepackage{algorithm}
\usepackage{algorithmic}

\lstdefinestyle{mystyle}{
    basicstyle=\scriptsize\ttfamily,
    breakatwhitespace=false,
    breaklines=true,
    captionpos=b, 
    frame=tb, 
    framesep=2pt,
    framerule=0pt,
    aboveskip=0pt,
    belowskip=0pt,
    xleftmargin=0pt, 
    xrightmargin=0pt, 
    numbers=none, 
}

\lstdefinestyle{test-smell-listing}{
    language=Java,
    basicstyle=\scriptsize\ttfamily, 
    breakatwhitespace=false,
    breaklines=true,
    captionpos=b, 
    frame=single, 
    framesep=2pt,
    framerule=0.5pt, 
    aboveskip=0pt,
    belowskip=0pt,
    xleftmargin=10pt, 
    xrightmargin=10pt,
    numbers=left, 
    numberstyle=\tiny\color{gray}, 
    stepnumber=1,
    numbersep=5pt, 
    showstringspaces=false, 
    keywordstyle=\color{blue}\bfseries, 
    commentstyle=\color{green}, 
    stringstyle=\color{red}, 
    morecomment=[s][\color{magenta}]{/**}{*/}, 
    morecomment=[s][\color{magenta}]{/*}{*/}, 
    emph={public,private,protected,static,void,int,double,float,char,boolean}, 
    emphstyle=\color{blue}\bfseries, 
}

\newcommand{\highlight}[1]{\begin{tcolorbox}[leftrule=0mm,rightrule=0mm,toprule=0mm,bottomrule=0mm,left=2pt,right=2pt,top=2pt,bottom=2pt]
  #1
  \end{tcolorbox}
}

\definecolor{mylightgray}{RGB}{224,224,224}

\newboolean{showcomments}
\setboolean{showcomments}{true}
\ifthenelse{\boolean{showcomments}}
 { \newcommand{\mynote}[2]{
      \fbox{\bfseries\sffamily\scriptsize#1}      {\small$\blacktriangleright$\textsf{\emph{#2}}$\blacktriangleleft$}}}
        { \newcommand{\mynote}[2]{}}

\def\BibTeX{{\rm B\kern-.05em{\sc i\kern-.025em b}\kern-.08em
    T\kern-.1667em\lower.7ex\hbox{E}\kern-.125emX}}

\makeatletter
\DeclareRobustCommand\onedot{\futurelet\@let@token\@onedot}
\def\@onedot{\ifx\@let@token.\else.\null\fi\xspace}

\makeatother

\newcolumntype{b}{>{\color{blue}}r}  
\usepackage{cancel}

\begin{document}

\title{On the Diffusion of Test Smells in LLM-Generated Unit Tests
}


\author{Wendkûuni C. Ouédraogo}
\email{wendkuuni.ouedraogo@uni.lu}
\affiliation{
  \institution{University of Luxembourg}
 	\country{Luxembourg}
}

\author{Yinghua Li}\authornote{Corresponding author.}
\email{yinghua.li@njust.edu.cn}
\affiliation{
  \institution{Nanjing University of Science and Technology}
 	\country{China}
}

\author{Xueqi Dang}
\email{xueqi.dang@uni.lu}
\affiliation{
  \institution{University of Luxembourg}
 	\country{Luxembourg}
}

\author{Xunzhu Tang}
\email{xunzhu.tang@uni.lu}
\affiliation{
  \institution{University of Luxembourg}
 	\country{Luxembourg}
}

\author{Anil Koyuncu}
\email{anil.koyuncu@cs.bilkent.edu.tr}
\affiliation{
  \institution{Bilkent University}
 	\country{Turkey}
}

\author{Jacques Klein}
\email{jacques.klein@uni.lu}
\affiliation{
  \institution{University of Luxembourg}
  \country{Luxembourg}
}

\author{David Lo}
\email{davidlo@smu.edu.sg}
\affiliation{
  \institution{Singapore Management University}
  \country{Singapore}
}

\author{Tegawend\'e F. Bissyand\'e}
\email{tegawende.bissyande@uni.lu}
\affiliation{
  \institution{University of Luxembourg}
 	\country{Luxembourg}
}

\renewcommand{\shortauthors}{W. Ouédraogo et al.}

\begin{abstract}

LLMs promise to transform Java unit test generation from a manual burden into an automated solution. Yet, beyond metrics such as compilability or coverage, the \emph{quality} of LLM-generated Java tests (particularly their susceptibility to \emph{test smells}, design flaws that undermine readability and maintainability) remains insufficiently understood. Recent studies have begun to examine test smells and broader quality attributes in LLM-generated tests, but a comprehensive, multi-benchmark analysis of test smell diffusion in Java unit tests, comparing LLM-generated suites with SBST-generated and human-written tests under multiple detectors, is still missing.
This paper presents a multi-benchmark, large-scale analysis of test smell diffusion in LLM-generated Java unit tests. We contrast LLM outputs with human-written suites (as the reference for real-world practices) and SBST-generated tests from EvoSuite (as the automated baseline), disentangling whether LLMs reproduce human-like flaws or artifacts of synthetic generation. Our aim is to systematically characterise the prevalence, distribution, and co-occurrence of test smells (what we term their diffusion) in LLM-generated Java tests across models, prompting strategies, and generation granularities, and to compare these diffusion patterns with SBST-generated and human-written suites.
Our study draws on 20,505 class-level suites generated by four LLMs (GPT-3.5, GPT-4, Mistral 7B, Mixtral 8×7B) over three Java benchmarks (Defects4J, SF110, and a curated dataset, CMD), 972 method-level cases from TestBench generated by three LLMs (GPT-3.5, GPT-4, CodeLlama-13B-Instruct), 14,469 EvoSuite tests, and 779,585 human-written tests from 34,635 open-source Java projects (CAT-LM dataset, SF110, Defects4J). Using two complementary detection tools (TsDetect and JNose) together with a manually validated sample of 240 classes and 447 methods, we analyze prevalence, co-occurrence, and correlations with software attributes and generation parameters.
%
Results show that LLM-generated Java tests consistently manifest smells such as Assertion Roulette and Magic Number Test, with patterns strongly influenced by prompting strategy, context length, and model scale. 
Comparisons reveal overlaps with human-written tests, especially under certain detectors, which are consistent with shallow imitation of common testing idioms or possible training-data overlap, while EvoSuite exhibits distinct, generator-specific flaws.
These findings highlight both the promise and the risks of LLM-based Java test generation, and call for the design of smell-aware generation frameworks, prompt engineering strategies, and enhanced detection tools to ensure maintainable, high-quality test code.
\end{abstract}

\ccsdesc[500]{Software and its engineering~Software testing and debugging}
\ccsdesc[300]{Computer systems organization~Neural networks}

\keywords{Test Smells, Large Language Models, Unit Test, SBST, Data Leakage, EvoSuite, LLM, Empirical Study}

\maketitle

\section{Introduction} \label{intro}

Unit tests are foundational to modern software development, providing early feedback on correctness and supporting long-term maintainability~\cite{beck2000extreme,siddiqui2021learning,shore2021art}. However, writing effective unit tests is a labor-intensive and often introduces test smells—design flaws that hinder maintainability and understanding~\cite{bavota2012empirical,spadini2018relation,garousi2018smells,junior2020survey}. 

Traditional automated test generation tools like EvoSuite~\cite{fraser2011evosuite} reduce manual effort but prioritize structural coverage over readability, frequently producing tests with quality issues such as Assertion Roulette or Eager Test~\cite{palomba2016diffusion,panichella2020revisiting}.  The emergence of Large Language Models (LLMs) such as GPT-4~\cite{chatgpt2023openai} and Mistral~\cite{jiang2023mistral} promises a paradigm shift toward generating more human-like, readable test suites~\cite{jain2025testforge,pan2025aster,biagiola2025improving,deljouyi2024leveraging}.

Recent empirical studies have extensively evaluated the quality of LLM-generated unit tests along dimensions such as syntactic correctness, compilability, coverage, fault detection, and readability. Yuan et al.~\cite{yuan2023no} assess ChatGPT-generated tests on Java projects focusing on syntactic and execution correctness, coverage, and perceived readability and usability. Siddiq et al.~\cite{siddiq2024using} evaluate GPT-3.5-Turbo, StarCoder, and Codex on the Java version of HumanEval and on 47 projects from SF110, measuring compilation rates, test correctness, branch and line coverage, and the incidence of a set of test smells (e.g., Duplicated Asserts, Empty Tests). They primarily compare LLM-generated tests against EvoSuite on both benchmarks, and additionally report coverage of the original HumanEval reference tests as a baseline, but do not analyse or compare test smells in human-written tests. 
Ouédraogo et al.~\cite{ouedraogo2024large} perform a large-scale evaluation of LLM-based and Search-Based Software Testing (SBST)–based test generation on three Java datasets (SF110, Defects4J, and a curated Custom Mini Dataset, CMD) designed, among other goals, to mitigate potential training data leakage. They compare multiple LLMs and prompting strategies against EvoSuite in terms of structural correctness, multi-level coverage (line, instruction, and method), readability, and a catalogue of test smells, but do not quantify or compare test smells in human-written test suites. Comparative works such as Tang et al.~\cite{tang2024chatgpt} and Yang et al.~\cite{yang2024evaluation} further position LLM-based test generation with respect to SBST tools (e.g., EvoSuite), focusing on syntactic validity, coverage, and defect detection effectiveness. More recently, TestBench~\cite{zhang2024testbench} and MuTAP~\cite{dakhel2024effective} have started to employ mutation testing to assess or guide LLM-generated tests: TestBench uses mutation score to evaluate the fault-detection capability of class-level tests produced by LLMs, while Dakhel et al.~\cite{dakhel2024effective} use mutation testing as a feedback signal to steer LLM-based test generation towards higher mutation scores. 
Complementary work has also begun to examine the quality of LLM-generated tests beyond Java and from a broader quality-assessment perspective: Alves et al.~\cite{alves2025quality} analyse Python tests generated by LLMs, assessing structural quality and test smells, while Melo et al.~\cite{melo2025agentic} explore agentic LMs that actively hunt down test smells in existing test suites.

Overall, existing work on LLM-based test generation has mainly focused on functional adequacy, such as compilability, coverage, mutation score, and defect detection, with structural quality typically treated as one dimension among several. For example, Siddiq et al.~\cite{siddiq2024using} and Ouédraogo et al.~\cite{ouedraogo2024large} also report test smells when comparing LLM-generated tests with EvoSuite, but smells are not the primary focus of their analyses, and they do not study smell distributions or co-occurrence patterns at scale, nor their similarity to human-written test suites. Closer to our focus, Alves et al.~\cite{alves2025quality} evaluate the quality and smells of LLM-generated Python tests, while a complementary line of work investigates test smell detection itself, using either (small) language models~\cite{lucas2025investigating,melo2025agentic} or machine-learning-based detectors~\cite{pontillo2024machine}. These studies focus on detection capability and local quality assessment rather than on large-scale diffusion patterns across generators and benchmarks. Building on this line of work, we make test smells a central object of study: we analyse the diffusion of a broad smell catalogue in LLM-generated tests, and we contrast their smell profiles and co-occurrences with those observed in large human-written and EvoSuite-generated test suites.

\textit{This paper presents a multi-benchmark, large-scale empirical study of test smell diffusion in LLM-generated Java unit tests}. Our contribution is threefold: (1) we evaluate LLM-generated tests across diverse models, prompts, and benchmarks; (2) we analyze both class- and method-level test generation; and (3) we conduct complementary comparisons with SBST-generated tests (to assess automated paradigms) and human-written tests (to evaluate originality and alignment with real-world practices, and to explore smell-profile similarity as an indirect, correlational signal of possible training-data overlap or stylistic imitation).

We leverage two complementary benchmarks. 
Benchmark~1 targets class-level generation on three established datasets: Defects4J~\cite{just2014defects4j}, SF110~\cite{fraser2014large}, and CMD~\cite{ouedraogo2024llms}, and comprises 20,505 test suites generated by four LLMs (GPT-3.5, GPT-4, Mistral 7B, Mixtral 8×7B) using five structured prompting strategies, together with 14,469 EvoSuite-generated tests on the same set of classes~\cite{ouedraogo2024llms}. Benchmark~2 is TestBench~\cite{zhang2024testbench}, a method-level benchmark that includes 972 LLM-generated test cases from three LLMs (GPT-3.5, GPT-4, and CodeLlama-13B-Instruct) for 108 real-world functions from nine Java projects under varying contextual conditions. Across both benchmarks, we compare these generated tests against 779,585 human-written tests from 34,635 Java projects (CAT-LM dataset~\cite{rao2023cat}, SF110, and Defects4J), which serve as our real-world reference.

For methodological robustness, we employ two independent test smell detection tools—TsDetect~\cite{peruma2020tsdetect} and JNose~\cite{virginio2020jnose}—enabling cross-validation and reducing tool-specific bias. We analyze how software characteristics (e.g., KLOC (thousands of lines of code), Coupling Between Objects (CBO), Response For a Class (RFC), and Cyclomatic Complexity (CC)) and LLM parameters influence smell diffusion using multiple statistical measures (Pearson/Spearman correlations, Mutual Information (MI), Kruskal–Wallis tests).
Our analysis focuses on Java and JUnit-style unit tests. This choice is motivated by the availability of mature, widely used Java benchmarks (e.g., Defects4J~\cite{just2014defects4j}, SF110~\cite{fraser2014large}) that combine human-written tests, SBST-generated tests, and code amenable to LLM-based generation, as well as by the presence of two independent test-smell detectors (TsDetect~\cite{peruma2020tsdetect} and JNose~\cite{virginio2020jnose}) that specifically target Java/JUnit tests. Although many test smells we consider (such as Assertion Roulette, General Fixture, and Lazy Test) have analogues in other ecosystems, differences in language and framework idioms (e.g., pytest\footnote{\url{https://github.com/pytest-dev/pytest/}} fixtures, NUnit\footnote{\url{https://nunit.org/}} parameterised tests, Jest snapshots\footnote{\url{https://jestjs.io/docs/snapshot-testing}}) mean that our findings should be read as evidence about LLM-generated Java tests rather than as general statements about all languages. Understanding whether similar diffusion patterns emerge in other ecosystems is beyond the scope of this study.

Our aim is to systematically characterise how test smells diffuse in LLM-generated Java tests across models, prompting strategies, and generation granularities, and to compare these diffusion patterns to SBST-generated and human-written suites.

\vspace{0.2cm} \noindent\textbf{Contributions:} 
\begin{itemize} 

    \item \textbf{Multilevel Analysis of Test Smell Diffusion:} We quantify and compare the prevalence, distribution, and co-occurrence of test smells in LLM-generated Java test suites, analyzing both class-level and method-level benchmarks, combining large-scale automated detection with a manually validated sample to assess detector reliability.

    \item \textbf{Prompt Engineering and Context Effects:} We examine how structured prompting techniques and contextual variations affect smell patterns in LLM-generated tests, providing insights into the role of prompt design in test quality.
    
    \item \textbf{Comparison with SBST (RQ1–RQ3):} We contrast LLM-generated tests with EvoSuite-generated tests to investigate differences between modern LLM-based and traditional SBST-based test generation paradigms.

    \item \textbf{Comparison with Human-Written Tests (RQ4):} We compare LLM-generated tests with 779k+ human-written tests to evaluate originality, alignment with real-world testing practices, and to explore smell-profile similarity as a correlational indicator of possible training-data overlap or stylistic imitation, without making claims about strict memorization.

    \item \textbf{Cross-Tool Validation:} We leverage both TsDetect and JNose to triangulate smell detection, quantify divergences, and ensure robustness across independent implementations.
    
    \item \textbf{Reproducibility:} All artifacts, including generated tests, smell detection results, and analysis scripts, are publicly available\footnote{\url{https://github.com/Cwendkuuni/LLMTSDiff}} to support full replication. 

\end{itemize}

Following prior work on the diffusion and diffuseness of test smells in automatically generated tests~\cite{palomba2016diffusion,grano2019scented}, we use the term \emph{diffusion of test smells} in a purely cross-sectional sense. In this paper, diffusion denotes how widely and in what combinations different test smells occur across test suites, projects, and generation configurations, rather than any temporal propagation or evolution over time. Throughout the paper, we therefore use “diffusion” as a shorthand for the prevalence, distribution, and co-occurrence of smells in large corpora of LLM-, SBST-, and human-generated tests.
The remainder of the paper is structured as follows: \Cref{background} introduces key concepts and related work. \Cref{setup} details our experimental design and datasets. \Cref{results} presents findings from each analysis. \Cref{discussion} addresses implications and threats. \Cref{relatedwork} reviews prior work, and \Cref{conclusion} concludes.


\section{Background and Related Work}
\label{background}

This section provides the foundational background (definitions of test smells, detection tooling, and the software characteristics we use as contextual variables) and a concise review of related work on human-written, SBST-generated, and LLM-based unit tests, as well as data-leakage risks in LLM evaluation. We conclude by positioning our study and clarifying how our design addresses gaps in smell coverage, detector reliability, and cross-granularity analysis.

\subsection{Test Smells and Unit Testing}
\label{subsec:test-smell-unit-testing}

Test smells are design flaws in test code that negatively impact its maintainability, readability, and reliability~\cite{bavota2012empirical,spadini2018relation,garousi2018smells,junior2020survey,bavota2015test}. While these smells can be introduced by developers, they also frequently appear in automatically generated tests~\cite{bavota2015test,spadini2018relation}. The presence of test smells increases the long-term maintenance cost of test suites and complicates debugging when test failures occur.~\cite{palomba2016diffusion,peruma2019distribution}.
Consistent with this terminology, when we later refer to the \emph{diffusion} of a given smell, we mean its prevalence and spread across the large corpora of tests considered in our study (LLM-, SBST-, and human-generated), not a temporal process.

To further illustrate these test smells, we present two representative examples in Listings~\ref{lst:assertion-roulette} and~\ref{lst:magic-number}, showcasing how they manifest in practice. 
The first example (Listing~\ref{lst:assertion-roulette}) shows an instance of the \emph{Assertion Roulette} smell in a JUnit test from an open-source project. Lines~5,~7, and~8 contain three independent assertions on different aspects of the repository state, all grouped into a single test method and written without descriptive messages. \emph{The issue is not the sheer number of assertions, but the fact that they exercise distinct behaviours without any diagnostic messages.} If any of these assertions fails, the failure message does not reveal which behaviour broke, making the test harder to debug and maintain; this lack of diagnosability is exactly what TsDetect and JNose capture when they report this test as Assertion Roulette.

\vspace{3mm}
\begin{lstlisting}[style=test-smell-listing, caption=Example of Assertion Roulette Test Smell, label=lst:assertion-roulette]
@MediumTest
public void testCloneNonBareRepoFromLocalTestServer() throws Exception {
    Clone cloneOp = new Clone(false, integrationGitServerURIFor("small-repo.early.git"), helper().newFolder());
    Repository repo = executeAndWaitFor(cloneOp);
    assertThat(repo, hasGitObject("ba1f63e4430bff267d112b1e8afc1d6294db0ccc"));
    File readmeFile = new File(repo.getWorkTree(), "README");
    assertThat(readmeFile, exists());
    assertThat(readmeFile, ofLength(12));
}
\end{lstlisting}

The second example (Listing~\ref{lst:magic-number}) illustrates the \emph{Magic Number} smell. Line~3 passes the literal value \texttt{15.5D} to \texttt{getLocalTimeAsCalendar}, and lines~4--5 assert against the hard-coded constants \texttt{15} and \texttt{30}. These unnamed numerical literals encode domain knowledge directly in the test, without explanatory constants or comments, which obscures the intent of the test and makes it harder to adapt if the underlying business rules change.



\vspace{3mm}
\begin{lstlisting}[style=test-smell-listing, caption=Example of Magic Number Test Smell, label=lst:magic-number]
@Test
public void testGetLocalTimeAsCalendar() {
    Calendar localTime = calc.getLocalTimeAsCalendar(BigDecimal.valueOf(15.5D), Calendar.getInstance());
    assertEquals(15, localTime.get(Calendar.HOUR_OF_DAY));
    assertEquals(30, localTime.get(Calendar.MINUTE));
}
\end{lstlisting}

\vspace{3mm}
These smells represent core obstacles to test comprehension and evolution. For further details and additional examples of test smells, refer to the TsDetect documentation~\footnote{\url{https://testsmells.org/pages/testsmellexamples.html}}~\cite{peruma2020tsdetect}. 
Detecting such smells has been the subject of extensive research, leading to several automated detectors. In our study, we rely on two widely used tools: TsDetect~\cite{peruma2020tsdetect} and JNose~\cite{virginio2020jnose}. However, relying on a single detector risks bias due to tool-specific heuristics or limitations in smell coverage. Recent work by Panichella et al.~\cite{panichella2022test} highlights concerns about the detectability, validity, and reliability of test smell detectors, emphasizing the need for cross-tool validation to enhance confidence in empirical results. Similarly, the systematic mapping study by Aljedaani et al.~\cite{aljedaani2021test} shows that detection tools often overlap only partially in the types of smells they support. This motivates our decision to use both detectors, which provide complementary coverage and granularity.

In this study, we focus on the most frequently observed test smells across LLM-generated, SBST-generated, and human-written tests. Table~\ref{tab:test-smells-overview} presents a summary of the most frequent smells identified in our analysis, along with their descriptions and the impact they have on test quality.

\begin{table}[ht]
\centering
\caption{Most Frequent Test Smells Observed in Our Study}
\label{tab:test-smells-overview}
\scalebox{0.75}{
\begin{tabular}{ll}
\toprule
\textbf{Test Smell} & \textbf{Description and Impact} \\
\midrule
\textbf{Assertion Roulette (AR)} & Multiple assertions without messages hinder fault diagnosis. \\ 

\textbf{Magic Number Test (MT)} & Numeric literals in assertions reduce clarity and increase maintenance. \\ 

\textbf{Empty Test (EmT)} & Tests without executable statements may falsely pass and hide coverage gaps. \\

\textbf{Exception Handling (EH)} & Manual exception logic disrupts automation; use framework support instead. \\

\textbf{Sensitive Equality (SE)} & Reliance on \texttt{toString()} causes brittle tests prone to break. \\

\textbf{Eager Test (EaT)} & Tests that call multiple production methods are harder to isolate and comprehend. \\

\textbf{Lazy Test (LT)} & Multiple tests for one method indicate redundancy and lack of focus. \\

\textbf{Unknown Test (UT)} & Tests without assertions lack observable behavior and reduce diagnostic power. \\

\textbf{Duplicate Assert (DA)} & Repeated assertions suggest poor modularization and dilute test intent. \\

\textbf{General Fixture (GF)} & Unused setup objects increase maintenance cost and slow execution. \\

\textbf{Conditional Logic Test (CLT)} & Control-flow in tests reduces determinism and hampers localization~\cite{peruma2020tsdetect}. \\
\bottomrule
\end{tabular}
}
\end{table}


\subsection{Software Characteristics}

Understanding software structure is essential when analyzing the prevalence of test smells. 
Prior studies have demonstrated that certain structural and object-oriented metrics serve as strong indicators of fault-proneness and maintenance complexity in software systems~\cite{jureczko2011significance, chidamber1994metrics, rebro2023source}. 
These characteristics have also been extensively used in defect prediction tasks~\cite{gyimothy2005empirical,jureczko2011significance,rebro2023source} and shown to correlate with software quality and test maintainability~\cite{briand2000exploring, briand2001controlled, bagheri2011assessing}. 
In particular, prior studies on test smells and test quality have linked smell incidence to broad system- and test-level characteristics such as size, age, and the amount of test code~\cite{bavota2015test,palomba2016diffusion,peruma2019distribution}, and have begun to exploit structural and textual metrics within detection techniques and machine-learning models~\cite{peruma2019distribution,pontillo2024machine}. Building on this line of work, and on the broader literature connecting object-oriented metrics (e.g., LOC, CBO, RFC, DIT, cyclomatic complexity) to fault-proneness and maintainability~\cite{jureczko2011significance,chidamber1994metrics,briand2000exploring,briand2001controlled,bagheri2011assessing}, we focus on a small set of size, complexity, and coupling metrics that are widely supported by analysis tools (e.g., \texttt{ck}, Lizard) and can be computed consistently across all projects in our benchmarks.
Building on these findings, we concentrate on a set of widely used structural characteristics (size, control-flow complexity, coupling, and inheritance depth) together with test scope, which we hypothesise to influence the likelihood and type of test smells:

\begin{itemize}
    \item \textbf{Lines of Code (LOC):} Larger codebases tend to exhibit more complexity and maintenance challenges~\cite{rebro2023source}. Such systems often require more extensive tests, increasing the chance of smells like \textit{General Fixture (GF)} or \textit{Lazy Test (LT)} due to oversized test setups.
    
    \item \textbf{Cyclomatic Complexity (CC):} High control-flow complexity~\cite{mccabe1976complexity} can encourage developers or automated tools to write tests with multiple assertions or branches, raising the likelihood of smells such as \textit{Assertion Roulette (AR)} or \textit{Conditional Logic Test (CLT)}.
    
    \item \textbf{Coupling Between Objects (CBO):} Strong coupling between classes~\cite{chidamber1994metrics} makes test isolation harder and may foster smells such as \textit{Eager Test (EaT)} or \textit{General Fixture (GF)} because tests must instantiate and configure many collaborators.
    
    \item \textbf{Response For a Class (RFC):} A high RFC~\cite{chidamber1994metrics} implies many possible execution paths, which may push test generators toward breadth over focus. This can yield \textit{Duplicate Assert (DA)} or \textit{Lazy Test (LT)} smells as tests attempt to cover numerous interactions in one place.
    
    \item \textbf{Depth of Inheritance Tree (DIT):} Deep hierarchies~\cite{chidamber1994metrics} can introduce subtle dependencies that are difficult to isolate, leading to smells such as \textit{Unknown Test (UT)} or brittle \textit{Exception Handling (EH)} logic in generated tests.
    
    \item \textbf{Test Scope (Method-level vs. Class-level):} The granularity of test generation directly affects smell patterns. Prior work has shown that class-level SBST (e.g., EvoSuite) tends to produce tests with frequent smells such as \textit{Assertion Roulette} and \textit{Eager Test}~\cite{palomba2016diffusion}, while recent studies on method-level LLM-based test generation report fewer structural issues~\cite{siddiq2024using}. Building on these observations, our study systematically analyzes how test smells vary across scopes.

\end{itemize}

By connecting these software characteristics to potential test smell manifestations, we provide the rationale for analyzing them in conjunction with our empirical results.

\subsection{Human-Written, Automated, and LLM-Based Unit Tests}
\label{subsec:human-automated-llmbased}

Human-written tests have been extensively analyzed for the presence of test smells, particularly with regard to their impact on test comprehension, maintainability, and effectiveness.
Bavota et al.\cite{bavota2015test} showed that smells like Assertion Roulette and Magic Number Test are common in JUnit tests, impacting comprehension and maintainability. While these tests reflect domain knowledge and are generally more readable, they are time-consuming and often miss edge cases\cite{daka2014survey,kochhar2013empirical,kochhar2013adoption,bavota2012empirical}. Datasets like Defects4J~\cite{just2014defects4j} and SF110~\cite{fraser2014large} highlight common issues like test smells and incomplete coverage~\cite{spadini2018relation,bacchelli2008effectiveness}.

In contrast to human-written test cases, tools like EvoSuite~\cite{fraser2011evosuite} automate test generation by focusing on maximizing code coverage. However, as Palomba et al.~\cite{palomba2016diffusion} showed, EvoSuite-generated tests are also prone to test smells such as \textit{Assertion Roulette} and \textit{Eager Test}, raising concerns about their maintainability. More recently, studies have applied LLMs to unit test generation, reporting promising diversity and readability, yet test smells have generally been treated as a secondary aspect rather than a primary focus~\cite{siddiq2024using,ouedraogo2024large}.

Beyond human-written and SBST-generated tests, a growing body of work evaluates LLM-based test generation from a quality and adequacy perspective. Yuan et al.~\cite{yuan2023no} study ChatGPT-generated tests in terms of syntactic and execution correctness, coverage, and perceived readability and usability. Siddiq et al.~\cite{siddiq2024using} and Ouédraogo et al.~\cite{ouedraogo2024large} further compare multiple LLMs against EvoSuite on HumanEval, SF110, Defects4J, and CMD, jointly analysing compilation, coverage, readability and understandability, and a catalogue of test smells in LLM-generated tests. Other studies, such as Tang et al.~\cite{tang2024chatgpt} and Yang et al.~\cite{yang2024evaluation}, contrast LLM-based generation with SBST tools using coverage and defect-detection metrics, while TestBench~\cite{zhang2024testbench} and MuTAP~\cite{dakhel2024effective} integrate mutation testing to assess or guide the fault-detection capability of LLM-generated suites. In contrast to these adequacy-oriented evaluations, our work treats test smells as a primary object of study and focuses on their diffusion and distribution across LLM-generated, SBST-generated, and human-written test suites, thus addressing structural quality in a dimension that is largely orthogonal to coverage and mutation.

Despite advances in prompt engineering, such as techniques like Chain-of-Thought (CoT)~\cite{wei2022chain} and Tree-of-Thought (ToT)~\cite{yao2024tree}, LLMs still encounter challenges in achieving the quality of manually written tests, particularly in terms of maintaining clarity and minimizing test smells~\cite{siddiq2024using, ouedraogo2024llms}. 
Recent studies further highlight that LLM-based test generation can suffer from instability across generations, overfitting to benchmarks, and sensitivity to prompt variations~\cite{xu2024benchmarking, he2024unitsyn, hartmann2023sok}. Although advanced prompting strategies can mitigate some issues, achieving comparable coverage, readability, and maintainability remains challenging. This gap is critical, as it directly impacts the long-term reliability of LLM-generated tests. Our study addresses this gap by systematically analyzing how prompt engineering shapes test smell prevalence and maintainability.

\noindent\textbf{Comparison.} 
Traditional SBST tools such as EvoSuite excel at maximizing structural coverage~\cite{fraser2011evosuite}, 
but the resulting tests often suffer from rigidity and poor readability, with smells like \textit{Assertion Roulette} and \textit{Eager Test} being especially common~\cite{palomba2016diffusion}. 
In contrast, LLM-based approaches have shown the ability to produce tests that appear stylistically closer to human-written code, offering greater diversity and generalizability in terms of test scenarios~\cite{siddiq2024using, ouedraogo2024large}. 
However, they also exhibit distinct weaknesses, including instability across generations, redundant or verbose patterns, and the emergence of smell types that are less prevalent in SBST or human-written suites. 
For example, while prior studies have shown that Assertion Roulette and Magic Number Test dominate in human-written tests~\cite{bavota2015test, spadini2018relation, peruma2019distribution}, 
and Assertion Roulette and Eager Test are most frequent in EvoSuite-generated tests~\cite{palomba2016diffusion}, 
LLM-based generation more often introduces smells such as \textit{General Fixture (GF)}, \textit{Duplicate Assert (DA)}, or \textit{Lazy Test (LT)}.
A key distinction is that, unlike SBST, LLMs are not explicitly optimized for coverage. 
Instead, they generate tests that often appear more ``human-like,'' which alters the distribution of smells observed and motivates a systematic investigation of these differences.

\subsection{Data Leakage in LLMs}
\label{subsec:dataleakage-llms}

Data leakage in LLMs refers to cases where models reproduce memorized patterns from their training data, rather than generating novel solutions. In test generation, this can manifest as recurring smells such as \textit{Assertion Roulette} or \textit{Magic Number Test}, which may be inherited from benchmarks like Defects4J or SF110~\cite{xu2024benchmarking,he2024unitsyn,siddiq2024using}. Such memorization reduces the effectiveness and maintainability of generated tests, and may also cause benchmark leakage—where models appear to perform well on datasets overlapping with their training data, thus inflating evaluation results~\cite{sallou2023breaking,hartmann2023sok}. Distinguishing genuine reasoning ability from pattern recall is therefore challenging, underscoring the need for careful evaluation when analyzing LLM-generated tests.

To address these concerns, frameworks like UniTSyn emphasize the importance of careful evaluation to mitigate the impact of data leakage~\cite{he2024unitsyn}. Our study builds on these insights by comparing test smell patterns between LLM-generated and human-written tests, investigating whether LLMs inadvertently replicate patterns from their training data, thus introducing biases and affecting test quality. \\

\noindent\textbf{Positioning.} Taken together, prior studies highlight four open issues: (i) partial coverage and limited reliability of existing smell detectors, (ii) the influence of test scope (method- vs class-level) on smell prevalence, 
(iii) the lack of systematic analyses of test smells in LLM-generated tests compared to SBST and human-written suites, and (iv) the limited understanding of how closely smell profiles of LLM-generated tests align with those of human-written suites, and what such distributional similarity implies about originality and potential training-data overlap. 
Our study addresses these issues by combining two complementary detectors, analyzing both SBST- and LLM-generated tests at multiple granularities, and comparing them against a large corpus of human-written tests, treating smell-profile similarity as a correlational signal that is compatible with possible training-data overlap or shallow imitation of common testing idioms, but not as direct evidence of data leakage.

\section{Experimental Setup}
\label{setup}
\subsection{Approach Overview}
\label{subsec:approach-overview}

\begin{figure}[th]
  \centering
  \includegraphics[width=0.8\textwidth]{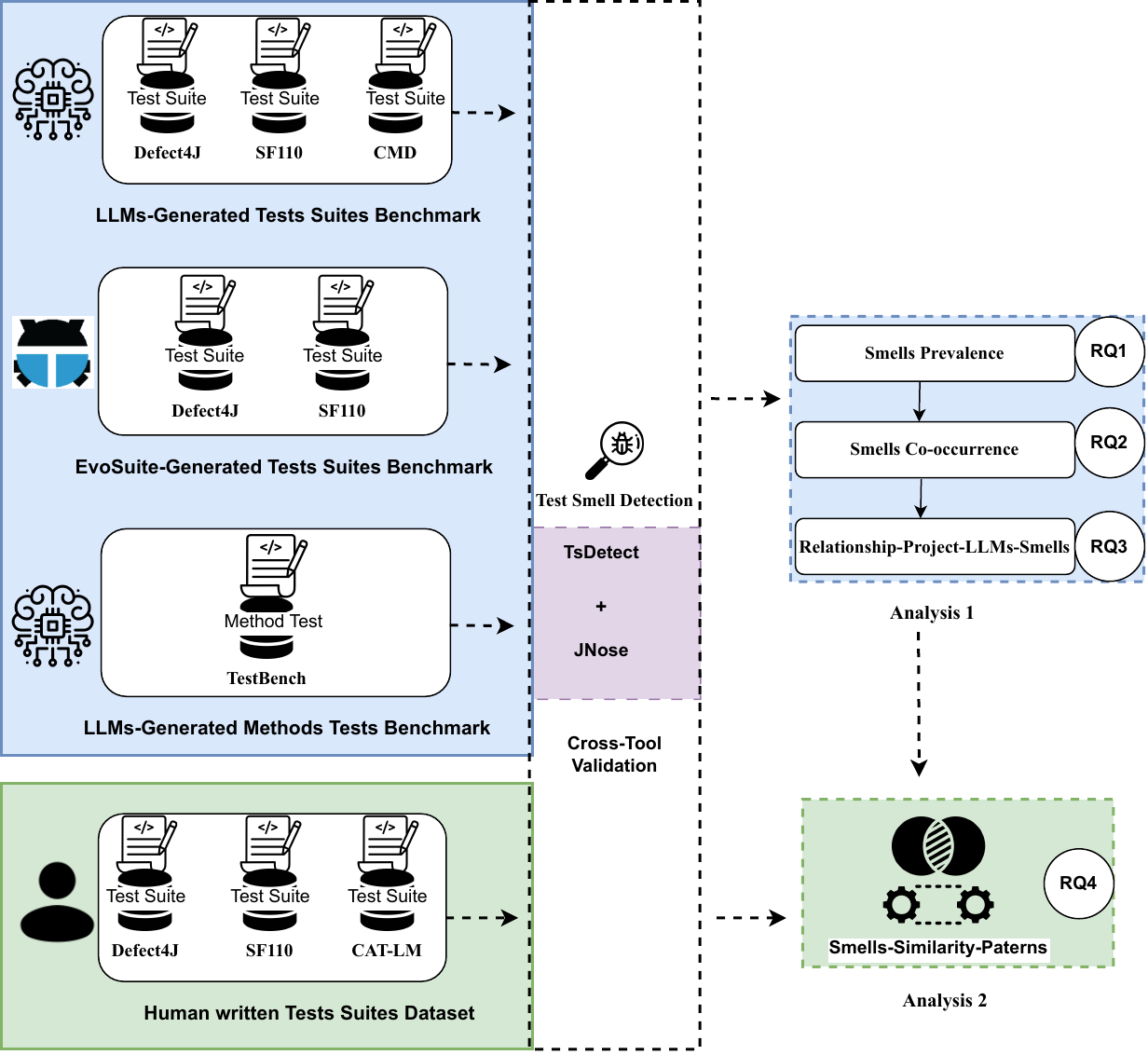} 
   \caption{Overview of the pipeline used to design the analysis. }
    \label{fig:analysis-overview}
\end{figure}

Our methodology, illustrated in Figure~\ref{fig:analysis-overview}, investigates the diffusion of test smells in LLM-generated test suites and contrasts them with SBST-generated and human-written counterparts. We leverage two benchmarks that are complementary in scope: Benchmark~1~\cite{ouedraogo2024llms} targets class-level test generation and analyzes the influence of five distinct prompt engineering strategies across multiple LLMs and datasets, whereas Benchmark~2~\cite{zhang2024testbench} focuses on method-level generation and studies the impact of prompt context by varying the amount of information provided to the model. While their generation settings differ, the two benchmarks offer complementary insights into how generation control—via prompting at the class level or contextual variation at the method level—shapes test smell profiles. In addition, we incorporate over 779,000 human-written test cases from 34,635 projects and 14,469 EvoSuite-generated test suites, allowing us to jointly consider human-written, SBST-generated, and LLM-generated tests in our assessments.

We conduct two primary analyses. In \textbf{Analysis 1}, we investigate how prompt strategies and generation settings influence the emergence of test smells in LLM-generated suites, comparing their prevalence and co-occurrence with EvoSuite’s SBST-generated tests. This analysis incorporates software attributes such as cyclomatic complexity and class coupling, as well as LLM-specific parameters like input size and context length. To uncover both linear and non-linear relationships, we apply Pearson correlation, Kruskal-Wallis, Spearman, and Mutual Information.

In \textbf{Analysis 2}, we compare LLM-generated suites with human-written ones to examine both the prevalence and distribution of test smells as well as their similarity. This dual perspective allows us to characterize how LLM-generated smell profiles align or diverge from those observed in human-written tests. Since LLMs are trained on human-produced code, human-written tests represent the most plausible source of inherited patterns. Beyond descriptive comparison, this analysis also investigates potential data leakage—whether LLMs inadvertently replicate test smells present in human-written tests, rather than producing them as novel generative artifacts. To this end, we measure smell prevalence and type distributions, assess overlap using Cosine Similarity and Jaccard Index, and evaluate frequency distribution differences with Euclidean and Hellinger Distances. Additionally, we use Pearson Correlation, Kruskal-Wallis, Spearman, and Mutual Information to analyze co-occurrence patterns and detect possible replication of human test behaviors in LLM-generated tests.

To ensure the robustness of our findings, we adopt a triangulated analysis strategy~\cite{runeson2009guidelines,kitchenham2002preliminary,flick2018designing,yin2017case}. We apply \textit{tool triangulation} via two independent smell detectors (TsDetect and JNose), \textit{data triangulation} using two complementary benchmarks—Benchmark~1 (class-level, varied prompting) and Benchmark~2 (method-level, contextual variation)—and \textit{analytical triangulation} by combining linear and non-linear statistics. These benchmarks offer distinct yet complementary insights into how generation control affects test smell profiles. We present results per tool, identify key trends, and discuss divergences in light of tool-specific heuristics and detectability concerns, as highlighted by Panichella et al.~\cite{panichella2022test}.

\subsection{Research Questions}
\label{subsec:research-questions}

\begin{itemize}[leftmargin=*]
\item \textbf{RQ1: What is the prevalence and distribution of test smells in LLM-generated unit tests, and how do they compare to tests generated by SBST?}

This research question assesses the presence of test smells in LLM-generated tests across both class-level (Benchmark~1) and method-level (Benchmark~2) generation. We explore: 
\ding{182} \emph{Prevalence}: how frequently test smells appear in LLM-generated test cases; 
\ding{183} \emph{Distribution}: the distribution of test smell types at class and method levels, considering both LLM- and SBST-generated tests, and how generation strategies influence them; \ding{184} \emph{Comparison with EvoSuite}: how LLM-generated test smells differ from those found in tests produced by SBST tools such as EvoSuite; and 
\ding{185} \emph{Detector reliability}: a manual validation of frequent and detector-divergent smells to estimate the precision and false-positive rate of automated detectors in this setting. This analysis aims to provide a comprehensive understanding of test smell diffusion across different test generation paradigms and levels of granularity.

\item \textbf{RQ2: How do test smells co-occur in LLM-generated test suites, and how do these patterns compare to tests generated by SBST?}

This research question investigates: \ding{182} \emph{Influence of prompt engineering techniques}: how structured prompting affects test smell prevalence and interaction; \ding{183} \emph{Effect of context variation}: whether the amount or structure of contextual information impacts test smell emergence;
\ding{184} \emph{Co-occurrence patterns}: whether certain test smells frequently appear together, and how these patterns differ between class-level and method-level generation; \ding{185} \emph{Comparison with SBST-generated tests}: whether LLM-generated tests exhibit distinct co-occurrence trends compared to SBST-generated tests. By identifying these trends, we aim to uncover structural dependencies and generative biases that may hinder test suite readability, maintainability, or fault localization.

\item \textbf{RQ3: How do software attributes and LLM parameters influence the prevalence of test smells?}

This research question investigates \ding{182} {\em software characteristics}: how project size (LOC, number of methods, classes), complexity metrics (cyclomatic complexity, class coupling), and test scope (method-level vs. class-level) impact test smell prevalence; \ding{183} {\em LLM parameters}: the influence of model size, training parameters, context window size, and prompting techniques on test smell patterns; and \ding{184} {\em comparative analysis}: how these factors affect test smells differently between LLM-generated and EvoSuite-generated tests. By exploring these relationships, we aim to understand how software attributes and generative parameters contribute to the emergence of test smells, ultimately informing more effective and targeted strategies for both LLM-based and SBST-based test generation.

\item \textbf{RQ4: Do LLM-generated tests exhibit similar test smells to human-written tests?}

This research question investigates \ding{182} {\em the extent of overlap in test smell types and frequencies} between LLM-generated and human-written tests, and \ding{183} {\em to what extent such similarities are consistent with possible training-data overlap or shallow stylistic imitation}, sometimes discussed under the umbrella of data leakage, where LLMs may unintentionally reproduce patterns from test suites seen during pre-training. This analysis aims to assess the originality and generalizability of LLM-generated test code, while explicitly treating smell-profile similarity as a correlational signal rather than direct evidence of memorization.

\end{itemize}

\subsection{Datasets and benchmarks}
\label{subsec:dataset}

Our study considers the following datasets and benchmarks: 
\textbf{Benchmark~1}~\cite{ouedraogo2024llms}, a class-level evaluation with LLM- and SBST-generated test suites across multiple datasets and prompting strategies; 
\textbf{Benchmark~2}~\cite{zhang2024testbench}, a method-level benchmark exploring different prompt context levels; 
in addition to a large corpus of human-written test suites collected from Defects4J, SF110, and Cat-LM~\cite{rao2023cat}. 
Together, these resources provide LLM-, SBST-, and human-generated tests for our comparative analyses.

\subsubsection{Prompting strategies and contexts.}
\label{subsec:prompt-strategies}
In this study, we analyse the LLM-generated tests produced by two existing benchmarks. For Benchmark~1, Ouedraogo et al.~\cite{ouedraogo2024llms} follow prior work on practical and structured unit test generation~\cite{siddiq2024using,yuan2023no,chen2023chatunitest,tang2024chatgpt} and define prompts using a standardised template with two components: a Natural Language Description (NLD) and a \emph{Source Code} placeholder. The NLD combines (i) a role-playing directive that asks the model to act as a professional tester~\cite{tang2024chatgpt,chen2023chatunitest}, (ii) a task-specific instruction to generate a JUnit~4 test file, and (iii) delimiters (e.g., \texttt{\#\#\#Test START\#\#\#}, \texttt{\#\#\#Test END\#\#\#}) that separate test code from explanatory text. The source-code part then injects either only the class under test or, in Few-Shot settings, an additional example class with its test suite, following earlier learning-based unit test generation approaches~\cite{tang2024chatgpt,chen2023chatunitest}. Building on recent evaluations of prompt engineering for test generation~\cite{yuan2024evaluating,yang2024evaluation}, Benchmark~1 instantiates this template with five techniques—Zero-Shot Learning (ZSL), Few-Shot Learning (FSL), Chain-of-Thought (CoT), Tree-of-Thought (ToT), and Guided Tree-of-Thought (GToT)—to vary the amount of guidance (no example vs.\ 1--2 examples) and the form of reasoning support (linear CoT vs.\ multi-branch ToT and collaborative GToT).

For Benchmark~2, we reuse the tests generated in TestBench~\cite{zhang2024testbench}, which systematically vary the amount of surrounding code provided to the LLM through three prompt contexts: Self-Contained Context (only the target function’s signature and body), Simple Context (the full class structure without method bodies), and Full Context (the entire class including the target method). Zhang et al.\ motivate these variants as reflecting common developer behaviours when interacting with LLM assistants, where a developer may paste only the function of interest, a structurally simplified class, or the full file. Similar patterns (combining a base testing instruction with more or less contextual information) also appear in recent work on retrieval-augmented test generation~\cite{shin2024retrieval}.

Our use of these two benchmarks is thus grounded in prompt designs that have already been justified in the literature as both methodologically sound and representative of realistic developer--LLM interactions for unit test generation.

\subsubsection{Benchmark 1}
\label{subsec:benchmark-1}
\textbf{Benchmark 1}~\cite{ouedraogo2024llms} targets class-level unit test generation and evaluates the impact of prompt engineering across multiple LLMs and datasets. It comprises 20,505 test suites generated using four models—GPT-3.5, GPT-4, Mistral 7B, and Mixtral 8x7B—on three datasets: Defects4J, SF110, and CMD (Table~\ref{tab:datasets-benchmark-1-2}).

\begin{table}[!htbp]
\centering
\caption{Datasets Used in Benchmark 1 and 2}
\label{tab:datasets-benchmark-1-2}
\resizebox{\columnwidth}{!}{
\begin{tabular}{llll}

\toprule
\textbf{Dataset} & \textbf{Description} & \textbf{\#Proj} & \textbf{\#Classes / Methods} \\ 
\midrule
SF110 & Java programs from academia and industry for SBST research. & 74 & 182 classes \\ 
Defects4J & Real-world bug-fix scenarios exposing faults. & 16 & 477 classes \\ 
CMD & Curated modern open-source Java repos (e.g., OceanBase). & 2 & 31 classes \\ 
TestBench & 108 Java methods from 9 open-source projects across domains. & 9 & 108 methods \\
\bottomrule
\end{tabular}
}
\end{table}

The CMD dataset includes modern, production-grade Java systems (e.g., Conductor OSS\footnote{\url{https://github.com/conductor-oss/conductor}}, OceanBase Developer Center (ODC)\footnote{\url{https://github.com/oceanbase/odc}}), complementing Defects4J and SF110 with recent modular codebases. Due to resource constraints, GPT-4 and Mixtral 8x7B were only applied to CMD, while EvoSuite was excluded from CMD due to Java compatibility issues~\footnote{\url{https://github.com/EvoSuite/evosuite/issues/433}}.

Each model was prompted using five structured strategies: ZSL, FSL, CoT, ToT, and GToT (Table~\ref{tab:prompt-strategies}). For comparison, 14,469 EvoSuite-generated test suites were used as a structural testing baseline.

\begin{table*}[!htbp]
\centering
\caption{Overview of Prompting Strategies and Contexts Used in Benchmark 1 and Benchmark 2}
\label{tab:prompt-strategies}
\resizebox{\textwidth}{!}{
\begin{tabular}{@{}l c p{7cm} p{7cm}@{}}
\toprule
\textbf{Strategy / Context} & \textbf{Benchmark} & \textbf{Description} & \textbf{Objective} \\
\midrule
Zero-Shot Learning (ZSL) & B1 & No examples or context; the model directly generates tests. & Serves as baseline to evaluate raw test generation ability. \\
Few-Shot Learning (FSL) & B1 & Provides 1-2 examples before test generation. & Helps the model infer correct structure and semantics of test code. \\
Chain-of-Thought (CoT) & B1 & Generates test logic through step-by-step reasoning. & Improves assertion clarity and logical flow. \\
Tree-of-Thought (ToT) & B1 & Explores multiple reasoning branches before generating tests. & Encourages structural diversity and robustness. \\
Guided Tree-of-Thought (GToT) & B1 & Simulates discussion among multiple “testers” refining test generation iteratively. & Aims to improve correctness, coverage, and variability. \\
Self-Contained Context (SCC) & B2 & Only the target function’s signature and body are provided. & Evaluates base-level capability without class context. \\
Simple Context (SC) & B2 & Full class structure excluding function bodies. & Preserves structural information with reduced length. \\
Full Context (FC) & B2 & Entire class including target method is given. & Tests whether full context leads to more accurate and meaningful test cases. \\
\bottomrule
\end{tabular}
}
\end{table*}

All LLMs used a temperature of 0.7, with model-specific token limits. Each class underwent 30 generation attempts. EvoSuite was executed using the DynaMOSA algorithm~\cite{panichella2017automated}, with 3 minutes per class and 30 iterations. Importantly, EvoSuite's post-processing—designed to reduce flaky tests and refactor away certain test smells—was deliberately disabled to ensure a fair comparison with the raw outputs of LLMs.

\begin{table*}[!htbp]
\centering
\caption{Overview of Large Language Models (LLMs) and SBST Tools Used Across Benchmarks}
\label{tab:llms-tools-overview}
\resizebox{\textwidth}{!}{
\begin{tabular}{@{}l c l c p{7cm}@{}}
\toprule
\textbf{Model / Tool} & \textbf{Benchmark(s)} & \textbf{Size} & \textbf{Type} & \textbf{Strengths} \\
\midrule
GPT-3.5-Turbo           & B1, B2 & $\sim$175B (estimated) & Proprietary (OpenAI) & Widely used for LLM-based test generation; good balance of cost, performance, and generalization. \\
GPT-4                  & B1, B2 & $>$175B & Proprietary (OpenAI) & Stronger reasoning, coherence, and test intent alignment; higher generation cost. \\
Mistral 7B             & B1     & 7B & Open-source (Mistral) & Lightweight and efficient; less capable for multi-step reasoning in test generation. \\
Mixtral 8x7B           & B1     & 8×7B (MoE) & Open-source (Mistral) & Mixture-of-Experts improves generation diversity with selective routing. \\
CodeLlama-13B-Instruct & B2     & 13B & Open-source (Meta) & Strong coding ability, but context handling weaker than proprietary models. \\
EvoSuite               & B1     & N/A & Open-source SBST Tool & Provides high structural coverage; readability and maintainability are not prioritized. \\
\bottomrule
\end{tabular}
}
\vspace{-3mm}
\end{table*}

\begin{table}[!htbp]
\centering
\caption{Summary of Collected Test Suites}
\label{tab:test-suite-summary}
\centering
\scalebox{0.5}
{
\resizebox{\columnwidth}{!}{
\begin{tabular}{@{}p{2.2cm}p{2.2cm}p{0.8cm}r r@{}}
\toprule
\textbf{Type} & \textbf{Model/Tool} & \textbf{Bench} & \textbf{\#Proj} & \textbf{\#Suites} \\
\midrule
\multirow{7}{*}{LLM} 
  & GPT-3.5       & B1 & 3  & 16,744 \\
  & GPT-4         & B1 & 1  & 421 \\
  & Mistral 7B    & B1 & 3  & 3,318 \\
  & Mixtral 8x7B  & B1 & 1  & 22 \\
  & GPT-3.5       & B2 & 9  & 324 \\
  & GPT-4         & B2 & 9  & 324 \\
  & CodeLlama-13B & B2 & 9  & 324 \\
\cmidrule{2-5}
  & \multicolumn{3}{r}{\textbf{Total LLM}} & \textbf{21,801} \\
\midrule
SBST & EvoSuite & B1 & 2  & 14,469 \\
\midrule
Human & — & All & 34,635 & 779,585 \\
\bottomrule
\end{tabular}
}
}
\vspace{-3mm}
\end{table}
\subsubsection{Benchmark 2}
\label{subsec:benchmark-2}
\textbf{Benchmark 2}~\cite{zhang2024testbench} targets method-level unit test generation, enabling a fine-grained evaluation across dimensions such as syntactic correctness, compilability, runtime validity, coverage, and mutation-based defect detection. It includes 972 test cases generated by three LLMs—GPT-3.5, GPT-4, and CodeLlama-13B-Instruct—under three prompt contexts: Self-Contained Context (SCC), Simple Context (SC), and Full Context (FC), with 108 test methods produced per model-context pair (Table~\ref{tab:datasets-benchmark-1-2}).

Prompting strategies and their context definitions are detailed in Table~\ref{tab:prompt-strategies}. The collected dataset enables the analysis of test smell diffusion across varying levels of prompt granularity and conditioning.

The 108 target functions span nine diverse open-source Java projects, covering domains such as concurrency, microservices, data visualization, and low-code platforms. All LLMs were configured with a temperature of 0.7; OpenAI models used top-p sampling, and for consistency, only top-p was used for CodeLlama. Token limits were adjusted to each model’s maximum context window (Table~\ref{tab:llms-tools-overview}).

\subsubsection{Human-written tests dataset}
\label{subsec:human-written-dataset}
In addition to LLM-generated tests, we created a combined dataset of human-written tests from SF110~\cite{fraser2014large}, Defects4J~\cite{just2014defects4j}, and the Cat-LM benchmark~\cite{rao2023cat}. Our dataset includes 34,635 well-maintained Java projects, with the vast majority (34,563) coming from the Cat-LM benchmark, complemented by 56 projects from SF110 and 16 from Defects4J (Table~\ref{tab:test-suite-summary}). The resulting corpus comprises 779,585 test cases, with Cat-LM contributing most significantly (776,847 tests), while SF110 and Defects4J provide 1,546 and 1,192 tests respectively. This comprehensive collection of human-written tests from diverse sources enables a robust comparison with LLM-generated tests, allowing us to systematically identify similarities and differences in test quality characteristics and smell patterns between human and AI-generated testing approaches.

\subsection{Software Attributes and LLM Parameters}
\label{subsec:characteristics}

Understanding the factors influencing test smell diffusion is crucial for optimizing test generation techniques. We analyze these factors across three sources of tests: LLM-generated, SBST-generated (EvoSuite), and human-written. The focus depends on the experiment: in \textbf{Analysis~1}, we contrast LLMs with EvoSuite to study how generation strategies and software attributes (e.g., project size, complexity, test scope, and class coupling) affect smell prevalence and distribution; in \textbf{Analysis~2}, we contrast LLMs with human-written tests to examine prevalence, distribution, and similarity, with a particular focus on potential data leakage. Human tests serve as the relevant baseline for inherited patterns, while SBST tests provide a baseline for alternative automated, coverage-driven generation. Table~\ref{tab:llm-software-characteristics} summarizes both software system attributes (e.g., project size, complexity, test scope, and class coupling) and LLM-specific parameters (e.g., model size, temperature, top\_p, and context length) considered in our evaluation.

\begin{table*}[ht]
\centering
\caption{Characteristics of LLMs and Software Artifacts in Benchmark 1 and 2.}
\label{tab:llm-software-characteristics}
\scalebox{0.65}{
\begin{tabular}{@{}l l c c c c c c c c c c c@{}}
\toprule
\textbf{Benchmark} & \textbf{Model} & \textbf{\#Params (B)} & \textbf{Ctx Len} & \textbf{Temp.} & \textbf{Top-p} & \textbf{\#Classes} & \textbf{\#Methods} & \textbf{KLOC} & \textbf{CBO} & \textbf{RFC} & \textbf{DIT} & \textbf{CyCo} \\
\midrule
\multirow{4}{*}{\makecell[l]{\textbf{Benchmark 1} \\ (Class-Level)}}
& GPT-3.5 Turbo  & 175   & 4096  & 0.7 & 1.0  & 690  & 8922 & 85.28 & 7.77  & 14.01 & 1.55 & 2.49 \\
& GPT-4         & 1750  & 8192  & 0.7 & 1.0  & 31   & 164  & 2.37  & 12.68 & 26.24 & 1.32 & 2.25 \\
& Mistral 7B    & 7     & 4096  & 0.7 & 0.95 & 690  & 8922 & 85.28 & 7.77  & 14.01 & 1.55 & 2.49 \\
& Mixtral 8x7B  & 46.7  & 8192  & 0.7 & 0.95 & 31   & 164  & 2.37  & 12.68 & 26.24 & 1.32 & 2.25 \\
& EvoSuite-Defects4J  & --  & --  & -- & -- & 477   & 5905  & 50.96  & 6.23 & 13.24 & 1.55 & 2.55 \\
& EvoSuite-SF110  & --  & --  & -- & -- & 182   & 2853  & 31.95  & 4.3 & 14.79 & 1.56 & 2.67 \\
\midrule
\multirow{3}{*}{\makecell[l]{\textbf{Benchmark 2} \\ (Method-Level)}}
& GPT-3.5 Turbo   & 175   & 4096  & 0.7 & 1.0  & -- & 108 & 1.38 & 1.98 & 4.75 & 1    & 4.99 \\
& GPT-4          & 1750  & 8192  & 0.7 & 1.0  & -- & 108 & 1.38 & 1.98 & 4.75 & 1    & 4.99 \\
& CodeLlama-13B  & 13    & 4096  & 0.7 & 0.95 & -- & 108 & 1.38 & 1.98 & 4.75 & 1    & 4.99 \\
\bottomrule
\end{tabular}
}
\begin{tablenotes}
\tiny
\centering \item[1]$^*$ \#Params (B): \#Parameters (Billion), Ctx Len: Context Length, Temp: Temperature. 
\end{tablenotes}
\vspace{-3mm}
\end{table*}

\subsubsection{Software System Attributes} 
\label{subsubsec:software-system-characteristics}
We describe the software attributes used to evaluate their influence on test smell prevalence, employing a set of well-established metrics: LOC, number of methods, number of classes, Cyclomatic Complexity (CyCo), Coupling Between Objects (CBO), Response for a Class (RFC), and Depth of Inheritance Tree (DIT). 
Following prior work~\cite{palomba2016diffusion, bavota2015test}, we compute total values for size metrics (e.g., LOC, classes, methods) and project-level means for structural metrics (e.g., CBO, RFC, DIT, CyCo) to ensure comparability.  These metrics are extracted using \texttt{ck}\footnote{\url{https://github.com/mauricioaniche/ck}} for object-oriented features and \texttt{Lizard}\footnote{\url{https://github.com/terryyin/lizard}} for complexity and method-level information.

\subsubsection{LLM-Specific Parameters} 
\label{subsubsec:llm-specific-characteristics}
LLMs differ in several key parameters—such as model size, context length, temperature, and top-p sampling—that may influence the structure and quality of generated test cases. Although we reused existing benchmarks and did not configure these parameters ourselves, our study takes into account their differences across models to explore potential associations with test smell diffusion. Open-source models support both top-k and top-p sampling, while OpenAI models expose only top-p for nucleus sampling~\cite{holtzman2019curious}. For consistency, all models were evaluated under top-p sampling. Table~\ref{tab:llm-software-characteristics} summarizes these parameters, including model size (number of parameters), context length (token window), temperature (diversity control), and top-p (sampling scope).

\subsection{Test Smell Detection Strategy}
\label{subsec:test-smell-tools}

To ensure comprehensive and reliable test smell detection, we employ both TsDetect~\cite{peruma2020tsdetect} and JNose~\cite{virginio2020jnose}. We rely on the most recent version of TsDetect (v2.2)\footnote{\url{https://github.com/TestSmells/TestSmellDetector/releases/tag/v2.2}}, which supports 21 test smells—the same number and categories as JNose. These 21 smells constitute the complete catalogue analysed in this study: all prevalence, co-occurrence, and diffusion results reported in the paper are restricted to smell types that are detectable by both tools. Although their coverage is aligned, the tools differ significantly in implementation and reporting granularity. TsDetect relies on static structural heuristics and provides smell-level classification across entire projects, whereas JNose enables more fine-grained localization at the method or class level, with multiple analysis modes (per class, per smell, per file).

Using both tools enables cross-validation of smell detection through independent implementations and offers complementary perspectives—leveraging TsDetect’s robust detection pipeline alongside JNose’s precise mapping of smell instances within test classes and lines of code.

To better understand the reliability of TsDetect and JNose in our setting, we complement these automated analyses with a manual oracle. We manually inspect 240 sampled test entities (120 class-level and 120 method-level) covering frequent smells and cases with strong detector disagreement. For each sampled instance, we label the reported smell as a true or a false positive and derive precision, recall, and F1 for both detectors with respect to this oracle. We use these measurements to interpret our large-scale diffusion results as detector-dependent, upper-bound estimates rather than perfect ground truth.

When TsDetect and JNose disagree on the presence of a given smell, we treat their outputs as two independent operationalisations rather than forcing a single consensus label. In all our analyses, we therefore (i) report smell prevalence and diffusion patterns separately for each detector, (ii) quantify cross-detector differences in prevalence and co-occurrence when addressing our research questions, and (iii) use the manual oracle to inspect a stratified sample of frequent smells and instances with strong detector disagreement. Detector disagreement is thus treated as an object of analysis and as a construct-validity threat, not as something to be automatically reconciled.

\subsection{Metrics}
\label{subsec:metrics}

\subsubsection{Co-occurrence} 
\label{subsubsec:co-occurrence}

In the context of this work, we apply co-occurrence~\cite{ross1976first,papoulis1965random,agrawal1993mining} to quantify how often two test smells appear together within the same project or class, relative to the total occurrences of one of the test smells. In this work, we use the following formula to compute the co-occurrence of two test smells \( t_i \) and \( t_j \):

\begin{equation}
\text{co-occurrence}_{t_i,t_j} = \frac{|t_i \land t_j|}{|t_i|}
\end{equation}

Where \( |t_i \land t_j| \) represents the number of times both test smell \( t_i \) and \( t_j \) appear together in the same project or class (joint occurrences), and \( |t_i| \) is the total number of occurrences of test smell \( t_i \).

By calculating the co-occurrence between various test smells, we can identify whether certain test smells tend to appear together more frequently than by random chance. This helps uncover potential relationships between test smells and provides insights into how they might affect software quality collectively.

\subsubsection{PMCC}
\label{subsubsec:pmcc}
The Pearson Product-Moment Correlation Coefficient (PMCC)~\cite{kendall1943advanced,Cohen_88} is a metric that measures the strength and direction of the linear relationship between two continuous variables. It is commonly denoted by \( r \) and can take values between -1 and 1. Cohen et al.~\cite{Cohen_88} offered a set of guidelines for interpreting the correlation coefficient \( r \). The formula for the Pearson correlation coefficient \( r \) between two variables \( X \) and \( Y \) is:

\begin{equation}
r = \frac{\sum_{i=1}^{n} (X_i - \bar{X})(Y_i - \bar{Y})}{\sqrt{\sum_{i=1}^{n} (X_i - \bar{X})^2 \sum_{i=1}^{n} (Y_i - \bar{Y})^2}}
\end{equation}

Where \( X_i \) and \( Y_i \) are the individual sample points, \( \bar{X} \) and \( \bar{Y} \) represent the means of \( X \) and \( Y \), respectively, and \( n \) denotes the number of data points.

We use PMCC to assess correlations between the presence of test smells and system characteristics(Table~\ref{tab:llm-software-characteristics}).

\subsubsection{Similarity metrics}
\label{subsubsec:similarity-metrics}
To compare the patterns of test smells between LLM-generated and human-written test suites, we use several similarity metrics, each designed to capture different aspects of the test smell distribution.

\begin{itemize}[leftmargin=*]
\item \textbf{Cosine Similarity}

We use Cosine Similarity~\cite{manning2008introduction,huang2008similarity} to compare the frequency of test smells between the LLM-generated and human-written test suites.

\begin{equation}
\text{Cosine Similarity} = \frac{\sum_{i=1}^{n} A_i B_i}{\sqrt{\sum_{i=1}^{n} A_i^2} \times \sqrt{\sum_{i=1}^{n} B_i^2}}
\end{equation}

Where \( A_i \) and \( B_i \) are vectors representing the frequency of test smells in the respective test suites.

\item \textbf{Jaccard Index}
We apply the Jaccard Index~\cite{jaccard1901etude,mining2006introduction} to assess the overlap in the frequency of test smells between the two types of test suites, focusing on how often specific smells appear.

\begin{equation}
J(A, B) = \frac{\sum_{i=1}^{n} \min(A_i, B_i)}{\sum_{i=1}^{n} \max(A_i, B_i)}
\end{equation}

Where \( A \) and \( B \) represent the frequencies of test smells in the LLM-generated and human-written test suites, respectively.

\item \textbf{Euclidean Distance}
We use this metric~\cite{shakhnarovich2008nearest,wang2016towards} to compare the frequency distribution of test smells between LLM-generated and human-written test suites. This allows us to quantify how the distribution of various test smells differs between the two sets.

\begin{equation}
d(A, B) = \sqrt{\sum_{i=1}^{n} (A_i - B_i)^2}
\end{equation}

Where \( A \) and \( B \) are vectors representing the frequency of test smells in the respective test suites.

\item \textbf{Hellinger Distance}
We use the Hellinger Distance~\cite{hellinger1909neue,kailath1967divergence} to compare the normalized frequency distributions of test smells across the LLM-generated and human-written test suites. This metric is particularly useful for understanding the divergence between test smell distributions across both types of test suites.

\begin{equation}
H(A, B) = \frac{1}{\sqrt{2}} \sqrt{\sum_{i=1}^{n} ( \sqrt{A_i} - \sqrt{B_i} )^2}
\end{equation}

Where \( A \) and \( B \) are the probability distributions of test smell frequencies.
\end{itemize}


Each of these metrics provides a different lens through which we compare the LLM-generated and human-written test suites, enabling a multi-faceted analysis of how similar or divergent their test smell patterns are.

\subsubsection{Non-Linear Statistical Tests}
\label{subsubsec:non-linear-statistical-tests}
While previous studies by Bavota et al.~\cite{bavota2015test} and Palomba et al.\cite{palomba2016diffusion} used Pearson correlation (PMCC) to analyze relationships between test smells and system characteristics in both human-written and generated tests, PMCC only detects linear correlations with constant rates of change. Test smell occurrences, however, may follow non-linear patterns with threshold effects where quality shifts at specific breakpoints.

To capture these complex relationships, we complement PMCC with three non-linear approaches: i) Threshold-Based Analysis (Kruskal-Wallis test)~\cite{kruskal1952use} to identify stepwise changes in test smell occurrences; ii) Spearman's Rank Correlation~\cite{spearman1961proof} to detect monotonic but not strictly linear relationships; and iii) Mutual Information (MI)~\cite{cover1999elements} to measure non-linear dependencies between variables when no clear monotonic trend exists.

\begin{itemize}[leftmargin=*]
\item \textbf{Threshold-Based Analysis}
Threshold-based analysis examines how test smell occurrences shift at specific breakpoints in key variables like model size, context length, and project complexity. Using the Kruskal-Wallis test~\cite{kruskal1952use}, we identify stepwise or piecewise non-linear patterns rather than smooth transitions, such as comparing test smell frequency before and after a 4096-token context length threshold to detect significant quality changes.

\begin{equation}
H = \frac{12}{n(n+1)} \sum_{i=1}^{g} \frac{R_i^2}{n_i} - 3(n+1)
\end{equation}

Where \( g \) is the number of groups, \( n \) is the total number of observations, \( n_i \) is the number of observations in group \( i \), and \( R_i \) is the sum of ranks in group \( i \).

\item \textbf{Spearman’s Rank Correlation}
Spearman's rank correlation~\cite{spearman1961proof} detects monotonic relationships (consistently increasing or decreasing trends) between variables, capturing non-linear patterns that Pearson correlation would miss, such as when test smells increase sharply beyond certain LLM model sizes without following a linear progression.

\begin{equation}
\rho = 1 - \frac{6 \sum d_i^2}{n(n^2 - 1)}
\end{equation}

Where \( d_i \) is the difference between the ranks of corresponding values, and \( n \) is the number of data points.

\item \textbf{Mutual Information (MI)}
Mutual Information (MI)~\cite{cover1999elements} quantifies the general dependence between variables, detecting non-linear, non-monotonic relationships that correlation metrics miss, such as when Magic Number Test smells show complex dependencies on LLM model size.

\begin{equation}
I(X;Y) = \sum_{x \in X} \sum_{y \in Y} p(x,y) \log \left( \frac{p(x,y)}{p(x)p(y)} \right)
\end{equation}

Where \( p(x,y) \) is the joint probability distribution of \( X \) and \( Y \), while \( p(x) \) and \( p(y) \) are the marginal probability distributions of \( X \) and \( Y \).

\end{itemize}

\subsubsection{Hypothesis Testing and Effect Sizes}
\label{subsubsec:hypothesis-tests}

Beyond correlation and similarity measures, we also perform formal hypothesis testing on smell prevalence to assess whether differences between groups are statistically significant. In particular, for comparisons between two conditions---such as LLM-generated vs.\ EvoSuite-generated tests on Benchmark~1, or class-level vs.\ method-level LLM tests across benchmarks---we use the two-sided Mann--Whitney U test, a non-parametric alternative to the $t$-test that does not assume normality of the underlying distributions. Unless otherwise stated, we adopt a significance level of $\alpha = 0.05$ and adjust for multiple comparisons using the Benjamini--Hochberg procedure to control the false discovery rate.
To complement $p$-values with a notion of practical significance, we report Cliff's delta ($\delta$) as an effect size for pairwise comparisons. Cliff's delta measures the probability that a randomly chosen observation from one group exceeds a randomly chosen observation from the other group, minus the reverse probability:

\begin{equation}
\delta =
\frac{
\left|\{(x,y) \in X \times Y \mid x > y\}\right|
-
\left|\{(x,y) \in X \times Y \mid x < y\}\right|
}{
|X| \cdot |Y|
}.
\end{equation}

Here, $X$ and $Y$ denote the two samples being compared. Values of $\delta$ range from $-1$ (all observations in $Y$ greater than those in $X$) to $1$ (all observations in $X$ greater than those in $Y$), with $0$ indicating no stochastic dominance.

\subsection{Implementation and Configuration}
\label{subsec:implementation_and_Configuration}

We conducted our test smell analysis using Python 3.10 and Java 17. To detect smells, we relied on two complementary tools: \emph{TsDetect v2.2}, which supports automated batch execution, and \emph{JNose 2.2.0}, which operates through a web interface. Since JNose lacks CLI support, we developed an automation layer based on \emph{Playwright v1.51.0}\footnote{\url{https://github.com/microsoft/playwright}} and \emph{BeautifulSoup v4.13.3}\footnote{\url{https://www.crummy.com/software/BeautifulSoup/}} to simulate UI interaction and parse the resulting CSV reports programmatically.

We relied on \texttt{ck}\footnote{\url{https://github.com/mauricioaniche/ck}}~0.7.0 for extracting object-oriented metrics (e.g., CBO, RFC, DIT) and on \texttt{Lizard}~1.17.23 for complexity and method-level statistics (e.g., Cyclomatic Complexity, number of methods). These metrics were then used in correlation and distributional analyses to examine the relationship between software structure, generation settings, and test smell diffusion.

All experiments were executed on a machine equipped with an Intel Core i9-14900K CPU (32 threads, 6.0GHz), 64\,GB RAM, and an NVIDIA RTX 5000 Ada GPU (32\,GB VRAM). The full pipeline was fully automated to support scalable, reproducible analysis across all benchmarks.

\section{Experimental Results}
\label{results}

\subsection{[RQ1]: Prevalence and Comparison of Test Smells in LLM-vs. SBST-Generated Tests}

\noindent\textbf{[Experiment Design]:} 
We conduct this comparison at two levels of test generation granularity: class-level and method-level, based on Benchmark~1 and Benchmark~2 (described in Section~\ref{subsec:dataset}). 
To assess consistency between detectors, we compute the mean and standard deviation of absolute delta values per smell type, and highlight the top-5 largest deltas to reveal key divergences.
In addition, we perform a small manual validation study to estimate the precision of the detectors for frequent and detector-divergent smells. For both benchmarks, we first identify the most frequent smells across LLM-generated and EvoSuite-generated tests, and we also flag smells that exhibit strong disagreement between detectors, defined as cases where one detector reports the smell for at least 50\% of test entities (classes or methods) in a given setting, while the other reports it rarely or not at all. For these selected smells, we then randomly sample 20 instances per smell and generator from all test cases where at least one detector reports the smell; for detector-divergent smells, we additionally draw samples from strata where only one of the two detectors reports the smell. In total, this procedure yields 120 manually inspected class-level test cases (Benchmark~1) and 120 manually inspected method-level test cases (Benchmark~2). For each sampled instance, we label the reported smell as a true positive or a false positive. This manual validation allows us to approximate detector false positive rates for key smells and to interpret prevalence estimates as upper bounds on actual smell incidence in our setting.

\noindent\textbf{[Experiment Results]:}

\vspace{0.5em}
\noindent\textbf{Prevalence of Test Smells.}  
Assertion Roulette (AR) emerges as the most dominant smell across all settings (full per-smell distributions in Appendix~\ref{app:additional-tables}, Tables~\ref{tab:test-smells-benchmark1}–\ref{tab:test-smells-benchmark2}). In Benchmark~1 (class level), AR affects 38–49\% of LLM-generated suites according to TsDetect, and 50–100\% according to JNose, with EvoSuite also showing high prevalence (44–73\% depending on the tool). In Benchmark~2 (method level), AR is detected at consistently high rates across tools (92.8–99\%), confirming it as the most pervasive smell across models and granularities.
This dominance is consistent with prior empirical work on both industrial/open-source systems and educational settings, where Assertion Roulette also emerges as one of the most frequently detected smells~\cite{bavota2015test,bai2022assertion}. Bai et al.~\cite{bai2022assertion} further show, in a controlled study with students, that while AR is common, its measurable impact on objective code-quality outcomes is limited, suggesting that its harmfulness is more nuanced than raw prevalence alone would indicate.

\vspace{0.5em}
\noindent\colorbox{gray!15}{{\parbox{0.98\linewidth}{
\textbf{Finding 1:} \textit{Assertion Roulette is pervasive across all LLMs and granularities, exceeding 90\% prevalence at method level and remaining high at class level, with EvoSuite exhibiting similar vulnerability.}
}}}

\vspace{0.5em}
Beyond AR, LLM-generated tests also suffer from recurring weaknesses in logic depth and exception coverage. At class level (Table~\ref{tab:test-smells-benchmark1}), Lazy Test (LT) is widespread, ranging from 7–56\% across LLMs, while EvoSuite shows even higher prevalence (77–82\%). Unknown Test (UT) reaches up to 77\% in LLMs but remains almost absent in EvoSuite (0–7\%), owing to its structurally enforced assertions. Exception Handling (EH) is similarly underrepresented in LLM outputs (7–23\%), in stark contrast to EvoSuite (93\%). Method-level generation (Table~\ref{tab:test-smells-benchmark2}) alleviates some of these issues: LT decreases to 15–50\% (TsDetect) and below 20\% (JNose), while UT falls to 0–23\% (TsDetect) and 4–9\% (JNose). However, EH remains virtually absent across all LLMs at method level, suggesting that finer-grained generation improves modularity and assertion clarity but fails to address exception scenarios.

\vspace{0.5em}
\noindent\colorbox{gray!15}{{\parbox{0.98\linewidth}{
\textbf{Finding 2:} \textit{LLM-generated tests frequently exhibit Lazy Test, Unknown Test, and missing Exception Handling. While method-level generation reduces LT and UT prevalence, exception coverage remains absent—contrasting sharply with SBST (EvoSuite).}
}}}

\vspace{0.5em}
\noindent\textbf{Distribution Across Smells and Granularity.}  
Smell distribution differs by test granularity. While Assertion Roulette (AR) dominates at both levels, Lazy Test (LT) and Exception Handling (EH) are less frequent at method level: LT drops from 7–56\% at class level (LLMs; Table~\ref{tab:test-smells-benchmark1}) to 15–50\% (TsDetect) and 0–19\% (JNose) at method level (Table~\ref{tab:test-smells-benchmark2}), and EH falls to 0\% across LLMs in Benchmark~2. Cross-tool disagreement (full deltas in Appendix~\ref{app:additional-tables}, Table~\ref{tab:delta-distribution-all}, and summary in Table~\ref{tab:top5-deltas-all}) further shapes the observed distributions: AR is highly consistent at method level ($\Delta = 1.21$), but several fine-grained smells show large deltas between TsDetect and JNose. In Benchmark~1 (LLMs), the largest deltas occur for Magic Number Test (MT, $\Delta = 72.55$), Unknown Test (UT, $\Delta = 35.97$), and Empty Test (EmT, $\Delta = 17.28$) (Table~\ref{tab:delta-distribution-all}). In Benchmark~2, the largest deltas are for Dependent Test (DpT, $\Delta = 57.10$), Duplicate Assert (DA, $\Delta = 22.69$), and Lazy Test (LT, $\Delta = 19.04$), followed by Sleepy Test (ST, $\Delta = 14.64$) and Conditional Logic Test (CLT, $\Delta = 12.34$).

\vspace{0.5em}
\noindent\colorbox{gray!15}{{\parbox{0.98\linewidth}{
\textbf{Finding 3:} \textit{Smell distribution varies with granularity and detector: AR remains dominant across levels, but LT and EH are less frequent at method level, while fine-grained smells (e.g., MT, DpT, DA) exhibit large cross-tool variability (up to $\Delta = 72.55$), as shown in Tables~\ref{tab:delta-distribution-all} and~\ref{tab:top5-deltas-all}.}
}}}

\vspace{0.5em}
\noindent\textbf{Comparison with EvoSuite.}  
EvoSuite shares AR and LT with LLMs but differs on UT and EH. At class level (Benchmark~1), UT reaches 8–77\% in LLMs yet remains 0–7\% in EvoSuite; EH is 7–23\% in LLMs versus 93\% in EvoSuite according to TsDetect (66–80\% in JNose) (Table~\ref{tab:test-smells-benchmark1}). AR is high for both (LLMs: 38–49\% TsDetect and 50–100\% JNose; EvoSuite: 38–51\% TsDetect and 56–90\% JNose). 
Cross-tool variability, summarised in Table~\ref{tab:delta-stats-distribution}, is also lower for EvoSuite: the mean $\Delta$ is 6.51 with std.~dev.~14.06 in Benchmark~1, compared to 8.64 and 17.68 for LLMs. At method level (Benchmark~2), LLM tests still show notable deltas for several smells (e.g., DpT, DA, LT) even though AR is stable across tools (Tables~\ref{tab:delta-distribution-all} and~\ref{tab:delta-stats-distribution}), and the top-5 largest deltas for each generator are reported in Table~\ref{tab:top5-deltas-all}.

\vspace{0.5em}
\noindent\colorbox{gray!15}{{\parbox{0.98\linewidth}{
\textbf{Finding 4:} \textit{Compared to EvoSuite, LLM-generated tests not only exhibit higher Unknown Test and weaker Exception Handling (while both share AR and LT), but also induce greater cross-tool variability between TsDetect and JNose (mean $\Delta$=8.64 vs.~6.51; std.~dev.=17.68 vs.~14.06). The largest $\Delta$ values for LLMs occur on MT and UT in Benchmark~1 and DpT, DA, and LT in Benchmark~2 (Tables~\ref{tab:test-smells-benchmark1}–\ref{tab:top5-deltas-all}).}
}}}

\begin{table}[ht]
\centering
\caption{Mean and Standard Deviation of Delta Values between TsDetect and JNose for Test Smell Distributions}
\label{tab:delta-stats-distribution}
\scalebox{0.7}{
\begin{tabular}{lcc}
\toprule
\textbf{Benchmark} & \textbf{Mean Delta} & \textbf{Std Deviation} \\
\midrule
Benchmark 1 - LLM & 8.641 & 17.678 \\
Benchmark 1 - EvoSuite & 6.506 & 14.055 \\
Benchmark 2 & 7.834 & 12.750 \\
\bottomrule
\end{tabular}
}
\vspace{-3mm}
\end{table}

\begin{table}[ht]
\centering
\caption{Top-5 Largest Deltas between TsDetect and JNose}
\label{tab:top5-deltas-all}
\scalebox{0.6}{
\begin{tabular}{lcc|lcc|lcc}
\toprule
\multicolumn{3}{c|}{\textbf{B1-LLM}} & 
\multicolumn{3}{c|}{\textbf{B1-EvoSuite}} & 
\multicolumn{3}{c}{\textbf{B2-LLM}} \\
\cmidrule(lr){1-3} \cmidrule(lr){4-6} \cmidrule(lr){7-9}
\textbf{Smell} & \textbf{$\Delta$} & \textbf{Rk} & 
\textbf{Smell} & \textbf{$\Delta$} & \textbf{Rk} & 
\textbf{Smell} & \textbf{$\Delta$} & \textbf{Rk} \\
\midrule
MT   & 72.55 & 1 & MT   & 61.11 & 1 & DpT & 57.10 & 1 \\
UT   & 35.97 & 2 & AR   & 28.27 & 2 & DA  & 22.69 & 2 \\
AR   & 32.84 & 3 & EH   & 19.76 & 3 & LT  & 19.04 & 3 \\
EmT  & 17.28 & 4 & EmT  & 6.44  & 4 & ST  & 14.64 & 4 \\
EH   & 13.44 & 5 & MG   & 4.44  & 5 & CLT & 12.34 & 5 \\
\bottomrule
\end{tabular}
}
\end{table}

\noindent\textbf{Cross-detector divergence.}
Table~\ref{tab:delta-distribution-all} summarises, for each smell and benchmark, 
the absolute difference in prevalence reported by TsDetect and JNose. On Benchmark~1, 
LLM-generated tests exhibit large cross-detector deltas for \emph{Assertion Roulette} (AR, 32.84\%), 
\emph{Empty Test} (EmT, 17.28\%), \emph{Exception Handling} (EH, 13.44\%), and especially 
\emph{Unknown Test} (UT, 35.97\%) and \emph{Magic Number Test} (MT, 72.55\%). EvoSuite shows 
a similar pattern, with substantial disagreement on MT (61.11\%) and, to a lesser extent, AR 
and EH. Benchmark~2 highlights even more extreme divergences for method-level tests: while AR 
remains relatively stable across tools (delta 1.21\%), smells such as \emph{Conditional Logic Test} 
(CLT, 12.34\%), \emph{Lazy Test} (LT, 19.04\%), \emph{Duplicate Assert} (DA, 22.69\%), and 
in particular \emph{Dependent Test} (DpT, 57.10\%) show large discrepancies. 
Table~\ref{tab:delta-stats-distribution} further indicates that the average delta between 
detectors is non-negligible (6.5--8.6 percentage points) with high variability, confirming 
that some smells are relatively stable across tools whereas others (notably MT and DpT) are 
highly detector-dependent.

\vspace{0.5em}
\noindent\colorbox{gray!15}{{\parbox{0.98\linewidth}{
\textbf{Finding 5:} TsDetect and JNose show substantial disagreement on several key smells, 
with average deltas around 7--9 percentage points and much larger gaps for specific cases. 
AR remains relatively stable across tools, whereas MT, UT, and DpT are highly detector-dependent, 
particularly at method level. These results confirm that cross-tool variability is a major factor 
in smell diffusion analyses, and that conclusions for highly divergent smells should be interpreted 
with particular caution.
}}}

\vspace{0.5em}

\noindent\textbf{Statistical comparison of key test smells prevalence.}
Beyond descriptive percentages (Tables~\ref{tab:test-smells-benchmark1} and~\ref{tab:test-smells-benchmark2}), we assess whether the observed differences in smell prevalence are statistically significant. Rather than testing all smells, we focus on a set of key smells that are frequent and central to our analysis (AR, LT, MT, UT, EH, EmT, GF, and DA). For each of these smells and for each configuration, we consider the per-configuration prevalence and apply two-sided Mann--Whitney U tests. Table~\ref{tab:rq1_llm_evo_stats_b1} summarises the comparisons between LLM-generated and EvoSuite-generated tests on Benchmark~1, while Table~\ref{tab:granularity_llm} reports class- vs.\ method-level differences for LLM-generated tests only (with $p$-values, Benjamini--Hochberg adjusted $p_{adj}$, and Cliff's $\delta$ effect sizes).

For Benchmark~1, Table~\ref{tab:rq1_llm_evo_stats_b1} indicates that, according to TsDetect, 
EvoSuite-generated tests contain significantly more \emph{Lazy Test} (LT) and \emph{Exception Handling} (EH) instances than LLM-generated tests ($p = 0.044$ and $p = 0.049$, respectively), while LLM-generated tests exhibit significantly more \emph{Unknown Test} (UT) and \emph{General Fixture} (GF) ($p = 0.044$ and $p = 0.049$). Differences for \emph{Assertion Roulette} (AR), \emph{Magic Number Test} (MT), and \emph{Duplicate Assert} (DA) are not statistically significant at $\alpha = 0.05$, despite directional trends (AR and MT slightly higher in EvoSuite, DA slightly higher in LLM tests). 
In contrast, the JNose-based comparisons do not yield statistically significant differences for any of the 
key smells (all $p \geq 0.05$), although the directions broadly mirror TsDetect for several smells 
(e.g., more EH in EvoSuite, more UT and DA in LLM tests). Overall, these results suggest that LLMs and 
EvoSuite differ in how they distribute certain smells (notably UT, GF, LT, and EH), but the strength and 
significance of these differences depend on the detector used.

\begin{table}[t]
\centering
\small
\caption{LLM vs.\ EvoSuite key test smells prevalence on Benchmark~1.}
\label{tab:rq1_llm_evo_stats_b1}

\begin{subtable}[t]{0.48\linewidth}
\centering
\caption{TsDetect}
\label{tab:rq1_llm_evo_stats_tsdetect}
\scalebox{0.8}{
\begin{tabular}{lll}
\toprule
Smell & $p$-value & Direction \\
\midrule
AR   & 0.533 & EvoSuite $>$ LLM \\
LT   & 0.044 & EvoSuite $>$ LLM \\
MT   & 0.086 & EvoSuite $>$ LLM \\
UT   & 0.044 & LLM $>$ EvoSuite \\
EH   & 0.049 & EvoSuite $>$ LLM \\
EmT  & 0.150 & LLM $>$ EvoSuite \\
GF   & 0.049 & LLM $>$ EvoSuite \\
DA   & 0.694 & LLM $>$ EvoSuite \\
\bottomrule
\end{tabular}
}
\end{subtable}
\hfill
\begin{subtable}[t]{0.48\linewidth}
\centering
\caption{JNose}
\label{tab:rq1_llm_evo_stats_jnose}
\scalebox{0.8}{
\begin{tabular}{lll}
\toprule
Smell & $p$-value & Direction \\
\midrule
AR   & 1.000 & LLM $>$ EvoSuite \\
LT   & 0.111 & EvoSuite $>$ LLM \\
MT   & 0.376 & EvoSuite $>$ LLM \\
UT   & 0.262 & LLM $>$ EvoSuite \\
EH   & 0.053 & EvoSuite $>$ LLM \\
EmT  & 1.000 & LLM $>$ EvoSuite \\
GF   & 0.384 & n.s. \\
DA   & 0.262 & LLM $>$ EvoSuite \\
\bottomrule
\end{tabular}
}
\end{subtable}

\begin{tablenotes}
\scriptsize
\item[1] $^*$ Mann--Whitney U test on per-configuration smell prevalence; Direction indicates which group (LLM or EvoSuite) has higher median prevalence; n.s.\ denotes a non-significant difference ($p \geq 0.05$).
\end{tablenotes}
\end{table}


Table~\ref{tab:granularity_llm} analyses how smell prevalence varies between class-level and method-level LLM-generated tests. Under TsDetect, class-level tests show significantly higher prevalence of \emph{Magic Number Test} (MT), \emph{Unknown Test} (UT, marginal after correction), \emph{Exception Handling} (EH), \emph{Empty Test} (EmT), and \emph{General Fixture} (GF), while \emph{Assertion Roulette} (AR) is significantly more prevalent at method level ($p = 0.012$, $p_{adj} = 0.019$, $\delta = -1.00$). \emph{Lazy Test} (LT) and \emph{Duplicate Assert} (DA) show no statistically significant differences between granularities. JNose reports the same directional trends for several smells (e.g., AR and MT more frequent at method level, LT, EmT, and GF more frequent at class level), but none of these differences remain significant after Benjamini--Hochberg correction ($p_{adj} \geq 0.267$). Taken together, TsDetect points to a clear granularity effect—with class-level tests accumulating more global-context smells (EH, EmT, GF, MT, UT) and method-level tests concentrating AR—while JNose provides weaker statistical evidence but broadly consistent directions.

\begin{table}[t]
\centering
\small
\caption{Class- vs.\ method-level prevalence of key smells in LLM-generated tests.}
\label{tab:granularity_llm}

\begin{subtable}[t]{0.48\linewidth}
\centering
\caption{TsDetect}
\label{tab:granularity-tsdetect}
\scalebox{0.8}{
\begin{tabular}{llllll}
\toprule
Smell & $p$ & $p_{adj}$ & $\delta$ & Direction \\
\midrule
AR  & 0.012 & 0.019 & -1.00 & method$>$class \\
LT  & 0.630 & 0.630 &  0.25 & class$>$method \\
MT  & 0.012 & 0.019 &  1.00 & class$>$method \\
UT  & 0.048 & 0.065 &  0.83 & class$>$method \\
EH  & 0.012 & 0.019 &  1.00 & class$>$method \\
EmT & 0.012 & 0.019 &  1.00 & class$>$method \\
GF  & 0.012 & 0.019 &  1.00 & class$>$method \\
DA  & 0.576 & 0.630 &  0.25 & class$>$method \\
\bottomrule
\end{tabular}
}
\end{subtable}
\hfill
\begin{subtable}[t]{0.48\linewidth}
\centering
\caption{JNose}
\label{tab:granularity-jnose}
\scalebox{0.8}{
\begin{tabular}{llllll}
\toprule
Smell & $p$ & $p_{adj}$ & $\delta$ & Direction \\
\midrule
AR  & 0.350 & 0.678 & -0.43 & method$>$class \\
LT  & 0.075 & 0.300 &  0.76 & class$>$method \\
MT  & 0.033 & 0.267 & -0.90 & method$>$class \\
UT  & 1.000 & 1.000 & -0.05 & method$>$class \\
EH  & 0.508 & 0.678 & -0.33 & method$>$class \\
EmT & 0.333 & 0.678 &  0.43 & class$>$method \\
GF  & 0.475 & 0.678 &  0.43 & class$>$method \\
DA  & 0.933 & 1.000 & -0.10 & method$>$class \\
\bottomrule
\end{tabular}
}
\end{subtable}

\begin{tablenotes}
\scriptsize
\item[1] $p$: two-sided $p$-value of the Mann--Whitney U test comparing
per-configuration smell prevalence between class- and method-level
LLM tests; $p_{adj}$: $p$-value adjusted for multiple comparisons
using the Benjamini--Hochberg procedure; $\delta$: Cliff's delta
(positive values indicate higher prevalence at class level, negative
values at method level); Direction: group with higher median
prevalence.
\end{tablenotes}
\end{table}

Across both comparisons, the statistical analysis refines our RQ1 findings. First, LLM-generated and EvoSuite-generated tests exhibit systematically different smell profiles: under TsDetect, LLM tests tend to contain more UT and GF, whereas EvoSuite tests show higher prevalence of LT and EH. Second, test granularity plays a substantial role in smell diffusion: class-level generation amplifies smells tied to shared fixtures and exception handling, whereas method-level generation exacerbates AR by concentrating assertions in single-purpose tests. These patterns, combined with our manual validation and cross-detector analysis, support the view that both the choice of generator (LLM vs.\ EvoSuite) and the generation 
granularity (class vs.\ method) systematically shape how key smells emerge in practice.

\vspace{0.5em}
\noindent\colorbox{gray!15}{{\parbox{0.98\linewidth}{
\textbf{Finding 6:} Statistical tests on key smells (AR, LT, MT, UT, EH, EmT, GF, DA) confirm that smell diffusion is shaped both by the generator and by granularity. Under TsDetect, EvoSuite-produced tests contain significantly more LT and EH, whereas LLM-generated tests contain more UT and GF on Benchmark~1, while class-level LLM tests accumulate more global-context smells (MT, UT, EH, EmT, GF) and method-level tests concentrate AR. JNose shows weaker statistical evidence but broadly consistent directions, highlighting that the strength of these effects is detector-dependent.
}}}

\vspace{0.5em}
\noindent\textbf{Manual validation of detector reliability.}
After analysing the smell distributions reported in Table~\ref{tab:test-smells-benchmark1} (Benchmark~1) and Table~\ref{tab:test-smells-benchmark2} (Benchmark~2), we designed a manual validation sample focused on a small set of representative smells. For Benchmark~1 (class level), we selected three smells that are both frequent and central to our study: Assertion Roulette (AR), Lazy Test (LT), and Magic Number Test (MT). For Benchmark~2 (method level), we considered two frequent smells (AR and Conditional Logic Test, CLT) and two smells where TsDetect and JNose diverge the most in the large-scale results (MT and Dependent Test, DpT). 
For each benchmark, we then randomly sampled 120 test entities for manual checking, stratified by smell and generator configuration. At class level (Benchmark~1), this yields 327 smell–class instances (AR, LT, and MT over tests generated by LLMs and by EvoSuite). At method level (Benchmark~2), the manual oracle covers 120 smell–method instances (AR, CLT, MT, and DpT for LLM-generated tests). These manually validated instances are used as a ground truth to assess the reliability of TsDetect and JNose in the remainder of this section, and the full sample is included in our replication package.
Tables~\ref{tab:detector-reliability-b1} and~\ref{tab:detector-reliability-b2} summarise how TsDetect and JNose perform against our manual oracle on the sampled instances. Table~\ref{tab:detector-reliability-b1} reports class-level results for Benchmark~1, broken down by smell (AR, LT, MT) and generator (LLM vs.\ EvoSuite), while Table~\ref{tab:detector-reliability-b2} reports method-level results for Benchmark~2, covering frequent smells (AR, CLT) and divergent smells (MT, DpT).

At class level (Tables~\ref{tab:detector-reliability-b1}), both detectors achieve very high recall across all three smells, but with markedly different precision profiles.
For \emph{Assertion Roulette} (AR), TsDetect and JNose both reach a recall of~1.0 for EvoSuite- and LLM-generated tests, i.e., they do not miss any class that our oracle labels as AR. However, they substantially over-report this smell: TsDetect attains only moderate precision (0.55 for EvoSuite and 0.60 for LLM), and JNose is even more aggressive on EvoSuite (precision 0.34, improving to 0.62 on LLM tests). In practice, AR detections should therefore be interpreted as \emph{candidates} for inspection rather than as high-confidence findings.
For \emph{Lazy Test} (LT), both tools are much more reliable. TsDetect is close to perfect (precision~$\geq$~0.91, recall~$\geq$~0.98 across generators), and JNose is similarly strong (precision~1.00/0.94 and recall~1.00/0.89 for EvoSuite/LLM, respectively). This suggests that, at class level, LT detections from either tool can be treated as trustworthy, with only a small number of borderline cases on LLM-generated tests.
For \emph{Magic Number Test} (MT), we observe a clear trade-off between coverage and conservativeness. TsDetect maximises recall (1.0 for both EvoSuite and LLM), but at the cost of lower precision on LLM tests (0.55 versus 1.0 for EvoSuite), i.e., it tends to flag additional classes that our oracle does not consider MT. JNose, in contrast, is perfectly precise on EvoSuite tests (precision 1.0) but misses a substantial fraction of true MT cases (recall 0.61), and shows a more balanced behaviour on LLM tests (precision 0.77, recall 0.91). Overall, TsDetect behaves as a high-recall, noise-tolerant detector for MT, while JNose provides a more conservative but potentially incomplete view of MT diffusion, especially on EvoSuite-generated tests.

\begin{table}[t]
\centering
\small
\caption{Detector reliability against the manual oracle for Benchmark~1.}
\label{tab:detector-reliability-b1}
\scalebox{0.8}{
\begin{tabular}{lllrrrrrrr}
\toprule
Tool & Smell & Generator & TP & FP & FN & TN & Prec. & Rec. & F1 \\
\midrule
\multirow{6}{*}{TsDetect}
  & \multirow{2}{*}{AR} & EvoSuite & 16 & 13 &  0 & 20 & 0.55 & 1.00 & 0.71 \\
  &                     & LLM      & 36 & 24 &  0 &  0 & 0.60 & 1.00 & 0.75 \\
  & \multirow{2}{*}{LT} & EvoSuite & 48 &  0 &  1 &  0 & 1.00 & 0.98 & 0.99 \\
  &                     & LLM      & 53 &  5 &  0 &  2 & 0.91 & 1.00 & 0.95 \\
  & \multirow{2}{*}{MT} & EvoSuite & 49 &  0 &  0 &  0 & 1.00 & 1.00 & 1.00 \\
  &                     & LLM      & 33 & 27 &  0 &  0 & 0.55 & 1.00 & 0.71 \\
\midrule
\multirow{6}{*}{JNose}
  & \multirow{2}{*}{AR} & EvoSuite & 16 & 31 &  0 &  2 & 0.34 & 1.00 & 0.51 \\
  &                     & LLM      & 36 & 22 &  0 &  2 & 0.62 & 1.00 & 0.77 \\
  & \multirow{2}{*}{LT} & EvoSuite & 49 &  0 &  0 &  0 & 1.00 & 1.00 & 1.00 \\
  &                     & LLM      & 47 &  3 &  6 &  4 & 0.94 & 0.89 & 0.91 \\
  & \multirow{2}{*}{MT} & EvoSuite & 30 &  0 & 19 &  0 & 1.00 & 0.61 & 0.76 \\
  &                     & LLM      & 30 &  9 &  3 & 18 & 0.77 & 0.91 & 0.83 \\
\bottomrule
\end{tabular}
}
\begin{tablenotes}
\scriptsize
\item[1] TP: true positives; FP: false positives; FN: false negatives; TN: true negatives; Prec.: precision; Rec.: recall.
\end{tablenotes}
\end{table}

At method level (Tables~\ref{tab:detector-reliability-b2} ), we observe patterns that are consistent with the class-level findings for frequent smells, and more extreme divergences for MT and DpT. For \emph{Assertion Roulette} (AR), both detectors perform well: TsDetect achieves precision~0.78 and recall~0.90, while JNose reaches a slightly higher F1 (0.87) with recall~1.00 at the cost of a few extra false positives (precision~0.77). For \emph{Conditional Logic Test} (CLT), the contrast is sharper: JNose again attains perfect recall (1.00) with reasonable precision (0.75), whereas TsDetect misses nearly half of the true CLT methods (recall~0.44), despite a comparable precision (0.73). The “divergent” smells highlight more pronounced discrepancies. For \emph{Magic Number Test} (MT), our oracle marks all 24 sampled methods as MT, and JNose aligns perfectly with this judgement (precision and recall both 1.00), while TsDetect fails to detect any of them (24 false negatives). Conversely, for \emph{Dependent Test} (DpT), our oracle does not classify any of the 30 sampled methods as truly dependent, yet TsDetect flags all 30 as DpT (i.e., only false positives), whereas JNose correctly leaves all of them unmarked. Together, these results suggest that at method level JNose is generally more reliable for CLT and MT, while TsDetect’s current configuration for DpT behaves as an over-approximating detector on our LLM-based benchmark.

\begin{table}[t]
\centering
\small
\caption{Detector reliability against the manual oracle for Benchmark~2.}
\label{tab:detector-reliability-b2}
\scalebox{0.8}{
\begin{tabular}{llrrrrrrr}
\toprule
Smell & Tool & TP & FP & FN & TN & Prec. & Rec. & F1 \\
\midrule
\multirow{2}{*}{AR}
  & TsDetect & 18 &  5 &  2 &  1 & 0.78 & 0.90 & 0.84 \\
  & JNose    & 20 &  6 &  0 &  0 & 0.77 & 1.00 & 0.87 \\
\midrule
\multirow{2}{*}{CLT}
  & TsDetect &  8 &  3 & 10 &  3 & 0.73 & 0.44 & 0.55 \\
  & JNose    & 18 &  6 &  0 &  0 & 0.75 & 1.00 & 0.86 \\
\midrule
\multirow{2}{*}{DpT}
  & TsDetect &  0 & 30 &  0 &  0 & 0.00 & 1.00 & 0.00 \\
  & JNose    &  0 &  0 &  0 & 30 & 1.00 & 1.00 & 1.00 \\
\midrule
\multirow{2}{*}{MT}
  & TsDetect &  0 &  0 & 24 &  0 & 1.00 & 0.00 & 0.00 \\
  & JNose    & 24 &  0 &  0 &  0 & 1.00 & 1.00 & 1.00 \\
\bottomrule
\end{tabular}
}
\begin{tablenotes}
\scriptsize
\item[1] TP: true positives; FP: false positives; FN: false negatives; TN: true negatives; Prec.: precision; Rec.: recall.
\end{tablenotes}
\end{table}

Overall, our manual validation confirms that relying on a single detector can substantially bias conclusions about test smell diffusion. At class level (Benchmark~1), both TsDetect and JNose achieve high recall for AR, LT, and MT, but AR is markedly over-reported by both tools, LT is detected very reliably, and MT exhibits a clear precision–recall trade-off between a high-recall (TsDetect) and a more conservative (JNose) detector (Table~\ref{tab:detector-reliability-b1}). At method level (Benchmark~2), JNose is generally more trustworthy for CLT and MT, while TsDetect tends to over-approximate DpT and completely misses the MT cases in our sample (Table~\ref{tab:detector-reliability-b2}). Taken together, these results also suggest that some of TsDetect’s heuristics may be better calibrated for class-level patterns (spanning multiple test methods) than for fine-grained, per-method smell detection. Our findings align with prior evidence that test smell detectors can suffer from limited detectability, validity, and agreement~\cite{panichella2022test}, and that different tools provide only partially overlapping coverage~\cite{aljedaani2021test}. In this sense, our results empirically support the need for cross-tool and manual validation when using automated detectors such as TsDetect~\cite{peruma2020tsdetect} and JNose~\cite{virginio2020jnose} to study test smell diffusion.

\vspace{0.5em}
\noindent\colorbox{gray!15}{{\parbox{0.98\linewidth}{
\textbf{Finding 7:} On our manually validated sample, TsDetect and JNose often disagree, especially at method level. Both tools are reliable for Lazy Test, but they systematically over-report Assertion Roulette and exhibit complementary behaviours for Magic Number Test (TsDetect favouring recall, JNose precision). At method level, TsDetect’s heuristics appear less suited for fine-grained detection (notably for MT and DpT), reinforcing the need for cross-tool and manual validation when analysing test smell diffusion.
}}}

\vspace{0.5em}
\noindent\textbf{Interpretation of RQ1 (prevalence, granularity, and detector effects).}
Taken together, our RQ1 results portray LLMs as favouring broad, monolithic tests over small, focused ones. The dominance of General Fixture, Lazy Test, and Unknown Test in LLM-generated suites is consistent with a generation strategy where the model builds a single, rich fixture and then accumulates loosely related assertions to “cover” multiple behaviours in one shot. This maximises apparent coverage under typical prompts but sacrifices isolation and clarity. In contrast, Exception Handling remains underrepresented, indicating that current models rarely reason explicitly about error paths or failure scenarios when synthesising tests. The observed granularity effect—class-level generation amplifying global-context smells (GF, UT, EH, EmT, MT), while method-level generation concentrates Assertion Roulette—reinforces the view that more context encourages LLMs to overpopulate tests with shared setup and assertions rather than decomposing responsibilities into multiple cohesive test cases. Our cross-detector and manual validation results refine this picture. Some smells, such as Lazy Test, are detected reliably across TsDetect and JNose, whereas others (notably Assertion Roulette, Magic Number Test, and Dependent Test) are heavily detector-dependent, especially at method level. This confirms that part of the observed diffusion is a property of the detectors themselves, and that generator- and granularity-related conclusions are most trustworthy for smells that are stable across tools and supported by the manual oracle.
Given this ubiquity and the mechanical simplicity of its detection rule (multiple assertions without diagnostic messages in a single test), we treat Assertion Roulette as a structural baseline and focus our comparative analysis on how other smells—such as General Fixture, Lazy Test, Unknown Test, Exception Handling, and Duplicate Assert—differentiate LLM-, EvoSuite-, and human-generated suites.

\vspace{0.5em}
\noindent\highlight{Summary of \textbf{RQ1:}
LLM-generated tests systematically suffer from smells such as Assertion Roulette, Lazy Test, Unknown Test, and missing Exception Handling. While method-level generation alleviates some issues (lower LT and UT), exception handling remains absent, unlike EvoSuite which consistently covers exceptions.
Smell distribution is strongly influenced by test granularity and, as confirmed by our manual validation, by the detection tool itself: AR is consistently over-reported, LT is detected very reliably, and fine-grained smells such as Magic Number Test (MT) and Dependent Test (DpT) exhibit extreme cross-tool variability, with TsDetect favouring high recall and JNose favouring precision at method level. Compared to EvoSuite, LLM outputs amplify UT, underuse EH, and induce greater instability between detectors, underscoring both the originality and the fragility of LLM-generated test structures.
}

\subsection{[RQ2]: Test Smell Co-occurrence and Prompt Influence in LLM vs. SBST Suites}

\noindent\textbf{[Experiment Design]:} 
We analyze test smell co-occurrence using two benchmarks. Benchmark~1 varies prompting strategies (ZSL, FSL, CoT, ToT, GToT) on class-level test suites, while Benchmark~2 varies contextual information (Self-Contained, Simple, Full) on method-level test cases. After measuring individual smell prevalence, we compute normalized co-occurrence matrices for each benchmark, capturing the frequency of smell pairs across generation settings. Analyses are performed independently with TsDetect and JNose to assess detector consistency. We quantify inter-tool disagreement by calculating the mean and standard deviation of delta values between the two matrices, highlighting the five smell pairs with the largest disagreements. For comparison, we apply the same analysis to EvoSuite-generated test suites from Benchmark~1, enabling assessment of smell interaction patterns across LLM and SBST paradigms.

\noindent\textbf{[Experiment Results]:}

\noindent\textbf{Influence of Prompt Engineering Techniques.}
In Benchmark~1, prompting style measurably affects smell prevalence. For \textit{Assertion Roulette (AR)}, TsDetect reports rates from \textbf{20.94\%} (GToT) to \textbf{54.77\%} (ZSL), with JNose generally higher; structured reasoning prompts (GToT, CoT) \textit{moderately} reduce AR (full per-prompt distributions are reported in Appendix~\ref{app:additional-tables}, Table~\ref{tab:test-smells-benchmark1-prompts}). However, other smells remain sensitive to prompt design: \textit{Lazy Test (LT)} spans large intervals across prompts (e.g., \textbf{26.44\%}–\textbf{73.50\%}), indicating that some structured prompts can also correlate with more verbose or redundant logic.

\vspace{0.5em}
\noindent\colorbox{gray!15}{\parbox{0.98\linewidth}{
\textbf{Finding 8:} \textit{Structured reasoning prompts (e.g., GToT, CoT) can \emph{moderately} reduce AR at class level, but other smells (e.g., LT) remain highly prompt-sensitive, showing wide variance across strategies (Table~\ref{tab:test-smells-benchmark1-prompts}).}
}}

\vspace{0.5em}
\noindent\textbf{Effect of Context Variation.}
In Benchmark~2, varying the amount/structure of context (self-contained, simple, full) has limited effect on the main smells: AR remains \textbf{consistently above 90\%} across tools (TsDetect \textbf{94.78–96.23\%}, JNose \textbf{91.74–93.14\%}), \textit{Dependent Test (DpT)} exceeds \textbf{95\%} in TsDetect regardless of context, and \textit{Unknown Test (UT)} persists across levels. LT is generally lower than in Benchmark~1, yet \textit{Exception Handling (EH)} remains absent (full per-context distributions are reported in Appendix~\ref{app:additional-tables}, Tables~\ref{tab:test-smells-benchmark2-prompts} and~\ref{tab:test-smells-benchmark2}).

\vspace{0.5em}
\noindent\colorbox{gray!15}{\parbox{0.98\linewidth}{
\textbf{Finding 9:} \textit{Context granularity has limited impact on smells at method level: AR stays $>90\%$, DpT remains high, UT persists, and EH is not recovered, indicating that more input context alone does not mitigate key smell patterns (Table~\ref{tab:test-smells-benchmark2-prompts}).}
}}

\begin{figure*}[ht]
\vspace{-3mm}
\centering

\subfloat[{\scriptsize TsDetect Co-occurrence Matrix (Benchmark 1 - LLM)}\label{fig:cooc-bench1-llm-tsdetect}]{
    \includegraphics[width=0.48\textwidth]{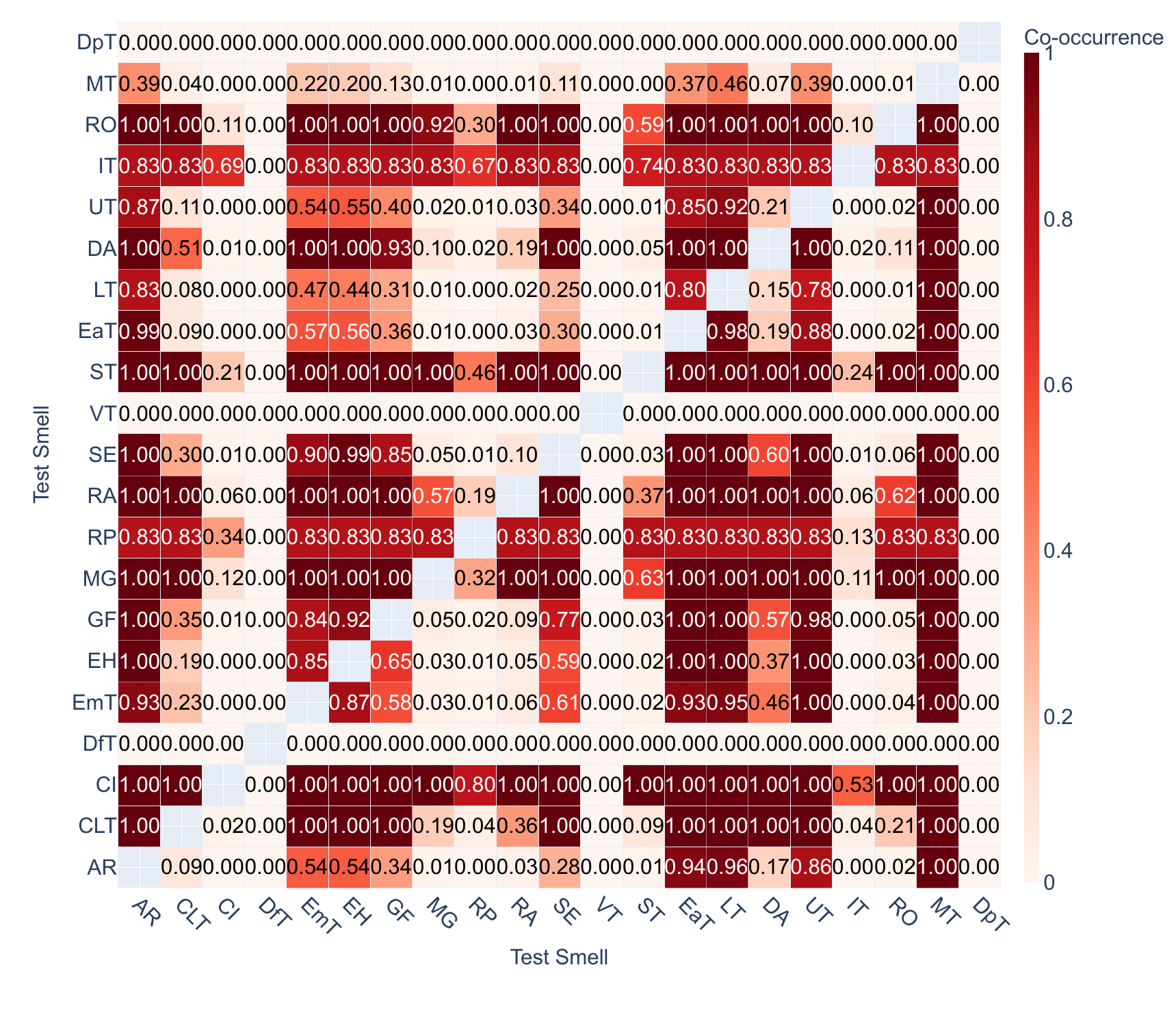}
}
\hfill
\subfloat[{\scriptsize JNose Co-occurrence Matrix (Benchmark 1 - LLM)}\label{fig:cooc-bench1-llm-jnose}]{
    \includegraphics[width=0.48\textwidth]{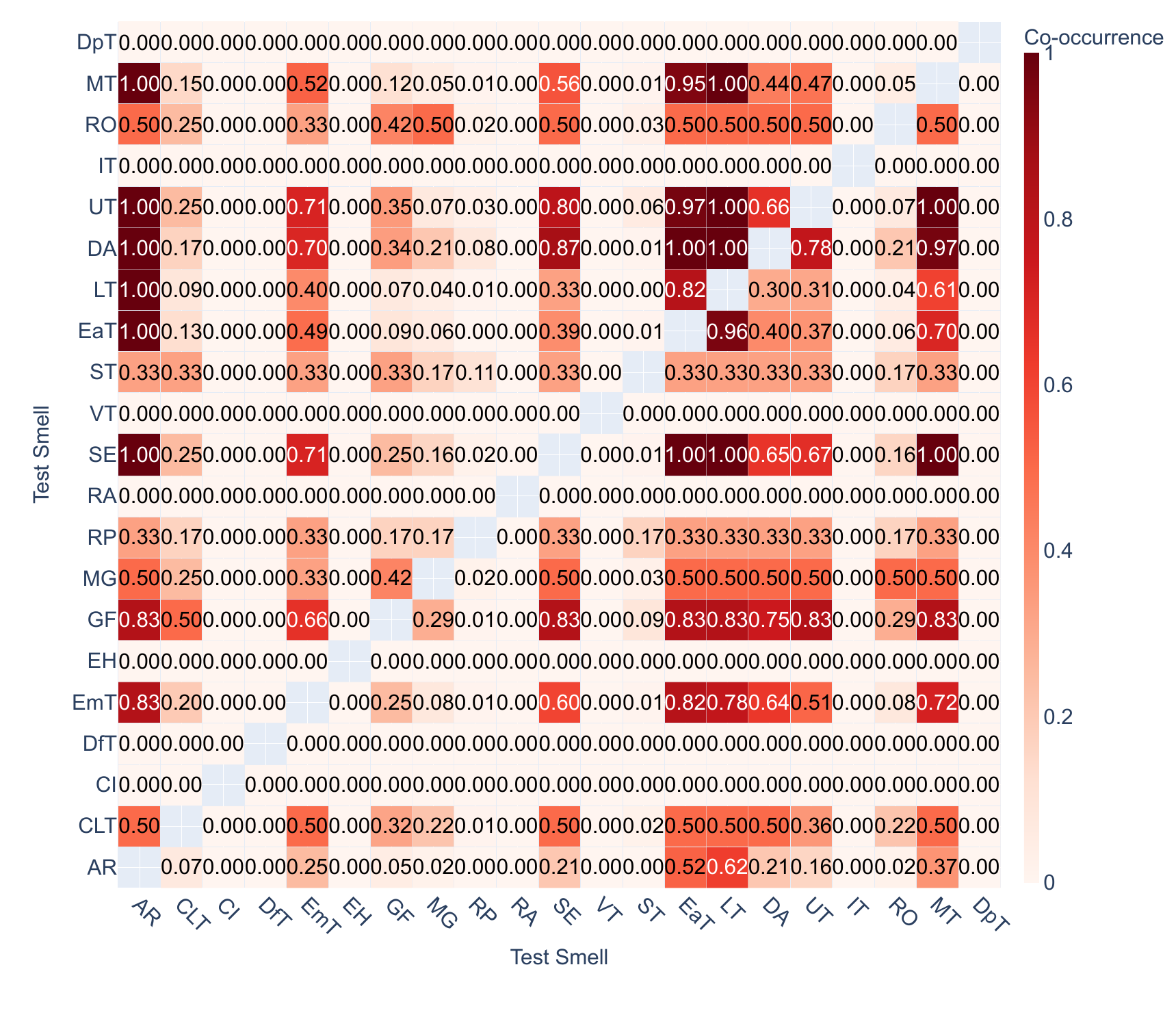}
}


\subfloat[{\scriptsize TsDetect Co-occurrence Matrix (Benchmark 1 - EvoSuite)}\label{fig:cooc-bench1-evosuite-tsdetect}]{
    \includegraphics[width=0.48\textwidth]{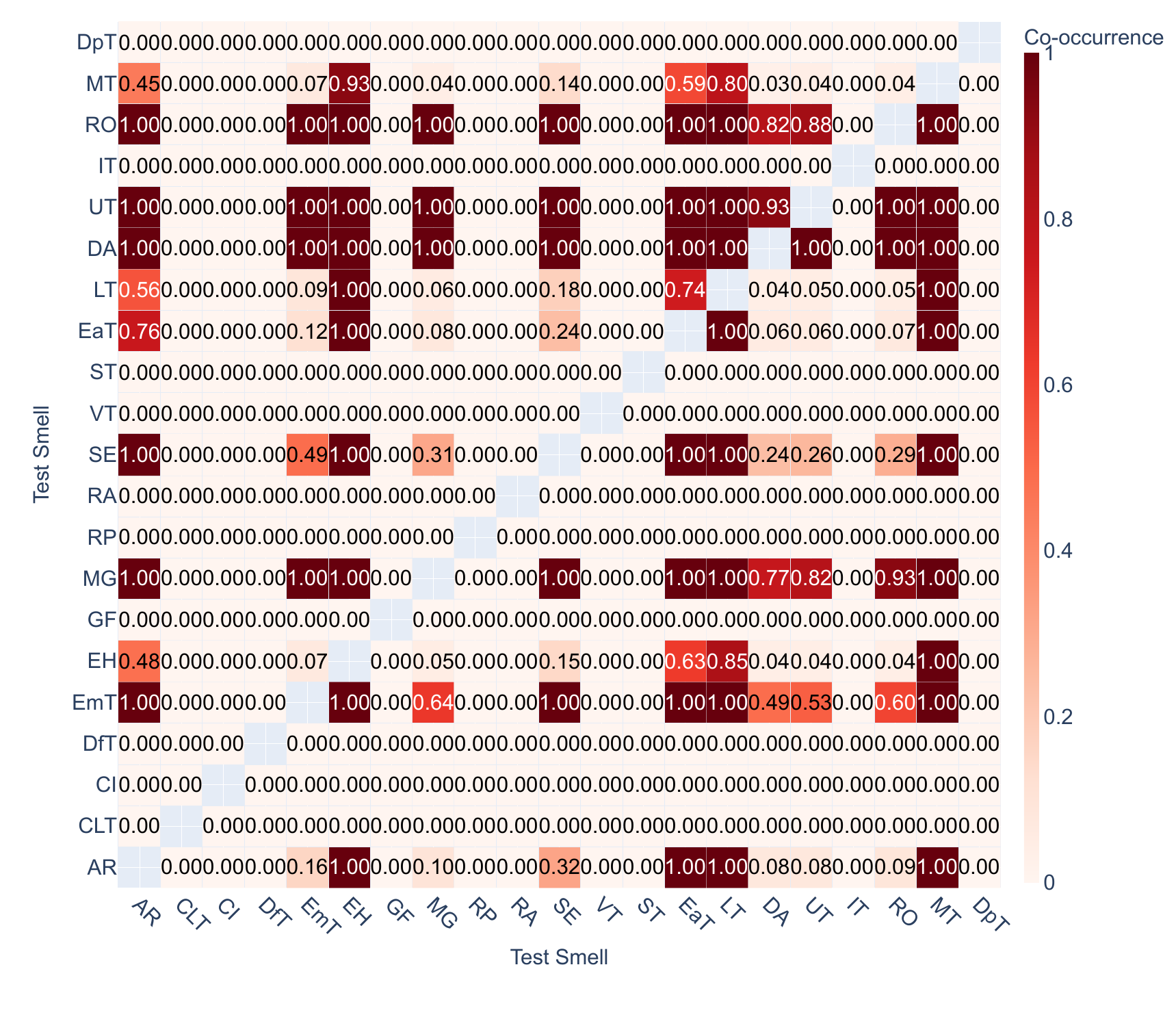}
}
\hfill
\subfloat[{\scriptsize JNose Co-occurrence Matrix (Benchmark 1 - EvoSuite)}\label{fig:cooc-bench1-evosuite-jnose}]{
    \includegraphics[width=0.48\textwidth]{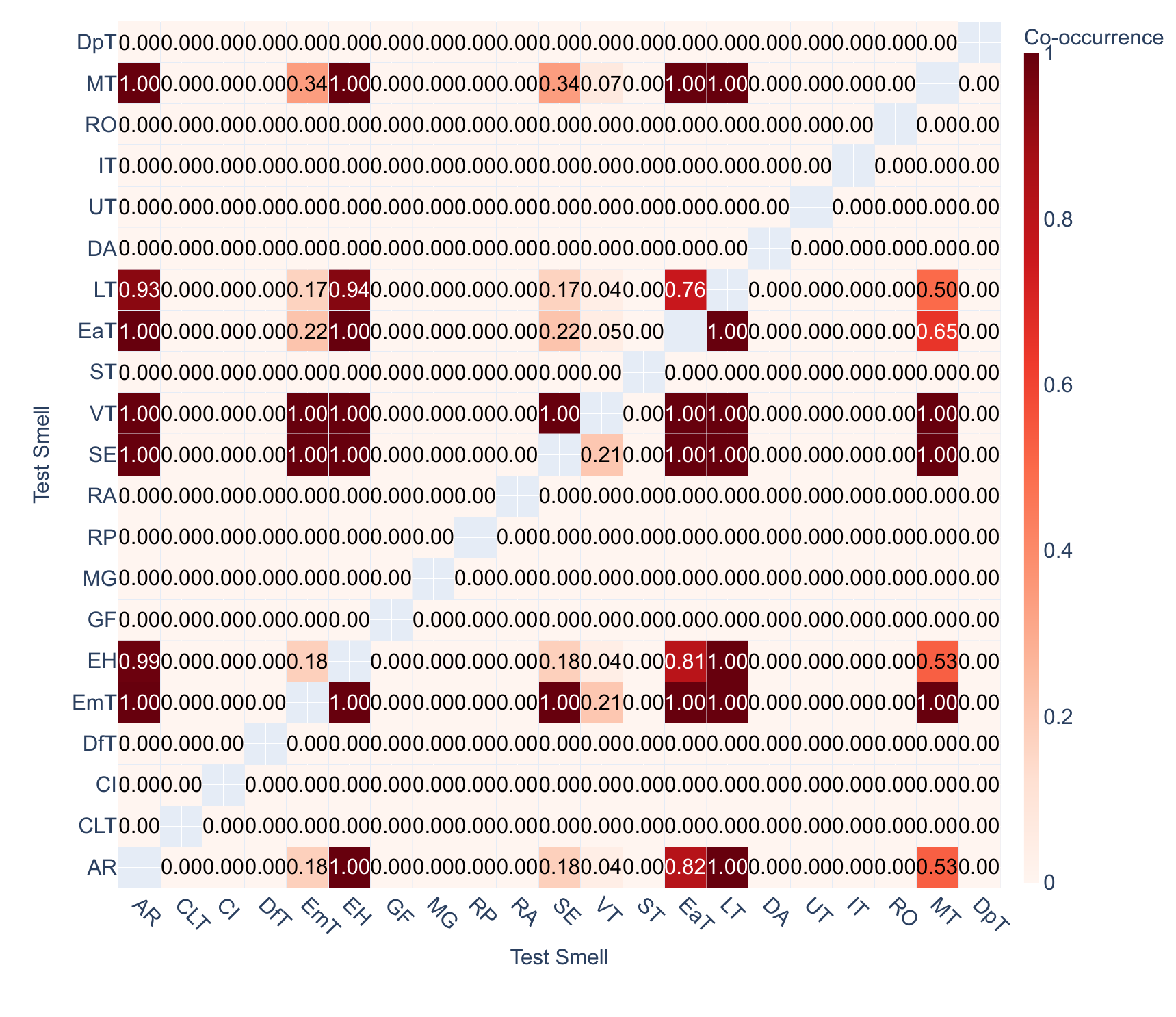}
}

\caption{Test Smell Co-occurrence Matrices detected by TsDetect and JNose for Benchmark 1 (LLMs and EvoSuite).}
\label{fig:cooc-benchmark1-combined}
\vspace{-3mm}
\end{figure*}

\begin{figure*}[ht]
\vspace{-3mm}
\centering
\subfloat[{\scriptsize TsDetect Co-occurrence Matrix}\label{fig:cooc-bench2-tsdetect}]{
    \includegraphics[width=0.48\textwidth]{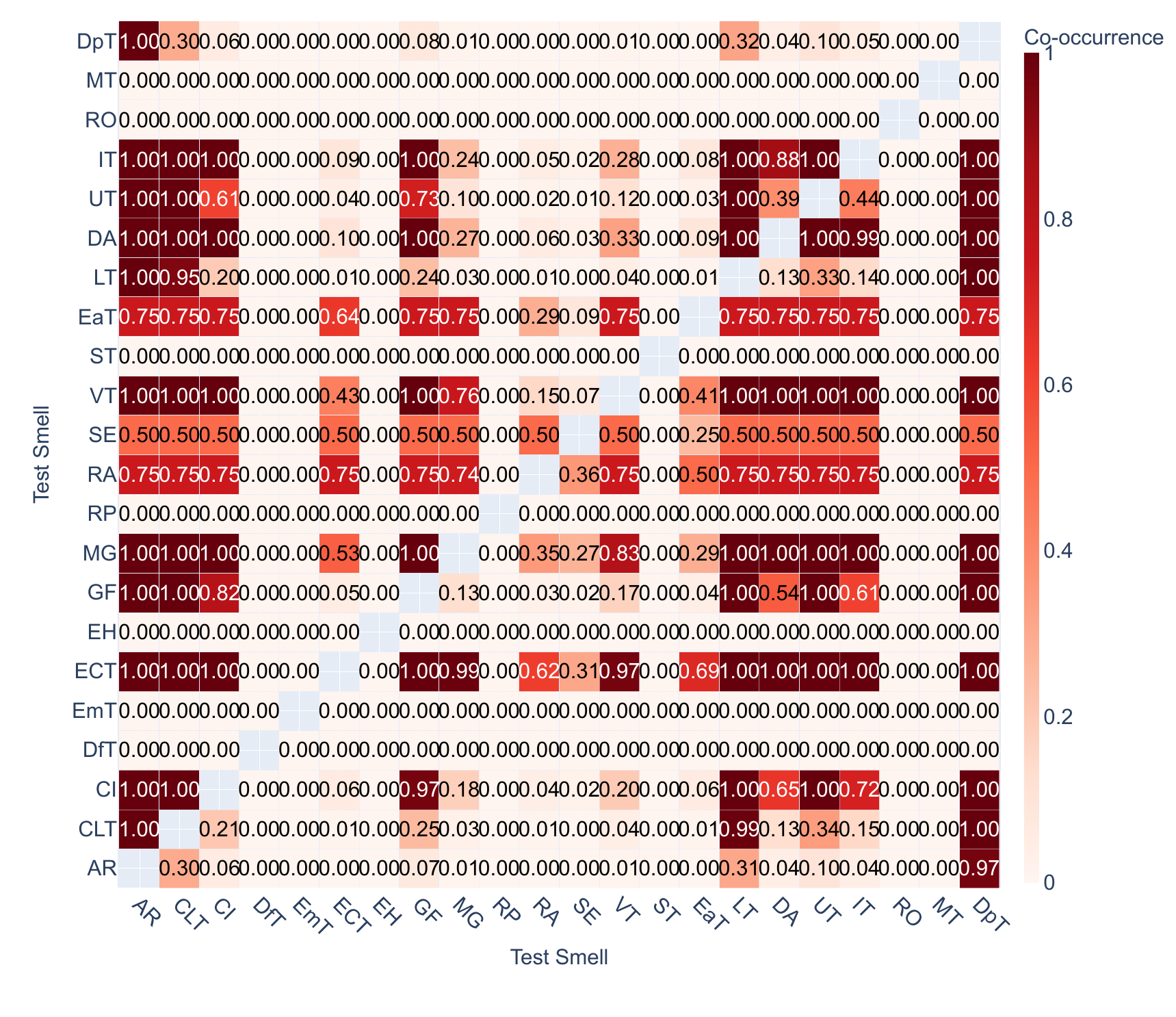}
}
\subfloat[{\scriptsize JNose Co-occurrence Matrix}\label{fig:cooc-bench2-jnose}]{
    \includegraphics[width=0.48\textwidth]{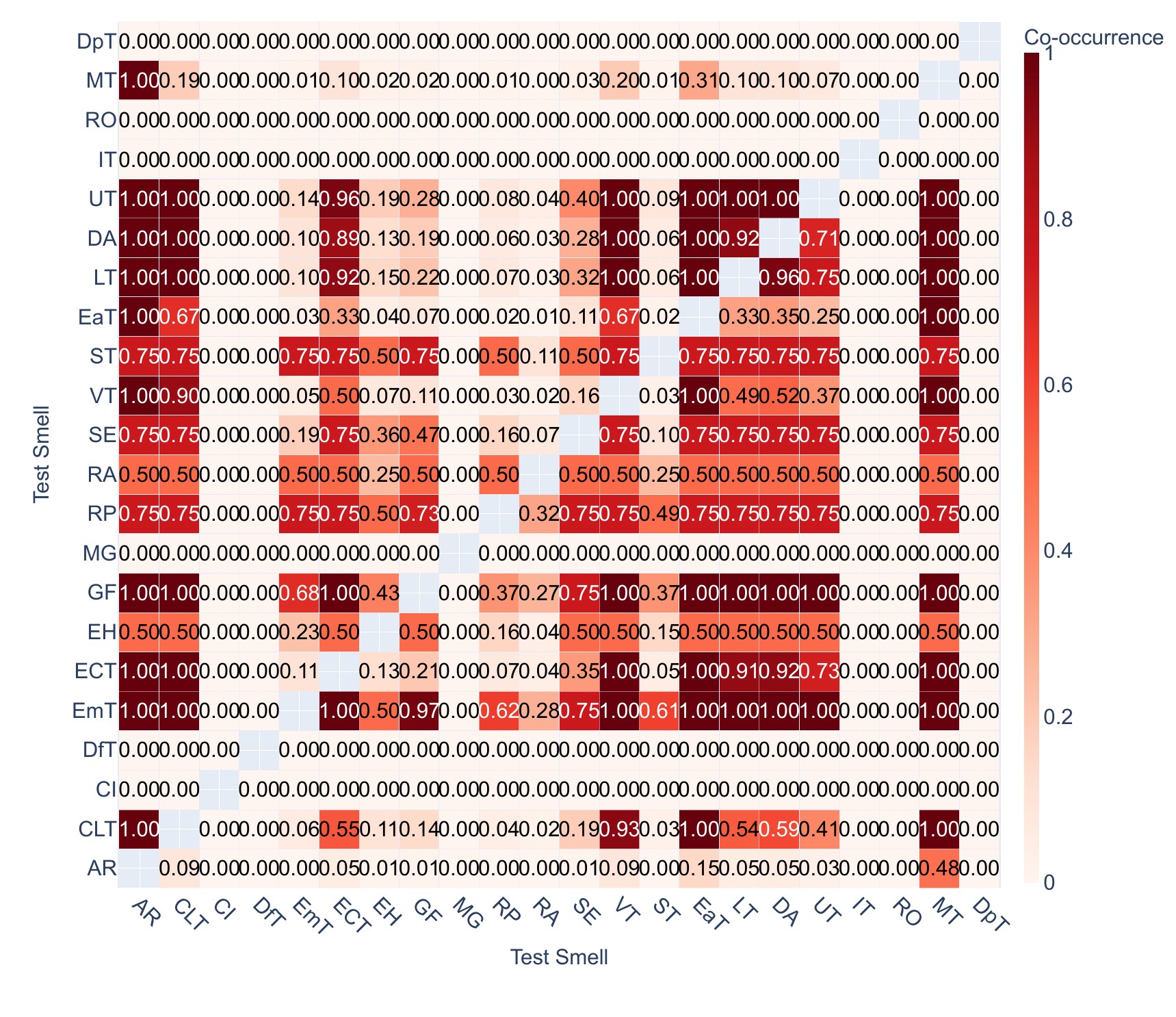}
}
\caption{Test Smell Co-occurrence Matrices for Benchmark 2 detected by TsDetect and JNose.}
\label{fig:cooc-benchmark2}
\vspace{-3mm}
\end{figure*}

\vspace{0.5em}
\noindent\textbf{Co-occurrence Patterns (Tools \& Granularity).}
The co-occurrence matrices that underpin our analysis are shown in Figure~\ref{fig:cooc-bench1-llm-tsdetect} and Figure~\ref{fig:cooc-bench1-llm-jnose} for class-level LLM-generated tests in Benchmark~1, in Figure~\ref{fig:cooc-bench1-evosuite-tsdetect} and Figure~\ref{fig:cooc-bench1-evosuite-jnose} for EvoSuite on the same benchmark, and in Figure~\ref{fig:cooc-bench2-tsdetect} and Figure~\ref{fig:cooc-bench2-jnose} for method-level LLM tests in Benchmark~2.
For LLM-generated tests, co-occurrence structures differ by detector and granularity. With TsDetect at class level (Benchmark~1), we observe dense structural clusters—\textit{GF, RO, RA, SE, AR}—often exceeding \textbf{90\%}, sometimes reaching \textbf{100\%}; JNose highlights more semantic pairings such as \textit{LT+DA} and \textit{UT+DA} (up to \textbf{100\%}) in the co-occurrence matrices of Figures~\ref{fig:cooc-bench1-llm-tsdetect}–\ref{fig:cooc-bench1-llm-jnose}. 
At method level (Benchmark~2), clusters become denser (e.g., \textit{CLT, CI, ECT} with TsDetect; \textit{VT} as hub with many smells in JNose) (Figures~\ref{fig:cooc-bench2-tsdetect}–\ref{fig:cooc-bench2-jnose}). 
Cross-tool disagreement on co-occurrence, summarised in Table~\ref{tab:delta-stats-cooc}, increases with granularity: mean $\Delta$ rises from \textbf{0.287} ($\sigma=\textbf{0.349}$) at class level to \textbf{0.323} ($\sigma=\textbf{0.370}$) at method level, and the top-5 most divergent smell pairs reach $\Delta=\textbf{1.00}$ (Table~\ref{tab:top5-delta-cooc}).
From a maintenance viewpoint, combinations such as \textit{LT+DA} and \textit{UT+DA} indicate tests that simultaneously lack focused scenarios and contain redundant assertions, forcing developers to inspect long, repetitive assertion sequences whose intent is only weakly specified. Similarly, TsDetect clusters around \textit{GF, RO, SE, AR, RA} correspond to tests that mix large, over-general fixtures, external resources, and unlabelled assertions, compounding debugging and refactoring effort compared to treating each smell in isolation.

\vspace{0.5em}
\noindent\colorbox{gray!15}{\parbox{0.98\linewidth}{
\textbf{Finding 10:} \textit{In LLM-generated tests co-occurrence depends on both detector and granularity: TsDetect emphasizes clusters centered on assertions and fixtures (e.g., GF, RO, RA, SE, AR), whereas JNose surfaces semantically grounded pairings (e.g., LT+DA, UT+DA). Disagreement grows at method level (mean $\Delta$ from 0.287 to 0.323), with several pairs reaching maximum divergence ($\Delta=1.00$).}
}}


\vspace{0.5em}
EvoSuite exhibits distinctive co-occurrence trends compared to LLMs. With TsDetect, \textit{Empty Test (EmT)} often anchors clusters that include \textit{AR, EaT, LT, DA} (frequently up to \textbf{100\%}); JNose shows strong exception-related groupings (e.g., \textit{EH+EaT}, \textit{EH+MT}, approaching \textbf{99–100\%}). These clusters largely reflect EvoSuite’s limited assertion synthesis and omission of exception handling. In contrast, LLM-generated clusters are shaped by verbose or repetitive logic (e.g., \textit{VT} as a hub; \textit{LT+DA}, \textit{UT+DA}), reflecting tendencies toward redundancy and loosely scoped assertions. Together, these differences highlight that EvoSuite clusters emerge from omission of logic and exception handling, whereas LLM clusters arise from overproduction and repetition of assertions (Figures~\ref{fig:cooc-bench1-evosuite-tsdetect}–\ref{fig:cooc-bench1-evosuite-jnose}). 
In practice, this means that automatically generated suites rarely exhibit isolated smells: EvoSuite often combines shallow tests (e.g., \textit{EmT}) with brittle exception patterns, while LLMs tend to accumulate verbosity- and redundancy-oriented smells, so that effective refactoring must address smell clusters rather than single instances.

\vspace{0.5em}
\noindent\colorbox{gray!15}{\parbox{0.98\linewidth}{
\textbf{Finding 11:} 
\textit{In our comparison between LLM-generated and EvoSuite-generated tests, co-occurrence patterns differ sharply. EvoSuite co-occurrence is anchored by \textit{Empty Test} and exception-related combinations (e.g., \textit{EmT} with \textit{AR}, \textit{EaT}, \textit{LT}, \textit{DA}, and pairs such as \textit{EH+EaT}, \textit{EH+MT}), reflecting omission of logic and exception handling. In contrast, LLM-generated tests accumulate clusters driven by redundancy, verbosity, and weakly specified logic (e.g., \textit{VT} as a hub and pairings such as \textit{LT+DA} or \textit{UT+DA}), which are rare or much less pronounced in EvoSuite, evidencing distinct generative biases.}
}}


\begin{table}[ht]
\centering
\caption{Detector disagreement on smell co-occurrence (Mean and Standard Deviation)}
\label{tab:delta-stats-cooc}
\scalebox{0.7}{
\begin{tabular}{lcc}
\toprule
\textbf{Benchmark} & \textbf{Mean Delta} & \textbf{Std Deviation} \\
\midrule
Benchmark 1 & 0.287 & 0.349 \\
Benchmark 2 & 0.323 & 0.370 \\
\bottomrule
\end{tabular}
}
\end{table}

\begin{table*}[ht]
\centering
\caption{Top-5 Largest Deltas between TsDetect and JNose for Co-occurrence Matrices}
\label{tab:top5-delta-cooc}
\resizebox{\textwidth}{!}{
\begin{tabular}{ccccc|ccccc|ccccc}
\toprule
\multicolumn{5}{c|}{\textbf{Benchmark 1 - LLM}} & \multicolumn{5}{c|}{\textbf{Benchmark 1 - EvoSuite}} & \multicolumn{5}{c}{\textbf{Benchmark 2}} \\
\midrule
\textbf{Smell Pair} & \textbf{TsDetect} & \textbf{JNose} & \textbf{Delta} & \textbf{Rank} &
\textbf{Smell Pair} & \textbf{TsDetect} & \textbf{JNose} & \textbf{Delta} & \textbf{Rank} &
\textbf{Smell Pair} & \textbf{TsDetect} & \textbf{JNose} & \textbf{Delta} & \textbf{Rank} \\
\midrule
RA-EH & 1.00 & 0.00 & 1.00 & 1 & SE-VT & 0.00 & 1.00 & 1.00 & 1 & EmT-EaT & 0.00 & 1.00 & 1.00 & 1 \\
CI-EmT & 1.00 & 0.00 & 1.00 & 2 & RO-MG & 1.00 & 0.00 & 0.93 & 2 & VT-IT & 1.00 & 0.00 & 1.00 & 2 \\
CI-RO & 1.00 & 0.00 & 1.00 & 3 & UT-RO & 1.00 & 0.00 & 0.88 & 3 & DA-MT & 0.00 & 1.00 & 1.00 & 3 \\
CLT-EH & 1.00 & 0.00 & 1.00 & 4 & UT-MG & 1.00 & 0.00 & 0.82 & 4 & MG-UT & 1.00 & 0.00 & 1.00 & 4 \\
RA-CLT & 1.00 & 0.00 & 1.00 & 5 & DA-MG & 1.00 & 0.00 & 0.77 & 5 & MG-DA & 1.00 & 0.00 & 1.00 & 5 \\
\bottomrule
\end{tabular}
}
\vspace{-3mm}
\end{table*}

\vspace{0.5em}
\noindent\textbf{Relation to Prior Work.}
These findings align with prior observations by Panichella et al.~\cite{panichella2022test}, who reported moderate disagreement between smell detectors on EvoSuite- and JTexpert-generated tests, and highlight that detector choice can substantially affect empirical conclusions. Our study extends this perspective by incorporating LLM-generated outputs and by introducing granularity (class- vs.\ method-level) as a key factor, showing that detector divergence is exacerbated for fine-grained LLM tests.
Complementary to detector disagreement, Santana et al.~\cite{santana2025empirical} conducted an empirical study of test smell co-occurrence in human-written tests, identifying recurrent smell clusters (e.g., patterns around Assertion Roulette and Magic Number) and arguing that smells often appear in combinations rather than in isolation. Our co-occurrence analysis generalises this line of work to automatically generated suites, contrasting LLMs and EvoSuite across two detectors and two granularities. We find that EvoSuite’s co-occurrence is anchored by \textit{Empty Test} and exception-related patterns, whereas LLM-generated tests concentrate redundancy- and verbosity-oriented clusters (e.g., \textit{LT+DA}, \textit{UT+DA}, and \textit{VT}-centred hubs), indicating that modern LLM-based generators introduce distinctive smell combinations compared to both SBST and previously studied human-written tests.

\vspace{0.5em}
\noindent\textbf{Interpretation of RQ2 (prompting strategies and input context).}
From a generative perspective, the RQ2 results indicate that more expressive prompting and richer context do not simply “clean up” LLM-generated tests, but instead shift the balance between smells. Structured reasoning prompts (e.g., CoT, GToT) can moderately reduce Assertion Roulette at class level, yet they often encourage verbose fixtures and long assertion sequences, reinforcing General Fixture, Lazy Test, and Duplicate Assert. Varying the amount of input context at method level has similarly limited impact on core defects: AR remains extremely high, Dependent Test and Unknown Test persist, and Exception Handling is not recovered. This suggests that providing more information or more elaborate reasoning instructions alone is insufficient to steer the model towards smell-free designs. The co-occurrence patterns further illustrate distinct generative biases. TsDetect emphasises clusters centred on assertions and fixtures (e.g., GF with RO, RA, SE, AR), whereas JNose surfaces pairings that reflect redundancy and weak specification (e.g., LT+DA, UT+DA), particularly at method level where detector disagreement increases. Compared to EvoSuite—whose co-occurrence is anchored by Empty Test and exception-related templates—LLM-generated tests accumulate smell clusters driven by verbosity and loosely specified logic. Overall, RQ2 shows that prompting mainly redistributes smells rather than eliminating them, and that LLMs tend to optimise for plausibility and “thoroughness” rather than for minimal, well-factored test suites.

\vspace{0.5em}
\noindent\highlight{Summary of \textbf{RQ2:}
Test smell co-occurrence in LLM-generated suites is influenced more by prompting strategies than by context granularity: reasoning-based prompts (e.g., CoT, GToT) moderately reduce AR at class level, but key smells (LT, UT, DpT) persist even with richer context. Detector choice further shapes observed clusters—TsDetect emphasizing assertion- and fixture-centered structures, while JNose surfaces semantic pairings such as LT+DA or UT+DA—with disagreement peaking at method level. Compared to EvoSuite, which clusters around omissions (Empty Test, missing exceptions), LLMs form redundancy- and verbosity-driven clusters, highlighting distinct generative biases.}

\subsection{[RQ3]: Impact of Software Attributes and LLM Parameters on Test Smell Prevalence}

\noindent\textbf{[Experiment Design]:} 
We analyze test smell prevalence in LLM-generated tests (Benchmarks~1 and~2) and EvoSuite-generated tests (Benchmark~1). As explanatory variables, we consider both software attributes—LOC, number of classes/methods, cyclomatic complexity, CBO, RFC, and DIT—and LLM parameters such as model size, context window, temperature, and top-p. To capture relationships between these features and smell prevalence, we adopt a hybrid correlation strategy: Pearson’s correlation (PMCC) is applied across all benchmarks to measure linear associations, while Spearman’s rank correlation, Mutual Information, and Kruskal–Wallis tests are selectively used to detect monotonic and non-linear effects, restricted to Benchmark~1 where variance is sufficient. Non-linear analyses are omitted when variable distributions are invariant or statistically unsuitable (e.g., method-level tests in Benchmark~2). Invariant smells and features are filtered to ensure statistical validity. This design enables robust yet scalable correlation analysis across diverse generation settings.

\noindent\textbf{[Experiment Results]:}

\vspace{0.5em}
\noindent\textbf{Non-Linear Influence Analysis (Benchmark 1).}  
We first assess whether software attributes or LLM parameters exert non-linear effects on smell prevalence. Using Kruskal--Wallis and Mutual Information (MI), we analyzed LLM-generated suites in Benchmark~1, which exhibits sufficient variance across features. Benchmark~2 was excluded from non-linear analysis due to highly invariant distributions (e.g., AR $>$90\%, DpT $>$95\%). Results (summarised in Appendix~\ref{app:additional-tables}, Tables~\ref{tab:kruskal-benchmark1-summary} and~\ref{tab:mi-benchmark1}) show no significant non-linear effects: all $p$-values $\geq 0.1$, and MI scores were zero or negligible ($\sim\!10^{-16}$, attributable to floating-point noise).

\vspace{0.5em}
\noindent\colorbox{gray!20}{\parbox{0.98\linewidth}{
\textbf{Finding 12:} \textit{Non-linear analyses (Kruskal--Wallis and MI) show no measurable influence of software attributes (size, complexity, coupling) or LLM-specific parameters (model size, context length, decoding settings) on test smell prevalence in Benchmark~1.}
}}

\begin{figure*}[ht]
\centering
\begin{subfigure}[b]{0.48\textwidth}
    \centering
    \includegraphics[width=\textwidth]{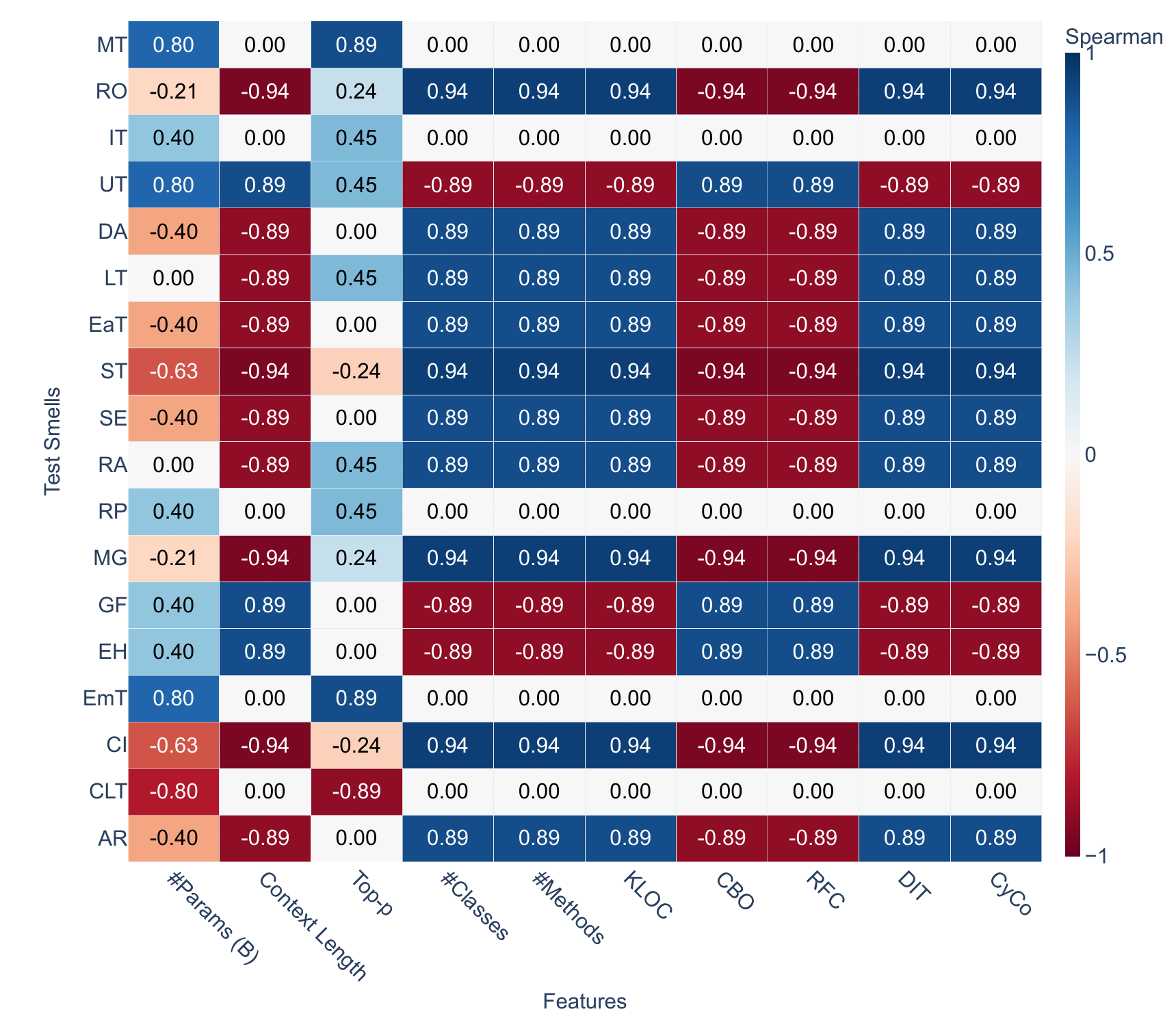}
    \caption{Spearman Correlation results from TsDetect (Benchmark 1.)}
    \label{fig:spearman-tsdetect-benchmark1}
\end{subfigure}
\hfill
\begin{subfigure}[b]{0.48\textwidth}
    \centering
    \includegraphics[width=\textwidth]{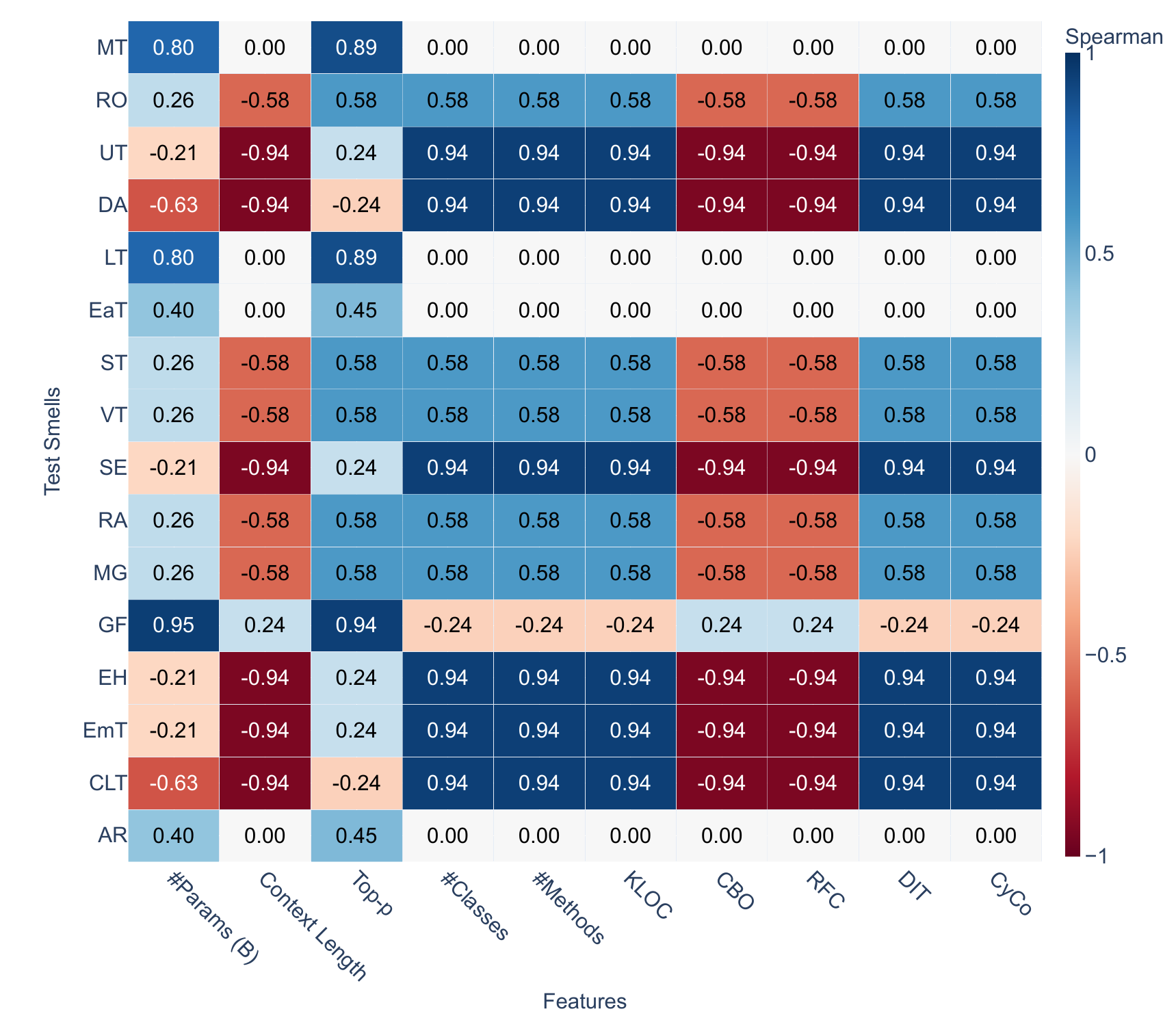}
    \caption{Spearman Correlation results from JNose (Benchmark 1.)}
    \label{fig:spearman-jnose-benchmark1}
\end{subfigure}
\caption{Spearman Correlation results between Project and LLM Characteristics and Test Smell Presence.}
\label{fig:spearman-combined-barplots}
\end{figure*}

\begin{figure*}[ht]
\centering
\begin{subfigure}[b]{0.48\textwidth}
    \centering
    \includegraphics[width=\textwidth]{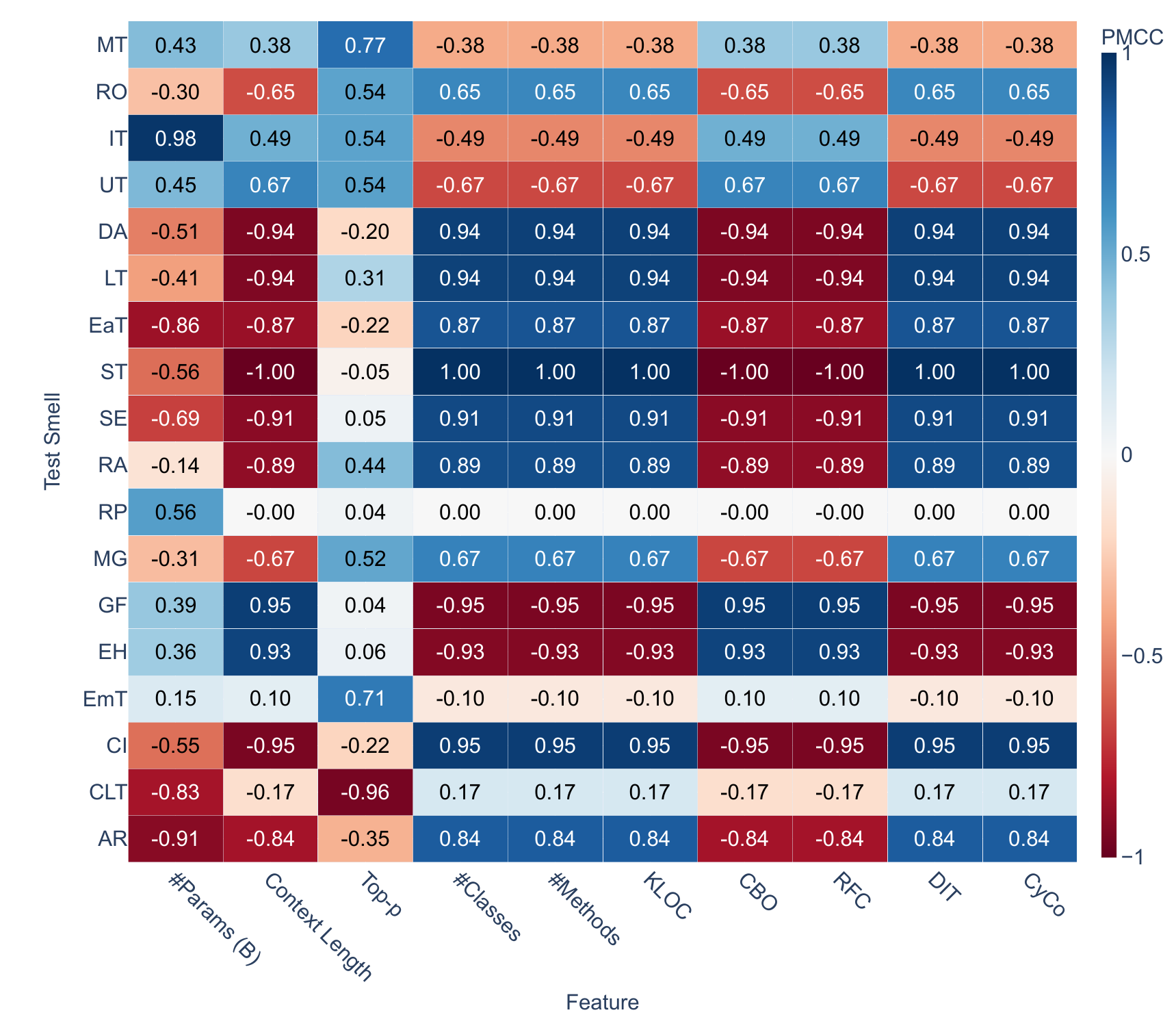}
    \caption{PMCC result from TsDetect (Benchmark 1-LLM).}
    \label{fig:pmcc-tsdetect-benchmark1}
\end{subfigure}
\hfill
\begin{subfigure}[b]{0.48\textwidth}
    \centering
    \includegraphics[width=\textwidth]{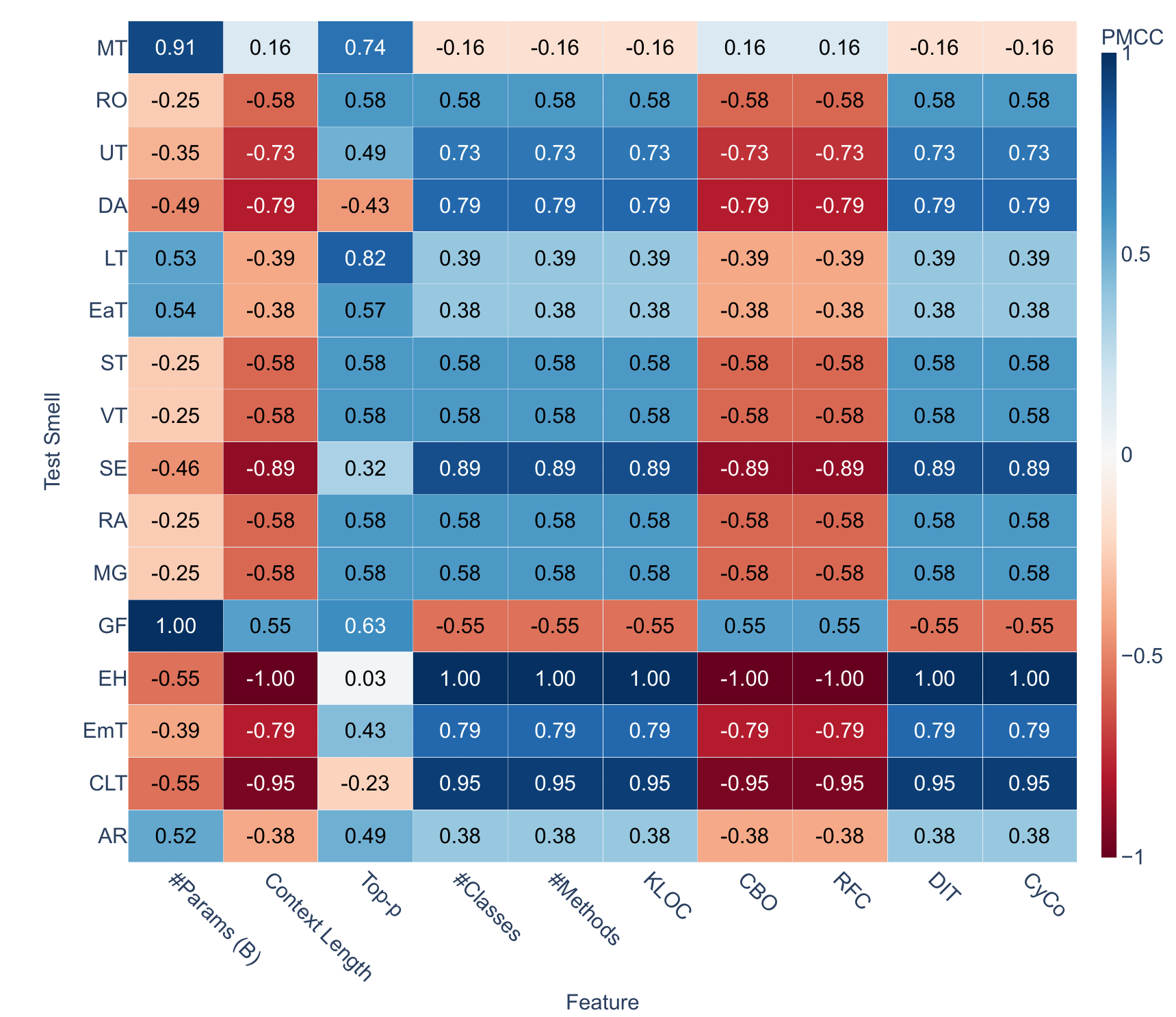}
    \caption{PMCC result from JNose (Benchmark 1-LLM).}
    \label{fig:pmcc-jnose-benchmark1}
\end{subfigure}
\hfill
\begin{subfigure}[b]{0.48\textwidth}
    \centering
    \includegraphics[width=\textwidth]{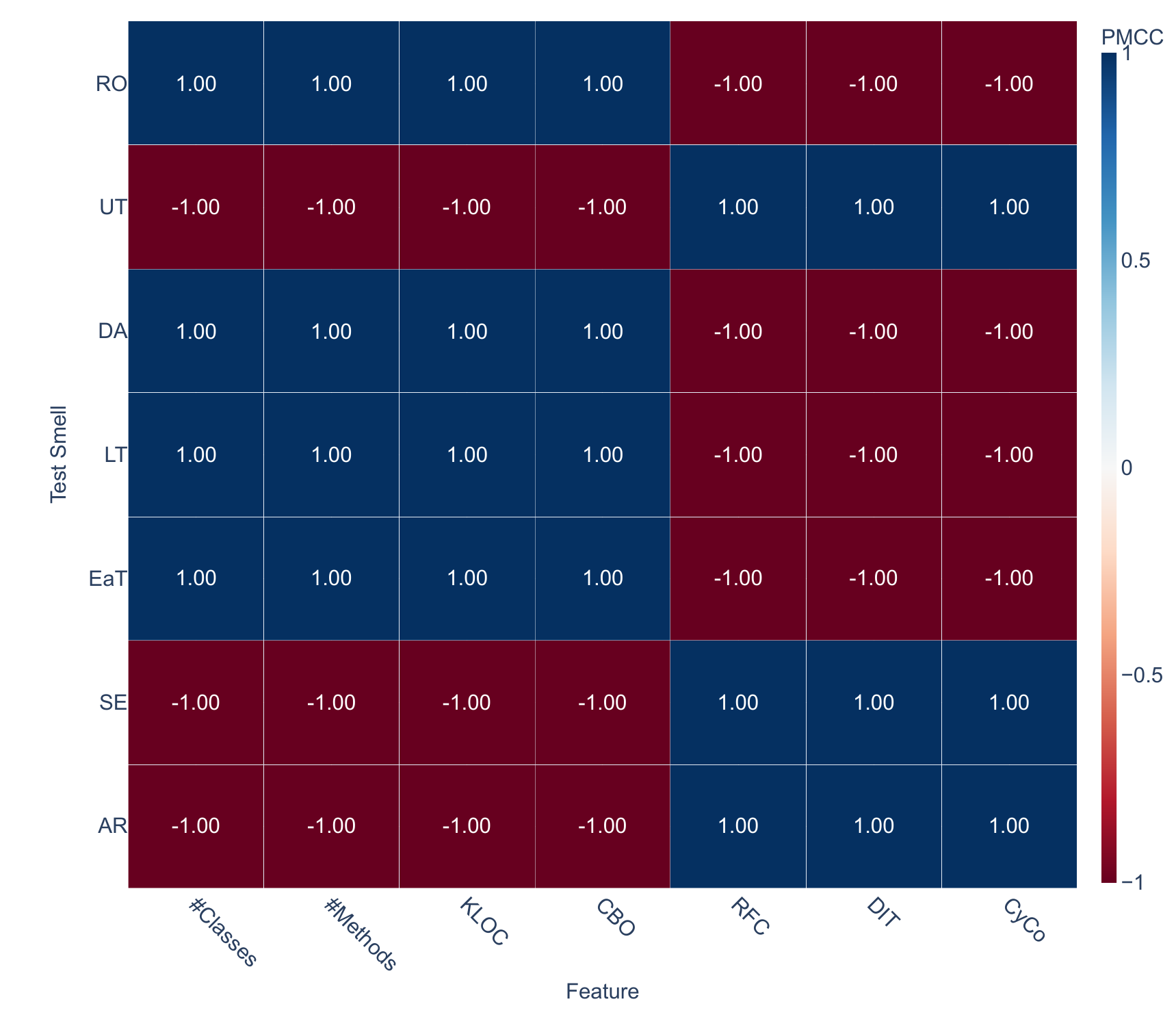}
    \caption{PMCC result from TsDetect (Benchmark 1-EvoSuite).}
    \label{fig:pmcc-tsdetect-benchmark1-evosuite}
\end{subfigure}
\hfill
\begin{subfigure}[b]{0.48\textwidth}
    \centering
    \includegraphics[width=\textwidth]{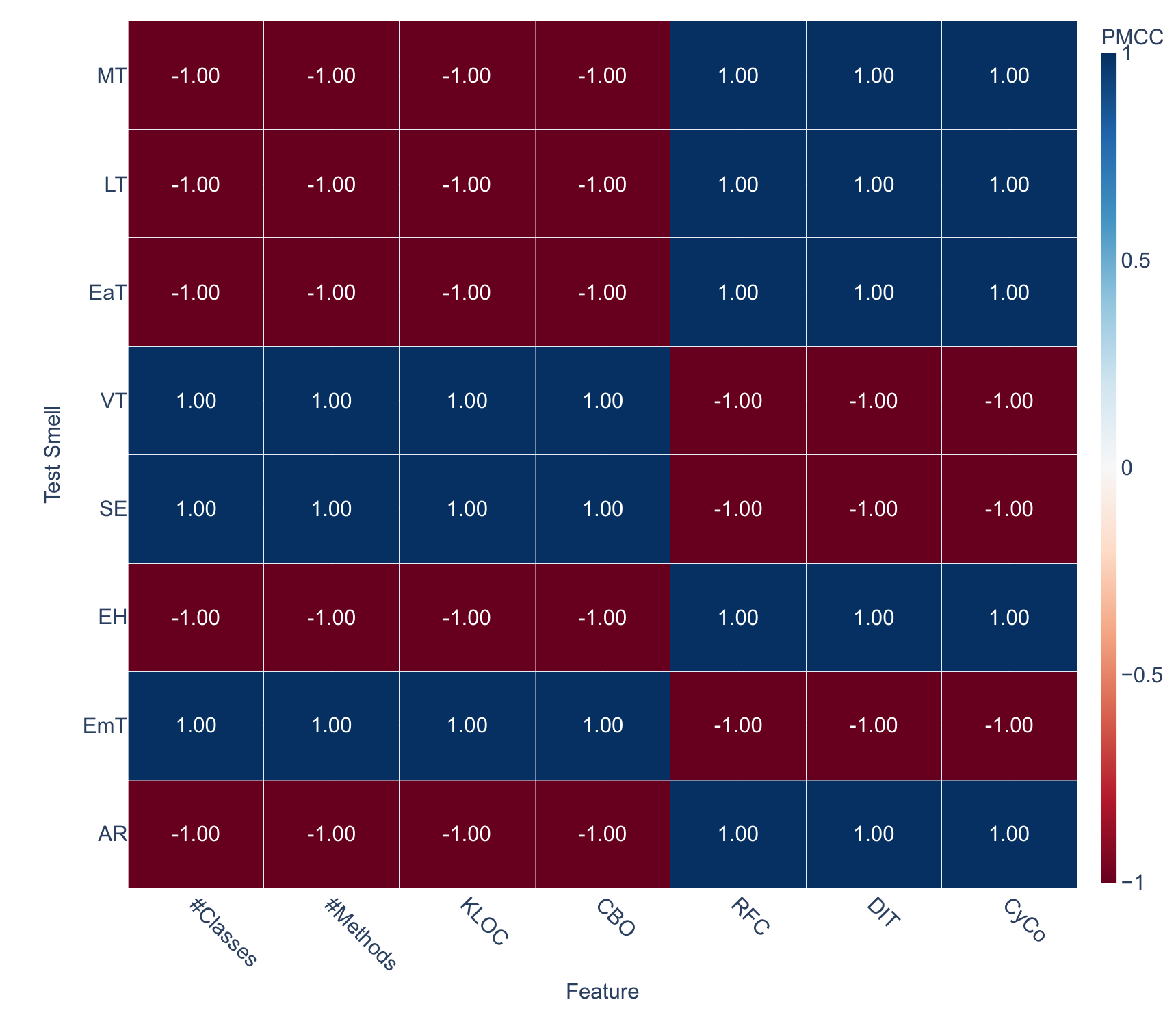}
    \caption{PMCC result from JNose (Benchmark 1-EvoSuite).}
    \label{fig:pmcc-jnose-benchmark1-evosuite}
\end{subfigure}
\caption{PMCC results between Project and LLM Characteristics and Test Smell Presence in Benchmark 1.}
\label{fig:pmcc-combined-heatmaps-benchmark1}
\end{figure*}

\vspace{0.5em}
\noindent\textbf{Impact of Software Attributes (Benchmark~1).}  
Figure~\ref{fig:spearman-combined-barplots} summarises the Spearman correlations between software attributes and smell presence across both detectors, while Figures~\ref{fig:pmcc-tsdetect-benchmark1} and~\ref{fig:pmcc-jnose-benchmark1} detail Pearson coefficients per tool. 
Correlation analysis (Spearman and Pearson) shows that project size and complexity/coupling exert distinct influences (Figures~\ref{fig:spearman-combined-barplots}, \ref{fig:pmcc-tsdetect-benchmark1}, \ref{fig:pmcc-jnose-benchmark1}).  
Larger systems (more classes, methods, LOC) amplify redundant and superficial smells. TsDetect reports strong positive correlations $\rho \approx +0.84$ for \texttt{AR}, \texttt{RA}, \texttt{SE}, \texttt{EaT}, \texttt{LT}, while JNose highlights \texttt{MG}, \texttt{RO}, and \texttt{VT} ($\rho \approx +0.58$).  
Higher complexity and coupling (CBO, RFC, CyCo) shift smell prevalence toward architectural flaws: TsDetect reports $\rho \geq +0.89$ for \texttt{EH}, \texttt{GF}, \texttt{CLT}, while JNose corroborates with \texttt{EmT} and \texttt{CLT}. Verbosity (\texttt{VT}) decreases slightly, suggesting a transition from shallow redundancy to deeper structural issues. 

\vspace{0.5em}
\noindent\colorbox{gray!20}{\parbox{0.98\linewidth}{
\textbf{Finding 13:} \textit{Software attributes shape smell prevalence: large-scale systems amplify redundant smells (AR, RA, MG), while higher complexity and coupling induce architectural flaws (EH, GF, CLT). This suggests a progression from superficial to structural defects as systems grow in size and interdependence.}
}}

\begin{figure*}[ht]
\centering
\begin{subfigure}[b]{0.48\textwidth}
    \centering
    \includegraphics[width=\textwidth]{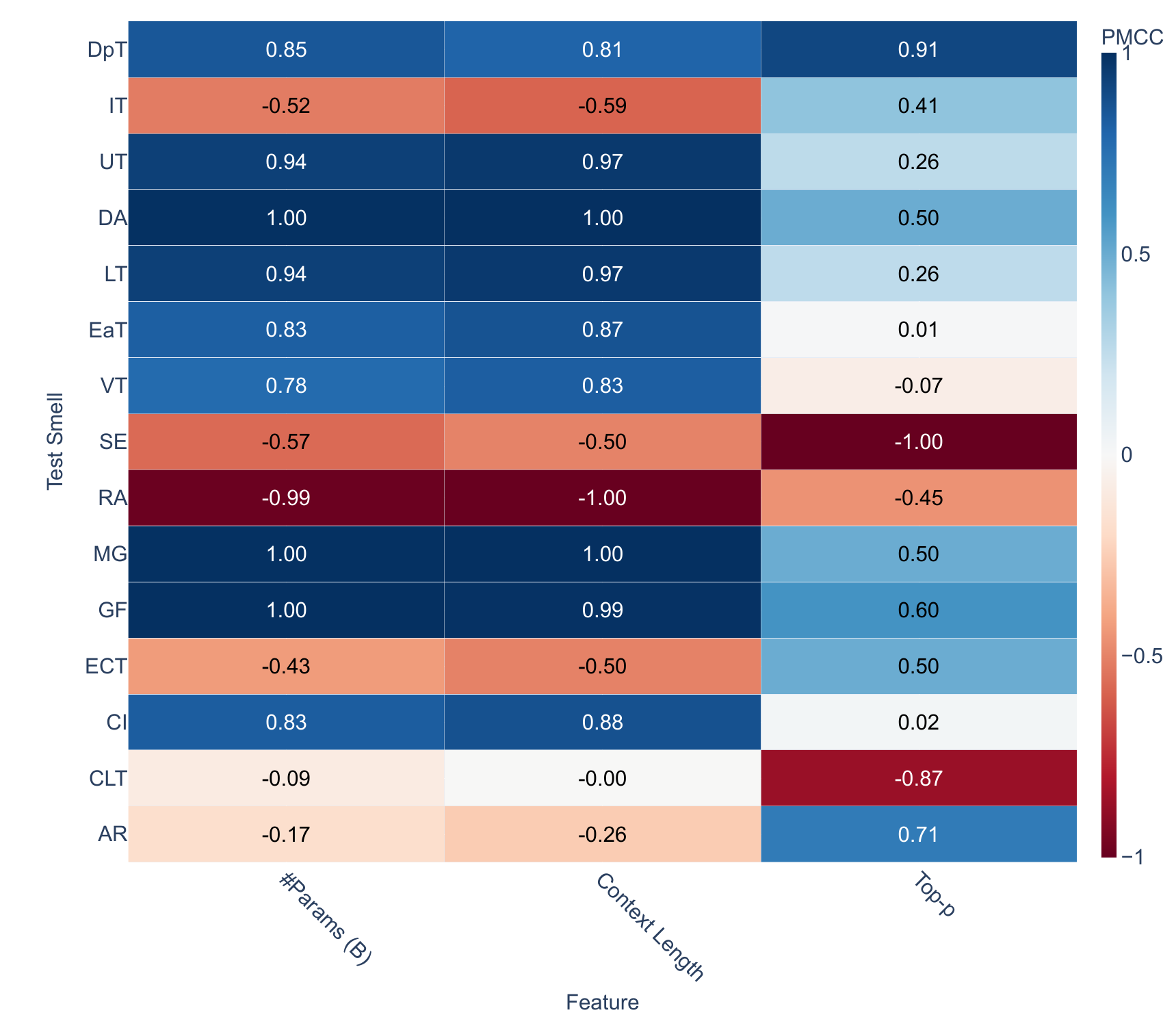}
    \caption{PMCC result from TsDetect (Benchmark 2-LLM).}
    \label{fig:pmcc-tsdetect-benchmark2}
\end{subfigure}
\hfill
\begin{subfigure}[b]{0.48\textwidth}
    \centering
    \includegraphics[width=\textwidth]{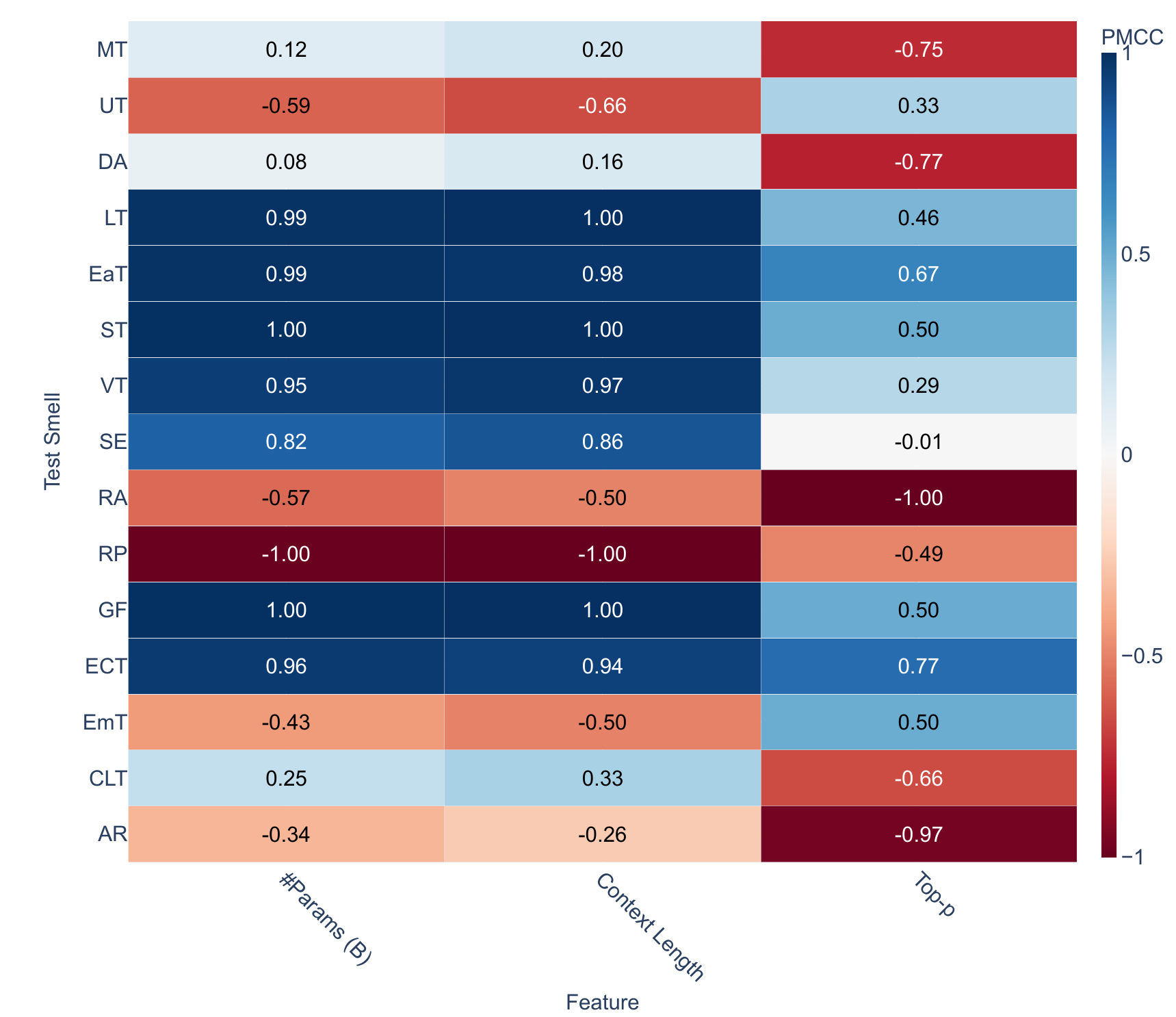}
    \caption{PMCC result from JNose (Benchmark 2-LLM).}
    \label{fig:pmcc-jnose-benchmark2}
\end{subfigure}
\caption{PMCC results between LLM Characteristics and Test Smell Presence in Benchmark 2.}
\label{fig:pmcc-combined-heatmaps-benchmark2}
\end{figure*}

\noindent\textbf{Influence of LLM Parameters.}  
LLM generation settings modulate smell behavior across both benchmarks (Figures~\ref{fig:spearman-tsdetect-benchmark1}--\ref{fig:spearman-jnose-benchmark1}, \ref{fig:pmcc-tsdetect-benchmark2}--\ref{fig:pmcc-jnose-benchmark2}).  
With Benchmark~1, longer contexts reduce assertion-level smells: TsDetect reports $r \in [-0.91,-0.51]$ for \texttt{AR}, \texttt{RA}, \texttt{SE}, \texttt{DA} (Figures~\ref{fig:spearman-tsdetect-benchmark1}); JNose shows $\rho \approx -0.94$ for \texttt{CLT}, \texttt{EmT}, \texttt{EH}, \texttt{SE}, \texttt{DA}, \texttt{UT} (Figures~\ref{fig:spearman-jnose-benchmark1}). At method level with Benchmark~2 (Figures~\ref{fig:pmcc-tsdetect-benchmark2}--\ref{fig:pmcc-jnose-benchmark2}), longer contexts instead inflate verbosity smells: \texttt{GF}, \texttt{MG}, \texttt{DA}, \texttt{UT}, \texttt{LT}, \texttt{VT} all reach $r \geq 0.97$ (both detectors), while \texttt{AR} decreases modestly ($r \approx -0.26$).  
Benchmark~1 shows a trade-off: higher top-$p$ reduces \texttt{CLT} ($\rho \approx -0.89$ in TsDetect) but increases \texttt{EmT}/\texttt{MT} ($\rho \approx +0.89$). Benchmark~2 highlights detector divergence: TsDetect finds top-$p$ increases \texttt{AR} ($r=+0.71$) while JNose shows the opposite ($r=-0.97$); both agree \texttt{DpT} grows with top-$p$ ($r\approx+0.91$).  
Model size effects are mild in Benchmark~1, but JNose highlights strong positives for \texttt{GF} ($\rho \approx +0.95$) and moderate rises for \texttt{LT}/\texttt{MT} ($\rho \approx +0.80$). In Benchmark~2, larger models amplify verbosity smells (\texttt{GF}, \texttt{UT}, \texttt{VT}) with $r \geq 0.97$.

\vspace{0.5em}
\noindent\colorbox{gray!20}{\parbox{0.98\linewidth}{
\textbf{Finding 14:} \textit{LLM parameters directly govern smell variability: constrained decoding (short context, low top-p) produces brittle and repetitive tests (CLT, EmT, UT), while permissive settings (long context, high top-p) suppress assertion smells but may inflate verbosity and ambiguous logic. These effects intensify at method level, where model size and context length linearly amplify redundancy-oriented smells (GF, MG, DA, UT, LT, VT; $r$ up to 1.00).}
}}

\vspace{0.5em}
\noindent\textbf{Comparative Analysis: LLMs vs. EvoSuite (Benchmark~1).}  
In contrast to LLMs, EvoSuite exhibits highly deterministic correlation profiles (Figures~\ref{fig:pmcc-tsdetect-benchmark1-evosuite}, \ref{fig:pmcc-jnose-benchmark1-evosuite}): $r = \pm 1.0$ for most attributes. Assertion smells (AR, EH, MT) correlate negatively with size but positively with complexity, while \texttt{SE}, \texttt{VT}, and \texttt{EmT} follow the opposite trend. Several smells (\texttt{CLT}, \texttt{GF}, \texttt{DpT}, \texttt{RO}) are absent altogether. These patterns reflect EvoSuite’s rigid, template-driven generation: predictable but lacking adaptability compared to LLMs.

\vspace{0.5em}
\noindent\colorbox{gray!20}{\parbox{0.98\linewidth}{
\textbf{Finding 15:} \textit{EvoSuite exhibits deterministic correlations ($r = \pm 1.0$) between smell prevalence and software attributes, reflecting rigid template-driven rules. LLMs instead show variance across $-0.9 \leq r \leq +1.0$, revealing greater variability but less stability.}
}}

\vspace{0.5em}
\noindent\textbf{Tool Sensitivity and Stochastic Smells.}  
Some smells—such as \texttt{RP}, \texttt{MT}, \texttt{IT}, and \texttt{UT}—exhibit weak or inconsistent correlations ($|r| < 0.4$), fluctuating across detectors and benchmarks. In Benchmark~2 and EvoSuite, many are absent altogether, suggesting structural homogenization or detector blind spots. JNose tends to capture fine-grained semantic noise (e.g., VT, ECT), while TsDetect favors rigid syntactic patterns. These inconsistencies suggest that certain smells emerge stochastically or evade systematic modeling.

\vspace{0.5em}
\noindent\colorbox{gray!20}{\parbox{0.98\linewidth}{
\textbf{Finding 16:} \textit{Some smells (RP, MT, IT, UT) emerge stochastically or remain undetected, highlighting detector-specific blind spots and the limitations of current smell modeling.}
}}

\vspace{0.5em}
\noindent\textbf{Interpretation of RQ3 (software attributes and LLM configuration).}
RQ3 reveals that smell diffusion in LLM-generated tests is shaped jointly by system structure and generation settings. On the one hand, larger and more complex systems tend to amplify redundant or strongly coupled smells (e.g., AR, RA, MG, GF, EH, CLT), suggesting that models struggle most precisely where high-quality tests are most needed. Faced with highly coupled (CBO) or deeply layered classes (RFC, CyCo), LLMs tend to expand the fixture and issue overlapping assertions rather than decomposing tests to mirror the underlying design, thereby reinforcing the General Fixture, Lazy Test, and Duplicate Assert patterns. On the other hand, LLM-specific parameters govern how this tendency manifests. Constrained decoding (short context, low top-$p$) yields brittle and repetitive tests that concentrate superficial smells (e.g., CLT, EmT, UT), while more permissive settings (longer context, higher top-$p$) suppress some assertion-related smells but can inflate verbosity and ambiguous test logic. These effects intensify at method level, where model size and context length linearly amplify redundancy-oriented smells (GF, MG, DA, UT, LT, VT). In contrast, EvoSuite exhibits almost deterministic correlations between smells and software attributes, reflecting rigid template-based rules. LLMs instead display a wider range of correlations, combining flexibility with less stability. Altogether, RQ3 suggests that tuning LLM parameters trades off different smell profiles rather than eliminating design defects.


\vspace{0.5em}
\noindent\highlight{Summary of \textbf{RQ3:}
Non-linear analyses show no significant effects of software or LLM parameters on smell prevalence (Finding~12). Instead, linear correlations reveal that project size amplifies redundant smells, while higher complexity and coupling induce architectural flaws (Finding~13). LLM parameters strongly modulate variability: constrained decoding (short context, low top-$p$) yields brittle and repetitive tests, whereas permissive settings reduce assertion smells but inflate verbosity and ambiguity, with effects intensifying at method level (Finding~14). Compared to LLMs, EvoSuite shows rigid template-driven correlations ($r=\pm1.0$), producing predictable but shallow patterns (Finding~15). Finally, some smells arise stochastically or remain undetected, reflecting detector blind spots and current modeling limits (Finding~16).}
%

\subsection{[RQ4]: Similarity of test smell patterns in LLM-generated tests to human-written tests.}

\noindent\textbf{[Experiment Design]:}
We assess the structural similarity between LLM-generated (Benchmark~1) and human-written test suites (Defects4J, SF110, CAT-LM) to characterise how closely LLM smell profiles align with those observed in real-world projects.. Our comparison involves two phases. First, we measure distributional similarity using Cosine and Jaccard indices for type overlap, and Euclidean and Hellinger distances for frequency differences across smell distributions. Second, we analyze co-occurrence and correlation patterns through Pearson and Spearman correlations between mean smell vectors, complemented by Mutual Information and Kruskal–Wallis tests to capture non-linear and group-based differences. To account for dataset imbalance (i.e., multiple LLM models vs. fewer human datasets), we compute metrics by averaging smell distributions within each group (LLM vs.~human) before comparison. Metrics are omitted when variance is insufficient for statistical validity. We interpret high similarity as a correlational signal that is \emph{consistent with} possible training-data overlap or convergence towards common testing idioms, but not as direct evidence of memorisation or data leakage. This design enables a principled, multi-dimensional comparison of smell profiles between LLM and human test suites.

\noindent\textbf{[Experiment Results]:}

\noindent\textbf{Overlap and Divergence in Smell Profiles.} 
Comparing human- and LLM-generated tests reveals both convergences and divergences across smell types and detectors. 
Table~\ref{tab:summary-llm-evosuite-human} provides a high-level side-by-side summary of key smells (AR, LT, GF, UT, EH, DA, DpT) for LLM-generated, EvoSuite-generated, and human-written tests, using qualitative labels (low / medium / high / none) that aggregate the detailed distributions reported in Appendix~\ref{app:additional-tables} (Tables~\ref{tab:test-smells-benchmark1}--\ref{tab:test-smells-benchmark2}, \ref{tab:tsdetect-tsmells-human}, and~\ref{tab:jnose-tsmells-human}.  
With TsDetect (Table~\ref{tab:tsdetect-tsmells-human}), human suites are dominated by Assertion Roulette (\texttt{AR}: 98.83\%) and Dependent Test (\texttt{DpT}: 81.22\%), whereas LLM tests reproduce AR but introduce verbosity-oriented smells (\texttt{GF}, \texttt{LT}, \texttt{DA}). 
EvoSuite (Table~\ref{tab:test-smells-benchmark1}), in contrast, suppresses diversity: several smells are absent entirely (CLT, GF, DpT), reflecting template-driven rigidity.  
Cross-tool discrepancies are substantial. TsDetect consistently reports higher prevalence for structural and behavioral smells (e.g., DpT 81.22\% in humans, 93\% in LLMs), while JNose emphasizes stylistic issues such as Magic Number Test (\texttt{MT}: 29.57\% in humans (Table~\ref{tab:jnose-tsmells-human}), up to 66.33\% in GPT-4 tests (Table~\ref{tab:test-smells-benchmark1})). 
Some categories show convergence across tools (e.g., CI: $\sim$25\%), while others remain tool-dependent (GF, LT, MT). Certain smells (EH, EmT, DfT) are undetected altogether, suggesting rarity or structural invisibility.

\begin{table}[t]
\centering
\small
\caption{High-level comparison of key test smells across generators.}
\label{tab:summary-llm-evosuite-human}
\scalebox{0.8}{
\begin{tabular}{lccc}
\toprule
\textbf{Smell} & \textbf{LLM-generated} & \textbf{EvoSuite-generated} & \textbf{Human-written} \\
\midrule
Assertion Roulette (AR) 
  & high 
  & high 
  & high \\
Lazy Test (LT) 
  & medium 
  & high 
  & low \\
General Fixture (GF) 
  & high 
  & none / very low 
  & low / very low \\
Unknown Test (UT) 
  & high 
  & low 
  & low \\
Exception Handling (EH) 
  & low 
  & high 
  & low / rare \\
Duplicate Assert (DA) 
  & medium 
  & medium 
  & low \\
Dependent Test (DpT) 
  & high 
  & none 
  & high \\
\bottomrule
\end{tabular}
}
\begin{tablenotes}
\scriptsize
\item[1]$^*$Labels (low / medium / high / none) summarise the combined distributions reported for TsDetect and JNose (Tables~\ref{tab:test-smells-benchmark1}--\ref{tab:test-smells-benchmark2}, \ref{tab:tsdetect-tsmells-human}, and~\ref{tab:jnose-tsmells-human}).
\end{tablenotes}
\end{table}

\vspace{0.5em}
\noindent\colorbox{gray!20}{\parbox{0.98\linewidth}{
\textbf{Finding 17:} 
\textit{
Human and LLM tests share dominant structural issues (AR, DpT), but LLM-generated suites additionally introduce verbosity-oriented patterns (GF, LT, DA) that are largely absent or much less prevalent in human-written tests. 
EvoSuite diverges strongly, producing rigid but shallow tests whose smell profile is characterised by high Lazy Test and Exception Handling but almost no General Fixture, Unknown Test, or Dependent Test, as summarised in Table~\ref{tab:summary-llm-evosuite-human}.}
}}


\vspace{0.5em}
\noindent\textbf{Detector Complementarity.}  
To better understand these contrasts, we analyzed detector-specific deltas (Table~\ref{tab:delta-humanwritten-full}). TsDetect consistently reports higher prevalence for coupling-based and behavioral smells, while JNose emphasizes style- and token-level patterns. For instance, TsDetect exceeds JNose by +81.22 points for \texttt{DpT} and +43.47 for \texttt{GF}, while JNose excels on \texttt{MT} (+26.73 points). On average, deltas across all smells reach $18.69 \pm 23.99$ percentage points, underscoring the divergent detection philosophies. Neither tool alone captures the full spectrum, but their combined use provides more holistic coverage.

\vspace{0.5em}
\noindent\colorbox{gray!20}{\parbox{0.98\linewidth}{
\textbf{Finding 18:} \textit{TsDetect favors structural and behavioral coupling smells (large deltas for DpT, GF), while JNose captures stylistic patterns (e.g., MT). Cross-tool deltas (mean difference $\sim$19 percentage points, stdev $\sim$24 percentage points) confirm that tool complementarity is necessary for reliable smell profiling.}
}}

\vspace{0.5em}
\noindent\textbf{Originality and similarity to human-written tests.}
Similarity metrics provide contrasting pictures depending on the detector (Figures~\ref{fig:radar-similarity-metrics}--\ref{fig:barplot-non-linear-metrics}).  
Under TsDetect, LLM-generated test suites diverge sharply from human tests: Cosine similarity is low (0.34), distances are high (Euclidean 169.41, Hellinger 0.68), and correlations are near-zero or negative (Pearson –0.02, Spearman –0.12; MI = 0). This suggests that LLM tests exhibit behaviorally novel and structurally distinct smell patterns.  
Under JNose, however, similarities are more pronounced: Cosine similarity rises to 0.77, correlations are positive (Pearson 0.68, Spearman 0.45), and MI is non-zero (0.21). These results indicate that LLM outputs partially align with human smell profiles, which is consistent with either shallow imitation of common testing idioms or possible overlap with patterns present in public Java test suites (e.g., Defects4J, SF110). Importantly, this alignment is inherently correlational: similarity in smell distributions alone does \emph{not} prove memorisation of specific tests or direct data leakage, and more targeted analyses (e.g., near-duplicate detection, training-data audits) would be needed to establish such claims.

\vspace{0.5em}
\noindent\colorbox{gray!20}{\parbox{0.98\linewidth}{
\textbf{Finding 19:} \textit{Under TsDetect, LLM-generated smell patterns appear largely distinct from human-written suites, whereas JNose reveals partial alignment between LLM and human smell distributions. This alignment is a correlational signal that is compatible with both convergence towards common testing idioms and possible training-data overlap, but does not by itself demonstrate memorization. It underlines the need for multi-tool triangulation and more direct leakage-focused analyses.}
}}

\begin{table*}[ht]
\centering
\caption{Delta Values between TsDetect and JNose for Test Smell Distributions (Human-Written Test Suites)}
\label{tab:delta-humanwritten-full}
\scalebox{0.55}{
\begin{tabular}{lccccccccccccccccccccccc}
\toprule
\textbf{Dataset} & \textbf{AR} & \textbf{CLT} & \textbf{CI} & \textbf{DfT} & \textbf{EmT} & \textbf{ECT} & \textbf{EH} & \textbf{GF} & \textbf{MG} & \textbf{RP} & \textbf{RA} & \textbf{SE} & \textbf{VT} & \textbf{ST} & \textbf{EaT} & \textbf{LT} & \textbf{DA} & \textbf{UT} & \textbf{IT} & \textbf{RO} & \textbf{MT} & \textbf{DpT} \\
\midrule
Human-Written & 35.05 & 33.17 & 3.72 & 22.13 & 6.18 & 14.02 & 0.00 & 43.47 & 5.22 & 0.47 & 0.27 & 1.49 & 3.23 & 1.53 & 0.44 & 41.02 & 31.24 & 1.66 & 23.60 & 3.72 & 26.73 & 81.22 \\
\textbf{Mean ± StdDev} & \multicolumn{22}{c}{\textbf{18.69} ± \textbf{23.99}} \\
\bottomrule
\end{tabular}
}
\end{table*}

\begin{figure}
  \centering
  \includegraphics[width=0.7\textwidth]{Images/radar_similarity_metrics.pdf}
   \caption{Similarity Metrics between Human and LLM Distributions.}
    \label{fig:radar-similarity-metrics}
    \vspace{-3mm}
\end{figure}

\begin{figure*}[ht]
\centering
\begin{subfigure}[b]{0.48\textwidth}
    \centering
    \includegraphics[width=\textwidth]{Images/barplot_correlation_metrics.pdf}
    \caption{PMCC and Pearson Metrics between Human and LLM Test Smell Distributions.}
    \label{fig:barplot-correlation-metrics}
\end{subfigure}
\hfill
\begin{subfigure}[b]{0.48\textwidth}
    \centering
    \includegraphics[width=\textwidth]{Images/barplot_non_linear_metrics.pdf}
    \caption{Mutual Information and Kruskal-Wallis Metrics between Human and LLM Test Smell Distributions.}
    \label{fig:barplot-non-linear-metrics}
\end{subfigure}
\caption{Comparison of Different Correlation Metrics between Human and LLM Test Smell Distributions.}
\label{fig:pmcc-pearson-mi-kruskal-wallis-combined-barplots}
\end{figure*}

\vspace{0.5em}
\noindent\textbf{Interpretation of RQ4 (alignment with human-written tests).}
The RQ4 results show that LLM-generated tests occupy an intermediate space between human-written suites and EvoSuite. On the one hand, LLM and human tests share dominant structural issues such as Assertion Roulette and Dependent Test, while LLMs additionally introduce verbosity-driven patterns (General Fixture, Lazy Test, Duplicate Assert) that are largely absent from human suites. EvoSuite, by contrast, produces more rigid but shallow tests with a distinct smell signature. This suggests that LLMs partially imitate human-like testing habits but also exaggerate certain redundant structures. Similarity metrics and cross-detector analyses further nuance the picture. Under TsDetect, LLM smell distributions appear largely distinct from human ones, whereas JNose reveals partial alignment, especially on stylistic smells such as Magic Number Test. This alignment is a correlational signal that is compatible with both convergence towards common Java testing idioms and potential training-data overlap, but does not by itself demonstrate memorisation of specific test cases. It reinforces the need for multi-tool triangulation and more direct leakage-focused analyses when interpreting similarity between LLM- and human-generated tests, and suggests that current models may mix genuine generalisation with shallow imitation of prevailing—but not necessarily high-quality—testing practices.

\vspace{0.5em}
\noindent\highlight{Summary of \textbf{RQ4:}  
LLM-generated tests share dominant structural issues with human-written ones (e.g., Assertion Roulette, Dependent Test), but also introduce verbosity-driven patterns (GF, LT, DA) that are absent from human suites (Finding~17). Detector deltas confirm that TsDetect emphasizes structural and behavioral coupling, while JNose focuses on stylistic patterns such as Magic Number Test, reinforcing the need for cross-tool triangulation (Finding~18). Finally, while TsDetect highlights that LLM smell profiles remain largely distinct from human tests, JNose reveals partial alignment in smell distributions; this similarity is a correlational signal compatible with convergence towards common testing idioms and possible training-data overlap, but does not by itself demonstrate memorization (Finding~19).}


\section{Discussion}
\label{discussion}

\subsection{Generative biases behind LLM test smell patterns.}
Our results also point to systematic generative biases in how LLMs produce tests. First, models exhibit a clear preference for broad, monolithic test cases: when asked to “thoroughly test” a class or method, they tend to initialise large, shared fixtures and attach multiple loosely related assertions to a single test. This behaviour aligns with the overrepresentation of General Fixture, Lazy Test, and Duplicate Assert in our RQ1–RQ2 results, and suggests that models prioritise apparent behavioural coverage within each test case over cohesion and isolation across the test suite. Second, LLMs rarely introduce explicit Exception Handling, even when APIs expose rich error behaviour and prompts do not discourage it. Combined with the persistence of Unknown Test, this indicates that error paths and failure scenarios are not treated as first-class citizens in the generated tests, either because they are under-emphasised in training data or not sufficiently foregrounded by typical prompts.

Third, the correlations observed in RQ3 between structural complexity (e.g., higher coupling and response metrics) and smell prevalence indicate that, when confronted with complex classes, models tend to expand fixtures and assertions rather than refactoring tests into smaller, modular units. In practice, this amplifies exactly those smells—GF, LT, DA, AR—that make maintenance hardest in large systems. Taken together, these biases are consistent with the next-token prediction objective: the model is rewarded for producing sequences that resemble plausible testing idioms seen during training, not for optimising long-term maintainability or adhering to principled test design guidelines.

\subsection{Mitigating test smells in LLM-generated tests.}
\label{sec-discuss-mitigation}

\subsubsection{Prompt- and retrieval-based mitigations.}
Our findings suggest that many LLM-induced smells (e.g., General Fixture, Lazy Test, Duplicate Assert) are tightly coupled to how prompts frame “thoroughness” and contextual scope. One natural line of mitigation is therefore to steer LLMs away from smell-prone patterns at prompt time. Recent work on retrieval-augmented test generation shows that Retrieval-augmented generation (RAG) can improve line coverage by enriching the model’s context with API documentation, GitHub issues, and StackOverflow Q\&As, even though it does not consistently increase syntactic or dynamic correctness. In particular, Shin et al.~\cite{shin2024retrieval} report that RAG-based prompting over Machine Learning/Deep Learning libraries improves average line coverage by about 6.5\%, with GitHub issues being especially useful for surfacing edge cases and revealing new bugs. 
Building on these results, our smell-diffusion analysis points to complementary prompt-level mitigations. First, prompts could explicitly encode “anti-smell” guidelines (e.g., discouraging broad fixtures and unrelated assertions, encouraging exception-aware tests) or include few-shot examples of smell-poor tests. Second, RAG pipelines could retrieve tests from curated, smell-aware corpora rather than arbitrary examples, biasing the model towards cohesive fixtures and clearer oracles. Third, retrieval can be used to inject human-authored refactoring patterns (e.g., clean variants of Assertion Roulette or Magic Number tests), so that the model is exposed to high-quality idioms even when the underlying code base is complex. We see these prompt- and retrieval-level strategies as orthogonal to our empirical analysis: they do not change the underlying smell catalogue, but directly target the generative behaviours that our results identify as problematic.

\subsubsection{Fine-tuning, PEFT, and agentic repair pipelines.}
Beyond prompts and retrieval, several recent studies investigate how tuning LLMs or orchestrating them as agents can improve unit tests. Shang et al.~\cite{shang2025large} conduct a large-scale empirical study of 37 LLMs across three unit-testing tasks (test generation, assertion generation, and test evolution), showing that fine-tuned models generally outperform prior state-of-the-art techniques and that zero-shot prompt engineering can already be competitive for common unit-testing tasks.  Storhaug and Li~\cite{storhaug2024parameter} further demonstrate that parameter-efficient fine-tuning (PEFT) methods such as LoRA~\cite{hu2022lora} and prompt tuning can achieve performance comparable to full fine-tuning for unit test generation at a fraction of the training cost, albeit sometimes at the expense of executability (e.g., more calls to non-existent methods), while improving instruction and branch coverage and mutation scores.
From a test-smell perspective, these studies indicate that tuning strategies can “re-shape” the distribution of generated tests. If the tuning corpora or objectives were explicitly smell-aware—for instance, favouring smell-poor tests or penalising over-long fixtures and undocumented assertions—they could plausibly reduce the diffusion of the smells we observe. Complementary to tuning, agentic workflows have been proposed where LLMs act as smell detectors and refactoring agents over existing test suites. Melo et al.~\cite{melo2025agentic} show that small open models (e.g., Phi-4-14B, Gemma-2-9B) can detect nearly all instances of several common smells and automatically refactor a majority of them, with a subset of the refactorings being accepted and merged into open-source projects.
Our results therefore align well with a generate–detect–repair view of LLM-based test generation: LLMs can be used not only as test generators, but also as post-hoc smell detectors and refactoring agents. A promising direction for future work is to combine these elements into end-to-end pipelines that (i) generate tests, (ii) automatically flag and repair smell-prone designs using agentic LLMs, and (iii) optionally fine-tune or adapt models on smell-aware corpora so that their default behaviour drifts towards the cleaner side of the smell spectrum.

\subsection{Implications}
\label{discussion-of-findings-and-implications}

\subsubsection{Implications for Researchers} 

Our study offers new empirical foundations for understanding how test smells emerge and evolve in LLM-generated unit tests. The multi-perspective comparison between LLM- and human-written tests—combined with tool triangulation (TsDetect vs. JNose) and metric triangulation (correlation, mutual information, statistical testing)—highlights both structural divergences and superficial similarities in test smell profiles.

Importantly, our results challenge simplistic assumptions that LLMs merely replicate human behavior. As shown in our findings (e.g., Finding~17-19), TsDetect highlights originality in LLM smell patterns that diverge from human-written suites, whereas JNose reveals partial overlaps that are compatible with shallow stylistic imitation or with overlap in the underlying training data. These diverging signals underscore the need for caution when interpreting similarity as evidence of understanding: what appears as alignment may, in reality, stem from statistically reproducing common testing idioms, and our analyses remain correlational rather than direct proof of memorisation. This has direct implications for research on LLM reliability and evaluation. 

Evidence of surface-level similarity (e.g., high Pearson correlation under JNose, but null mutual information under TsDetect) is \emph{consistent} with possible training-data overlap or shallow reuse of patterns from public repositories like Defects4J and SF110, but does not by itself demonstrate training-set leakage or memorisation of specific tests. Establishing memorisation would require dedicated analyses (e.g., training data audits, membership-inference tests, or exact code-level matching), which are beyond the scope of this study.

Furthermore, the consistent presence of smells such as Assertion Roulette and Lazy Test—even under advanced prompting strategies—suggests that prompt engineering alone is insufficient to suppress smell diffusion. Future research should explore hybrid approaches combining LLM generation with automated repair or post-processing, and systematically evaluate smell mitigation at both generation and refinement stages. Our discussion of mitigation strategies (Section~\ref{sec-discuss-mitigation}) outlines several concrete directions, including prompt- and retrieval-based interventions, fine-tuning and parameter-efficient adaptation on smell-aware corpora, and agentic pipelines where LLMs act as smell detectors and refactoring agents.

\noindent\textbf{Key Takeaway for Researchers:} LLMs do not simply reproduce human testing habits—they exhibit a mix of novel, stylistic smell patterns, some of which may be influenced by the distributions present in their training data. These behaviors demand stronger interpretability, auditability, and debiasing mechanisms in empirical software engineering research.


\subsubsection{Implications for Developers} 

Our findings provide critical guidance for developers who adopt LLMs to assist in writing or maintaining unit tests. While LLMs significantly reduce manual effort, their output is not free from quality concerns. Pervasive smells such as Assertion Roulette and Lazy Test remain dominant, even across advanced models like GPT-4 and across both class- and method-level generations. This suggests that LLM-generated tests may compromise maintainability or robustness unless carefully reviewed.

Notably, LLMs tend to underrepresent behaviors like Exception Handling, which are crucial for robustness testing. At the same time, their tendency to produce verbose or repetitive assertions hints at a lack of contextual grounding and limited behavioral insight. 

These limitations are exacerbated by the fact that, for some detectors, similarity metrics indicate partial structural and stylistic alignment between LLM- and human-written suites. Such overlap is consistent with LLMs reproducing common testing idioms that are frequent in public codebases and possibly present in their training data, but our analyses are purely correlational and do not establish memorisation of specific tests. From a developer’s perspective, however, this still raises questions about the originality and long-term generalisation of generated tests.

To mitigate these issues, developers should integrate smell detection tools like TsDetect and JNose directly into the CI/CD pipeline. Manual inspection remains essential for critical code paths, particularly when adopting aggressive prompting strategies or larger models. Developers should also favor hybrid workflows, combining LLM-based generation with targeted manual refinement or automated repair systems that enforce test design best practices. Concretely, this may include enriching prompts with explicit testing guidelines or high-quality examples, using retrieval-augmented generation over curated, smell-aware test corpora, and leveraging emerging agentic tools that automatically detect and refactor common smells before tests are merged.

\noindent\textbf{Key Takeaway for Developers:} 
In the context of automated unit test generation using prompting techniques and varying input contexts, LLM-generated tests should be treated as draft-quality artifacts that require validation, even when they look structurally similar to human-written tests. Our findings underscore the need to integrate smell detection, human review, and guided refinement into test workflows to ensure maintainable and robust test code.

\subsection{Threats to Validity}
\label{threats-to-validity}

\noindent\textbf{External Validity.} 
Our study targets Java-based systems and relies on a smell catalogue and detectors (TsDetect and JNose) that are tailored to JUnit-style tests. This focus limits generalizability to other ecosystems (e.g., Python, C\#, JavaScript, TypeScript), where testing frameworks, design idioms, and even the manifestation of specific smells may differ. We therefore interpret our findings as evidence about the diffusion of test smells in LLM-generated \emph{Java} tests. The breadth and diversity of our datasets and LLM configurations strengthen the robustness of our results within the Java ecosystem, but they do not mitigate this language-specific threat. Consequently, further experiments are required to assess whether the trends we observe hold in other programming languages and testing paradigms.
To support robustness within Java, Benchmark~1 encompasses two well-established datasets—Defects4J (faulty real-world code), SF110 (diverse academic and industrial projects), and a curated dataset CMD (modern codebases)—which together cover a wide range of software styles and quality profiles. 
Benchmark~2 complements this with 108 real-world functions from nine projects across varied domains, including concurrency (JCTools), computer vision (JavaCV), data processing (Zxing), utility libraries (Commons-lang, Commons-math), and cloud microservices (Apollo, Jeecg Boot). 
Our evaluation also spans seven LLM configurations, combining proprietary (GPT-3.5, GPT-4) and open-source models (Mistral 7B, Mixtral 8x7B, CodeLlama 13B-Instruct), applied under diverse prompting paradigms (e.g., Zero-Shot, Few-Shot, Chain-of-Thought, Full Context). The tests produced reflect both method- and class-level contexts, enabling a multi-granularity perspective on test smell emergence. While future LLMs or test generation paradigms may evolve with different characteristics, expanding this analysis to other programming languages and non-LLM-based generators remains an essential avenue for replication and broader generalization.

\noindent\textbf{Internal Validity.} 
A key threat stems from the test smell detectors themselves. While TsDetect and JNose represent complementary detection philosophies (semantic-graph and syntactic-pattern, respectively), both are imperfect. TsDetect may miss stylistic smells like Magic Number Test, while JNose tends to over-detect template-based smells, as confirmed in prior evaluations reporting over 70\% false positive rates for certain smells in human tests~\cite{panichella2022test}. To limit detector-induced bias, we cross-compared results between the two tools and explicitly highlighted divergences throughout the analysis. 
In addition, we manually inspected 240 sampled test entities (120 class-level and 120 method-level), covering frequent smells and cases with strong detector disagreement, and derived precision, recall, and F1 for both detectors against this oracle. This validation confirms that some smells (e.g., Lazy Test) are detected very reliably, while others (e.g., Assertion Roulette, Magic Number Test, Dependent Test) are either over-reported or unevenly detected across tools. Consequently, our prevalence and similarity results should be interpreted as detector-dependent, upper-bound estimates on actual smell incidence, and our conclusions focus on relative differences between LLM-, EvoSuite-, and human-generated suites rather than on exact absolute rates.
When TsDetect and JNose disagree on specific instances, we therefore keep detector-specific results separate and treat these disagreements as part of the construct-validity threat that our manual oracle and cross-detector comparison are designed to triangulate, rather than attempting to enforce a single consensus label.

Another internal threat lies in the statistical assumptions made. Pearson correlation assumes linearity, which may not hold across all project and smell characteristics. To compensate, we incorporated non-parametric tests (Spearman, Kruskal--Wallis) and Mutual Information to capture non-linear or distributional relationships. 
Nevertheless, our manual oracle only covers a small fraction of all tests, and cannot fully validate every detected pattern. Extending this validation to the full corpus would require annotating tens of thousands of smell instances across hundreds of thousands of tests, with multiple annotators to control for subjectivity, which is beyond the scope and resources of the present study. A larger-scale manual annotation of LLM-generated tests therefore remains an important avenue for future work.

\noindent\textbf{Construct Validity.} 
Construct threats arise from the conceptual framing of smells and their interpretation.
In addition, our empirical analyses are restricted to the 21 test smells supported by TsDetect and JNose; while this catalogue covers all smell categories considered in this study, it does not exhaust the broader space of test smells discussed in the literature (e.g., interaction- or environment-related smells, or certain mocking-specific patterns). Our conclusions should therefore be interpreted as pertaining to this subset of smells.
A further construct threat concerns our choice of structural metrics (LOC, CC, CBO, RFC, DIT). We focus on widely used, tool-supported indicators that prior work has linked to software quality and test maintainability, but we do not capture other potentially relevant factors such as change history, developer activity, or domain-specific design attributes; exploring richer metric sets is therefore an avenue for future work.
While our study focuses on the presence and distribution of smells, this does not automatically imply harmfulness. As shown by Panichella et al.\cite{panichella2022test}, many detected smells in human tests may not degrade maintainability. Conversely, Bavota et al.\cite{bavota2015test} argue that even isolated smells can impair evolution and comprehension. Our findings fall between these positions: some smells like General Fixture or Duplicated Test may indicate critical maintenance risks in both human and LLM-generated tests, while others (e.g., Magic Number Test) may be contextually benign. This underscores the need to refine smell taxonomies and ground them in their practical impact—an open challenge in the test quality literature.
{Similarly, our observations regarding possible training-data overlap or shallow stylistic imitation are purely correlational and based on distributional alignment between LLM- and human-written smell profiles. Demonstrating actual memorisation of specific tests would require dedicated analyses (e.g., access to pre-training data, membership-inference experiments, or code-level near-duplicate detection on controlled “leaked vs.\ unleaked” benchmarks), which are beyond the scope of this study.

\noindent\textbf{Conclusion Validity.} 
Our statistical analyses used standard thresholds (e.g., $\alpha = 0.05$), but some p-values in Kruskal-Wallis tests were near significance boundaries (e.g., p = 0.1658). We thus interpret distributional similarities and differences with caution, favoring effect size and divergence metrics (e.g., Mutual Information) over binary significance claims. Additionally, correlations do not imply causation. Although strong associations were found between certain LLM types and smell profiles, especially in TsDetect, these patterns may reflect prompt artifacts, model memorization, or project-specific traits rather than inherent generalization failures.

\section{Related work}
\label{relatedwork}
Test smells, like code smells, are indicators of poor design or implementation practices that may hinder the maintainability, readability, or evolvability of software. Over the past two decades, research has extensively investigated the presence, detection, and implications of test smells in both human-written and automatically generated test suites.

\subsection*{Test Smells and Their Impact}

The concept of test smells was first introduced by Van Deursen et al.\cite{van2001refactoring}, who described test design flaws such as Assertion Roulette, Eager Test, and Mystery Guest. These smells signal structural weaknesses that reduce test maintainability. Subsequent empirical studies, including the foundational work by Bavota et al.\cite{bavota2015test}, demonstrated that test smells are widespread and negatively impact program comprehension and maintenance, even when occurring in isolation. Test smells were found to increase change-proneness and reduce code understandability across a wide range of software systems.

However, recent work by Panichella et al.~\cite{panichella2022test} critically revisited the assumptions and detection practices surrounding test smells. They observed that existing detection tools often produce a high rate of false positives and negatives, and that several canonical smells are not necessarily indicative of actual design flaws in developer-written or generated tests. Their findings call for re-evaluating the definitions and thresholds used by test smell detectors, and suggest the necessity for human-in-the-loop or context-aware evaluations.

\subsection*{Test Smells in Human-Written Tests}

Several studies have focused on analyzing test smells in manually written test suites. Bavota et al.\cite{bavota2015test} examined open-source Java projects and confirmed the persistent presence of smells such like Assertion Roulette and Magic Number Test, showing that these issues are not merely cosmetic but correlate with higher maintenance effort and long-term technical debt. Peruma et al.\cite{peruma2019distribution} extended this line of research to mobile applications, showing that mobile-specific smells (e.g., Conditional Test Logic, Constructor Initialization) often emerge early in a project's lifecycle and tend to persist. More recently, Pontillo et al.\cite{pontillo2024machine} proposed machine learning–based approaches to detect test smells, demonstrating that data-driven models can complement traditional rule-based detectors, but also highlighting challenges in generalising across projects and smell types. 
Complementing these studies, Santana et al.\cite{santana2025empirical} empirically analyse the co-occurrence of 19 test smells in 22 human-written Java projects, categorising smell pairs into structural, statistical, and ``vicious'' co-occurrences and proposing joint refactoring strategies. Their results confirm that smells rarely occur in isolation and that combinations such as \textit{LT+DA}, \textit{MG+RO}, or \textit{AR} with other smells are particularly frequent, reinforcing the need to reason about smell clusters rather than individual smells.
Bai et al.~\cite{bai2022assertion} revisit the Assertion Roulette smell in an educational context and find that, although AR is highly prevalent in student-authored tests, its impact on objective code-quality measures is limited, calling for a more nuanced interpretation of this smell's harmfulness.
These works collectively emphasise that (i) test smells are widespread in human-written suites and (ii) the choice of detection technique (rule-based vs.\ learning-based) has a strong impact on the resulting smell profiles.


\subsection*{Test Smells in Automatically Generated Tests}

Automatically generated tests have also been studied for their tendency to accumulate smells. EvoSuite, a widely used test generation tool, has been shown to produce high-coverage test suites that nonetheless suffer from smells such as Eager Test and Assertion Roulette~\cite{palomba2016diffusion,panichella2020revisiting}.
Palomba et al.\cite{palomba2016diffusion} and Grano et al.\cite{grano2019scented} analysed the \emph{diffusion} and \emph{diffuseness} of test smells in automatically generated suites, showing that specific smells can be highly pervasive and tightly clustered even when tests are syntactically well formed. Panichella et al.~\cite{panichella2022test} further demonstrated that the incidence of smells varies substantially depending on the generation tool and configuration, and that not all detected smells are semantically harmful: their hand-annotated benchmark revealed that certain smells, while prevalent, do not correlate well with real maintenance concerns.  
These studies provide the foundation for our use of smell diffusion as a cross-sectional notion of prevalence, distribution, and co-occurrence, and they highlight the need for large-scale, detector-aware analyses when assessing the quality of generated tests.

\subsection*{Test Smells in LLM-Generated Tests}

Recent studies have begun examining test smells in LLM-generated code. Siddiq et al.\cite{siddiq2024using} used TsDetect to analyze tests from models like GPT-3.5 and StarCoder\cite{li2023starcoder}, finding frequent issues such as Assertion Roulette, Magic Number Test, and Empty Test—especially in Codex outputs—despite syntactic correctness. Ouédraogo et al.~\cite{ouedraogo2024large} extended this line of work by evaluating the impact of prompting strategies (e.g., Zero-shot, Chain-of-Thought) on smell prevalence, showing that LLMs outperform EvoSuite in readability but still introduce smells, particularly under basic prompting. Their evaluation, however, had several key limitations: it employed only a single smell detector (TsDetect), focused exclusively on class-level test generation, and lacked cross-benchmark validation or cross-tool triangulation—limiting the generalizability of their findings despite the inclusion of multiple models and prompt strategies.

Beyond Java, Alves et al.~\cite{alves2025quality} conducted a quality assessment of Python tests generated by LLMs, combining structural metrics, mutation analysis, and smell detection to characterise strengths and weaknesses of generated suites. Their findings indicate that LLM-generated tests can achieve good coverage yet still exhibit design problems, reinforcing the need to look beyond simple adequacy metrics. Complementarily, Melo et al.~\cite{melo2025agentic} proposed agentic LMs for \emph{hunting down} test smells, showing that multi-agent workflows can detect and refactor smells but also depend strongly on prompt design and orchestration. Lucas et al.~\cite{lucas2025investigating} investigated the performance of small language models as test smell detectors for manual test cases, reporting competitive results for certain smells and projects but also notable variability across models and smell categories. Together with Pontillo et al.~\cite{pontillo2024machine}, these studies illustrate a growing interest in both (i) test smells \emph{in} LLM-generated tests and (ii) LLMs \emph{as} test smell detectors or refactoring tools.

In contrast, our study focuses on the diffusion of test smells in LLM-generated \emph{Java} tests across multiple benchmarks and generators, using traditional detectors rather than LLMs as the primary instrumentation. We adopt a dual-benchmark approach spanning class-level (Benchmark~1) and method-level (Benchmark~2) generation, enabling multi-granularity analysis. Our model set includes recent open-source LLMs—Mistral 7B, Mixtral 8$\times$7B, and CodeLlama 13B-Instruct—alongside GPT-3.5 and GPT-4, offering a broader view beyond single-model evaluations. We evaluate diverse prompt configurations: Benchmark~1 tests structured reasoning techniques (ZSL, FSL, CoT, ToT, GToT), while Benchmark~2 varies prompt context (Self-Contained, Simple, Full). Crucially, we use two complementary rule-based detection tools (TsDetect and JNose), complemented by a manual oracle, to cross-validate findings and mitigate tool bias. Beyond frequency counts, we apply co-occurrence analysis, correlation metrics, and distributional similarity (e.g., cosine, mutual information, Kruskal–Wallis), offering a multidimensional view of smell diffusion in LLM-generated tests and positioning our work as complementary to recent quality and detection studies in other languages and settings~\cite{alves2025quality,melo2025agentic,lucas2025investigating,pontillo2024machine}.
This study offer a systematic, tool-diverse, and statistically grounded analysis of test smell diffusion in LLM-generated Java tests, with comparisons to over 770k human-written tests and structured examination of prompt, model, and software characteristics. Our findings indicate a dual behavior: under TsDetect, LLM-generated smell distributions tend to diverge from those of human-written tests, suggesting generation-specific artifacts, whereas under JNose we observe partial overlaps that are consistent with shallow stylistic imitation or the reuse of common testing idioms. These distributional similarities are correlational and do not by themselves prove memorization of specific test cases, but they do raise important questions about tool reliability, evaluation fairness, and the possibility of training-data overlap that warrant further, dedicated investigation.

\subsection*{Summary}

Overall, while prior work has established the prevalence and potential harm of test smells in both human-written and automatically generated tests, the emergence of LLMs necessitates a paradigm shift. Our results indicate that LLM-generated tests neither simply mirror existing smell distributions nor define an entirely new regime: under some detectors they exhibit structurally divergent smell profiles, whereas under others they show partial stylistic alignment with human-written suites. This combination of novelty and overlap raises new questions about the adequacy of existing smell taxonomies and detection tools, which were originally designed for human-written code. As AI-assisted development becomes mainstream, reassessing the semantics, detection boundaries, and contextual relevance of test smells is no longer optional—it is essential for ensuring the reliability and maintainability of future software systems.

\section{Conclusion}
\label{conclusion}

This paper presents the first comprehensive analysis of test smell diffusion in LLM-generated Java unit tests, examining over 20,000 test suites across four models and multiple prompting strategies. Through systematic comparison with SBST tools and human-written tests, we reveal crucial quality characteristics of LLM-generated Java test code.

Our findings show that LLM-generated Java tests consistently exhibit specific smell signatures—particularly Assertion Roulette, Magic Number Test, and Empty Test—with prevalence strongly influenced by model scale, context length, and prompting strategy. While larger models reduce some structural smells, they paradoxically amplify others, revealing complex trade-offs in generation parameters. Software complexity metrics further exacerbate smell emergence, suggesting LLMs struggle with complex codebases where high-quality tests are most critical.

Comparative analysis reveals distinct paradigmatic differences: EvoSuite produces predictable template-based patterns, while LLM-generated Java tests show only partial alignment with human-written smell profiles, and primarily under certain detectors (e.g., JNose). These overlaps are inherently correlational and are consistent with shallow imitation of common testing idioms or possible training-data overlap, but they do not by themselves prove memorization of specific tests. Nevertheless, they highlight the risk that LLMs may reproduce flawed testing practices present in public code repositories.

These findings highlight the need for smell-aware generation techniques, robust detection methods for LLM-specific quality patterns, and comprehensive evaluation frameworks that balance functional correctness with maintainability. As LLMs become integral to software testing workflows, ensuring generated test code quality is essential for sustainable software development.
Our results are based on automated smell detectors whose behaviour we partially validated through a manual oracle, and should therefore be interpreted as detector-dependent, upper-bound estimates of smell diffusion rather than exact ground truth.

Finally, our analysis is restricted to Java test suites and Java-specific smell detectors; an important direction for future work is to replicate our study on other ecosystems (e.g., Python, C\#, TypeScript) with language- and framework-specific smell catalogues and detectors, to assess to what extent the diffusion patterns observed here generalise beyond Java.

\section*{Acknowledgements}{
This research was funded in whole, or in part, by the Luxembourg National Research Fund (FNR), grant reference AFR PhD bilateral, project reference 17185670. This work was also supported by the European Research Council (ERC) under the European Union’s Horizon 2020 research and innovation program (grant agreement No. 949014) and Fundamental Research Funds for the Central Universities (AE89991/478). For the purpose of open access, and in fulfilment of the obligations arising from the grant agreement, the author has applied a Creative Commons Attribution 4.0 International (CC BY 4.0) license to any Author Accepted Manuscript version arising from this submission.
}

\bibliographystyle{ACM-Reference-Format}
\bibliography{TOSEM/references}

\appendix

\section{Additional Tables}
\label{app:additional-tables}

\begin{table*}[!htbp]
\centering
\caption{Test Smells Distribution by Model and Datasets from Benchmark 1 Detected by TsDetect and JNose.}
\label{tab:test-smells-benchmark1}

\subfloat[Detected by TsDetect\label{tab:rq1-1-tsdetect}]{
\centering
\scalebox{0.55}{
\begin{tabular}{lllllllllllllllllllllll}
\toprule
\multicolumn{1}{c}{\textbf{Model}} & \multicolumn{1}{c}{\textbf{Dataset}} & \multicolumn{1}{c}{\textbf{AR}} & \multicolumn{1}{c}{\textbf{CLT}} & \multicolumn{1}{c}{\textbf{CI}} & \multicolumn{1}{c}{\textbf{DfT}} & \multicolumn{1}{c}{\textbf{EmT}} & \multicolumn{1}{c}{\textbf{EH}} & \multicolumn{1}{c}{\textbf{GF}} & \multicolumn{1}{c}{\textbf{MG}} & \multicolumn{1}{c}{\textbf{RP}} & \multicolumn{1}{c}{\textbf{RA}} & \multicolumn{1}{c}{\textbf{SE}} & \multicolumn{1}{c}{\textbf{VT}} & \multicolumn{1}{c}{\textbf{ST}} & \multicolumn{1}{c}{\textbf{EaT}} & \multicolumn{1}{c}{\textbf{LT}} & \multicolumn{1}{c}{\textbf{DA}} & \multicolumn{1}{c}{\textbf{UT}} & \multicolumn{1}{c}{\textbf{IT}} & \multicolumn{1}{c}{\textbf{RO}} & \multicolumn{1}{c}{\textbf{MT}} & \multicolumn{1}{c}{\textbf{DpT}} \\
\midrule
\multirow{3}{*}{GPT 3.5-Turbo} & Defects4J & 46.94 & 3.35 & 0.10 & 0.00 & 31.64 & 19.26 & 4.57 & 0.89 & 0.00 & 1.08 & 18.78 & 0.00 & 0.65 & 41.16 & 48.67 & 6.97 & 35.16 & 0.00 & 01.01 & 99.78 & 0.00 \\
 & CMD & 17.08 & 0.14 & 0.00 & 0.00 & 62.40 & 12.81 & 15.15 & 0.00 & 0.00 & 0.28 & 0.83 & 0.00 & 0.00 & 16.12 & 15.70 & 0.14 & 77.27 & 0.00 & 0.00 & 100.00 & 0.00 \\
 & SF110 & 48.53 & 2.14 & 0.06 & 0.00 & 17.96 & 12.91 & 07.09 & 2.44 & 0.05 & 1.98 & 20.39 & 0.00 & 0.30 & 49.82 & 51.73 & 8.41 & 28.70 & 0.02 & 2.70 & 99.79 & 0.00 \\
\cmidrule(r){2-23}
 & Average & 37.51 & 1.87 & 0.05 & 0.00 & 37.33 & 14.99 & 8.93 & 1.11 & 0.01 & 1.11 & 13.33 & 0.00 & 0.31 & 35.70 & 38.70 & 5.17 & 47.04 & 0.01 & 1.23 & 99.85 & 0.00 \\
 \midrule
GPT 4 & CMD & 22.80 & 0.48 & 0.00 & 0.00 & 27.55 & 19.95 & 18.53 & 0.00 & 0.24 & 0.48 & 1.19 & 0.00 & 0.00 & 27.79 & 32.54 & 0.48 & 50.83 & 0.24 & 0.00 & 98.34 & 0.00 \\
\midrule
\multirow{3}{*}{Mistral 7B} & Defects4J & 54.54 & 8.49 & 0.20 & 0.00 & 5.81 & 15.68 & 0.85 & 0.20 & 0.13 & 0.91 & 13.00 & 0.00 & 0.85 & 44.68 & 56.04 & 17.57 & 7.84 & 0.07 & 0.13 & 92.95 & 0.00 \\
 & CMD & 14.29 & 0.00 & 0.00 & 0.00 & 28.57 & 7.14 & 7.14 & 0.00 & 0.00 & 0.00 & 0.00 & 0.00 & 0.00 & 14.29 & 7.14 & 0.00 & 42.86 & 0.00 & 0.00 & 71.43 & 0.00 \\
 & SF110 & 41.51 & 4.51 & 0.06 & 0.00 & 7.78 & 10.43 & 2.20 & 0.23 & 0.56 & 1.69 & 13.99 & 0.00 & 0.17 & 44.16 & 44.90 & 7.73 & 12.92 & 0.06 & 0.23 & 91.88 & 0.00 \\
 \cmidrule(r){2-23}
 & Average & 36.78 & 4.33 & 0.08 & 0.00 & 14.05 & 11.08 & 3.39 & 0.14 & 0.23 & 0.86 & 8.99 & 0.00 & 0.34 & 34.37 & 36.02 & 8.43 & 21.20 & 0.04 & 0.12 & 85.42 & 0.00 \\
 \midrule
Mixtral 8x7B & CMD & 31.82 & 4.55 & 0.00 & 0.00 & 27.27 & 22.73 & 22.73 & 0.00 & 0.00 & 0.00 & 4.55 & 0.00 & 0.00 & 31.82 & 31.82 & 0.00 & 50.00 & 0.00 & 0.00 & 95.45 & 0.00 \\
\midrule
\multirow{2}{*}{EvoSuite} & Defects4J & 38.24 & 0.00 & 0.00 & 0.00 & 6.90 & 93.10 & 0.00 & 4.44 & 0.00 & 0.00 & 10.71 & 0.00 & 0.00 & 62.95 & 82.18 & 6.73 & 0.38 & 0.00 & 4.34 & 100.00 & 0.00 \\
& SF110 & 50.78 & 0.00 & 0.00 & 0.00 & 6.90 & 93.10 & 0.00 & 4.44 & 0.00 & 0.00 & 17.53 & 0.00 & 0.00 & 54.67 & 76.94 & 0.07 & 6.90 & 0.00 & 3.96 & 100.00 & 0.00 \\
 \cmidrule(r){2-23}
& Average & 44.51 & 0.00 & 0.00 & 0.00 & 6.90 & 93.10 & 0.00 & 4.44 & 0.00 & 0.00 & 14.12 & 0.00 & 0.00 & 58.81 & 79.56 & 3.40 & 3.64 & 0.00 & 4.15 & 100.00 & 0.00 \\
\bottomrule
\end{tabular}
}
}

\subfloat[Detected by JNose\label{tab:rq1-1-jnose}]{
\centering
\scalebox{0.55}{
\begin{tabular}{lllllllllllllllllllllll}
\toprule
\textbf{Model} & \textbf{Dataset} & \textbf{AR} & \textbf{CLT} & \textbf{CI} & \textbf{DfT} & \textbf{EmT} & \textbf{EH} & \textbf{GF} & \textbf{MG} & \textbf{RP} & \textbf{RA} & \textbf{SE} & \textbf{VT} & \textbf{ST} & \textbf{EaT} & \textbf{LT} & \textbf{DA} & \textbf{UT} & \textbf{IT} & \textbf{RO} & \textbf{MT} & \textbf{DpT} \\
\midrule
\multirow{3}{*}{GPT 3.5}  & Defects4J & 71.15 & 1.92 & 0.00 & 0.00 & 30.77 & 15.38 & 5.77 & 1.92 & 0.00 & 1.92 & 21.15 & 1.92 & 1.92 & 40.38 & 57.69 & 3.85 & 7.69 & 0.00 & 1.92 & 28.85 & 0.00 \\
  & CMD       & 50.00 & 0.00 & 0.00 & 0.00 & 40.00 & 0.00 & 0.00 & 0.00 & 0.00 & 0.00 & 0.00 & 0.00 & 0.00 & 20.00 & 20.00 & 0.00 & 40.00 & 0.00 & 0.00 & 0.00 & 0.00 \\
  & SF110     & 84.62 & 3.85 & 0.00 & 0.00 & 15.38 & 7.69 & 3.85 & 3.85 & 0.00 & 0.00 & 19.23 & 3.85 & 0.00 & 50.00 & 69.23 & 11.54 & 15.38 & 0.00 & 3.85 & 34.62 & 0.00 \\
\cmidrule(r){2-23}
  & Average   & 68.59 & 1.92 & 0.00 & 0.00 & 28.72 & 7.69 & 3.21 & 1.92 & 0.00 & 0.64 & 13.46 & 1.92 & 0.64 & 36.79 & 48.97 & 5.13 & 21.02 & 0.00 & 1.92 & 21.16 & 0.00 \\
\midrule
GPT 4  & CMD       & 100.00 & 0.00 & 0.00 & 0.00 & 0.00 & 0.00 & 50.00 & 0.00 & 0.00 & 0.00 & 0.00 & 0.00 & 0.00 & 50.00 & 50.00 & 0.00 & 0.00 & 0.00 & 0.00 & 50.00 & 0.00 \\
\midrule
\multirow{3}{*}{Mistral 7B}  & Defects4J & 83.33 & 0.00 & 0.00 & 0.00 & 16.67 & 8.33 & 0.00 & 0.00 & 0.00 & 0.00 & 0.00 & 0.00 & 0.00 & 41.67 & 33.33 & 16.67 & 8.33 & 0.00 & 0.00 & 16.67 & 0.00 \\
  & SF110     & 100.00 & 6.25 & 0.00 & 0.00 & 0.00 & 6.25 & 0.00 & 0.00 & 0.00 & 0.00 & 12.50 & 0.00 & 0.00 & 43.75 & 31.25 & 18.75 & 0.00 & 0.00 & 0.00 & 18.75 & 0.00 \\
\cmidrule(r){2-23}
  & Average   & 91.67 & 3.12 & 0.00 & 0.00 & 8.34 & 7.29 & 0.00 & 0.00 & 0.00 & 0.00 & 6.25 & 0.00 & 0.00 & 42.71 & 32.29 & 17.71 & 4.17 & 0.00 & 0.00 & 17.71 & 0.00 \\
  \midrule
Mixtral 8x7B & CMD & 0.00 & 0.00 & 0.00 & 0.00 & 0.00 & 0.00 & 0.00 & 0.00 & 0.00 & 0.00 & 0.00 & 0.00 & 0.00 & 0.00 & 0.00 & 0.00 & 0.00 & 0.00 & 0.00 & 0.00 & 0.00 \\
\midrule
\multirow{2}{*}{EvoSuite}  & Defects4J & 55.56 & 0.00 & 0.00 & 0.00 & 16.67 & 66.67 & 0.00 & 0.00 & 0.00 & 0.00 & 16.67 & 5.56 & 0.00 & 38.89 & 66.67 & 0.00 & 0.00 & 0.00 & 0.00 & 27.78 & 0.00 \\
  & SF110     & 90.00 & 0.00 & 0.00 & 0.00 & 10.00 & 80.00 & 0.00 & 0.00 & 0.00 & 0.00 & 10.00 & 0.00 & 0.00 & 80.00 & 90.00 & 0.00 & 0.00 & 0.00 & 0.00 & 50.00 & 0.00 \\
\cmidrule(r){2-23}
  & Average   & 72.78 & 0.00 & 0.00 & 0.00 & 13.34 & 73.34 & 0.00 & 0.00 & 0.00 & 0.00 & 13.34 & 2.78 & 0.00 & 59.45 & 78.34 & 0.00 & 0.00 & 0.00 & 0.00 & 38.89 & 0.00 \\
\bottomrule
\end{tabular}
}
}
\begin{tablenotes}
\scriptsize
\item[1]$^*$ AR: Assertion Roulette, CLT: Conditional Logic Test, CI: Constructor Initialization, DfT: Default Test, EmT: Empty Test, EH: Exception Handling, GF: General Fixture, MG: Mystery Guest, RP: Redundant Print, RA: Redundant Assertion, SE: Sensitive Equality, VT: Verbose Test, ST: Sleepy Test, EaT: Eager Test, LT: Lazy Test, DA: Duplicate Assert, UT: Unknown Test, IT: Ignored Test, RO: Resource Optimism, MT: Magic Number Test, DpT: Dependent Test.
\end{tablenotes}
\end{table*}
%

\begin{table*}[!htbp]
\centering
\caption{Test Smells Distribution from Benchmark 2 Detected by TsDetect and JNose.}
\label{tab:test-smells-benchmark2}
\subfloat[Detected by TsDetect\label{tab:testbench-tsdetect}]{
\centering
\scalebox{0.55}{
\begin{tabular}{lcccccccccccccccccccccc}
\toprule
\textbf{Model} & \textbf{AR} & \textbf{CLT} & \textbf{CI} & \textbf{DfT} & \textbf{EmT} & \textbf{ECT} & \textbf{EH} & \textbf{GF} & \textbf{MG} & \textbf{RP} & \textbf{RA} & \textbf{SE} & \textbf{VT} & \textbf{ST} & \textbf{EaT} & \textbf{LT} & \textbf{DA} & \textbf{UT} & \textbf{IT} & \textbf{RO} & \textbf{MT} & \textbf{DpT} \\
\midrule
GPT 3.5-Turbo   & 98.98 & 16.20 & 1.05  & 0.00 & 0.00 & 1.08 & 0.00 & 4.68  & 0.00 & 0.00 & 0.34 & 0.00 & 0.37 & 0.00 & 0.00 & 14.75 & 0.00  & 0.00  & 6.78  & 0.00 & 0.00 & 93.60 \\
GPT 4           & 94.43 & 28.24 & 10.79 & 0.00 & 0.00 & 0.00 & 0.00 & 12.76 & 2.97 & 0.00 & 0.00 & 0.00 & 1.98 & 0.00 & 0.65 & 49.51 & 11.14 & 22.95 & 2.64  & 0.00 & 0.00 & 99.03 \\
CodeLlama 13B   & 92.75 & 40.37 & 5.71  & 0.00 & 0.00 & 0.00 & 0.00 & 3.49  & 0.00 & 0.00 & 0.32 & 0.32 & 1.27 & 0.00 & 0.32 & 24.02 & 0.00  & 6.03  & 3.12  & 0.00 & 0.00 & 85.66 \\
\midrule
\textbf{Average} & 95.39 & 28.27 & 5.85 & 0.00 & 0.00 & 0.36 & 0.00 & 6.98 & 0.99 & 0.00 & 0.22 & 0.11 & 1.21 & 0.00 & 0.32 & 29.43 & 3.71 & 9.66 & 4.18 & 0.00 & 0.00 & 92.76 \\
\bottomrule
\end{tabular}
}
}

\subfloat[Detected by JNose\label{tab:testbench-jnose}]{
\centering
\scalebox{0.55}{
\begin{tabular}{lcccccccccccccccccccccc}
\toprule
\textbf{Model} & \textbf{AR} & \textbf{CLT} & \textbf{CI} & \textbf{DfT} & \textbf{EmT} & \textbf{ECT} & \textbf{EH} & \textbf{GF} & \textbf{MG} & \textbf{RP} & \textbf{RA} & \textbf{SE} & \textbf{VT} & \textbf{ST} & \textbf{EaT} & \textbf{LT} & \textbf{DA} & \textbf{UT} & \textbf{IT} & \textbf{RO} & \textbf{MT} & \textbf{DpT} \\
\midrule
GPT 3.5-Turbo & 94.85 & 3.09 & 0.00 & 0.00 & 3.09 & 8.25 & 0.00 & 0.00 & 0.00 & 1.03 & 0.00 & 1.03 & 0.00 & 0.00 & 17.53 & 0.00 & 0.00 & 9.28 & 0.00 & 0.00 & 45.36 & 0.00 \\
GPT 4 & 95.96 & 20.20 & 0.00 & 0.00 & 0.00 & 12.12 & 0.00 & 7.07 & 0.00 & 0.00 & 0.00 & 3.03 & 38.38 & 1.01 & 25.25 & 19.19 & 11.11 & 4.04 & 0.00 & 0.00 & 59.60 & 0.00 \\
CodeLlama 13B  & 98.98 & 24.49 & 0.00 & 0.00 & 0.00 & 6.12 & 0.00 & 0.00 & 0.00 & 1.02 & 1.02 & 2.04 & 9.18 & 0.00 & 15.31 & 1.02 & 17.35 & 5.10 & 0.00 & 0.00 & 66.33 & 0.00 \\
\midrule
\textbf{Average} & 96.60 & 15.93 & 0.00 & 0.00 & 1.03 & 8.83 & 0.00 & 2.36 & 0.00 & 0.68 & 0.34 & 2.03 & 15.85 & 0.34 & 19.36 & 6.74 & 9.49 & 6.14 & 0.00 & 0.00 & 57.10 & 0.00 \\
\bottomrule
\end{tabular}
}
}
\begin{tablenotes}
\scriptsize
\item[1]$^*$ AR: Assertion Roulette, CLT: Conditional Logic Test, CI: Constructor Initialization, DfT: Default Test, EmT: Empty Test, ECT: Exception Catching Throwing, EH: Exception Handling, GF: General Fixture, MG: Mystery Guest, RP: Redundant Print, RA: Redundant Assertion, SE: Sensitive Equality, VT: Verbose Test, ST: Sleepy Test, EaT: Eager Test, LT: Lazy Test, DA: Duplicate Assert, UT: Unknown Test, IT: Ignored Test, RO: Resource Optimism, MT: Magic Number Test, DpT: Dependent Test.
\end{tablenotes}
\end{table*}

\begin{table*}[ht]
\centering
\caption{Delta Values between TsDetect and JNose for Test Smell Distributions (Benchmark 1 and 2)}
\label{tab:delta-distribution-all}
\scalebox{0.55}{
\begin{tabular}{lccccccccccccccccccccc}
\toprule
\textbf{Benchmark} & \textbf{AR} & \textbf{CLT} & \textbf{CI} & \textbf{DfT} & \textbf{EmT} & \textbf{EH} & \textbf{GF} & \textbf{MG} & \textbf{RP} & \textbf{RA} & \textbf{SE} & \textbf{VT} & \textbf{ST} & \textbf{EaT} & \textbf{LT} & \textbf{DA} & \textbf{UT} & \textbf{IT} & \textbf{RO} & \textbf{MT} & \textbf{DpT} \\
\midrule
Benchmark 1 (LLM Avg) & 32.84 & 1.55 & 0.03 & 0.00 & 17.28 & 13.44 & 0.10 & 0.17 & 0.12 & 0.45 & 2.09 & 0.48 & 0.00 & 0.04 & 1.96 & 2.19 & 35.97 & 0.07 & 0.14 & 72.55 & 0.00 \\
Benchmark 1 (EvoSuite) & 28.27 & 0.00 & 0.00 & 0.00 & 6.44 & 19.76 & 0.00 & 4.44 & 0.00 & 0.00 & 0.78 & 2.78 & 0.00 & 0.64 & 1.22 & 3.40 & 3.64 & 0.00 & 4.15 & 61.11 & 0.00 \\
Benchmark 2 & 1.21 & 12.34 & 5.85 & 0.00 & 1.03 & 8.47 & 0.00 & 4.62 & 0.99 & 0.68 & 0.12 & 1.92 & 14.64 & 0.34 & 19.04 & 22.69 & 5.78 & 3.52 & 4.18 & 0.00 & 57.10 \\
\bottomrule
\end{tabular}
}
\end{table*}
%

\begin{table*}[!htbp]
\centering
\caption{Test Smells Distribution by Prompt Techniques from Benchmark 1 Detected by TsDetect and JNose.}
\label{tab:test-smells-benchmark1-prompts}
\subfloat[Detected by TsDetect\label{tab:rq2-1-tsdetect}]{
\centering
\scalebox{0.55}{
\begin{tabular}{llllllllllllllllllllll}
\toprule
\textbf{Prompt} & \textbf{AR} & \textbf{CLT} & \textbf{CI} & \textbf{DfT} & \textbf{EmT} & \textbf{EH} & \textbf{GF} & \textbf{MG} & \textbf{RP} & \textbf{RA} & \textbf{SE} & \textbf{VT} & \textbf{ST} & \textbf{EaT} & \textbf{LT} & \textbf{DA} & \textbf{UT} & \textbf{IT} & \textbf{RO} & \textbf{MT} & \textbf{DpT} \\
CoT & 31.34 & 3.69 & 0.02 & 0.00 & 23.83 & 20.60 & 4.55 & 0.43 & 0.22 & 0.69 & 9.91 & 0.00 & 0.36 & 28.35 & 50.96 & 7.91 & 39.43 & 0.00 & 0.44 & 98.04 & 0.00 \\
FSL & 44.91 & 4.63 & 0.09 & 0.00 & 10.37 & 13.97 & 18.31 & 0.61 & 0.31 & 1.58 & 14.99 & 0.00 & 0.36 & 45.83 & 40.31 & 9.19 & 35.78 & 0.05 & 0.69 & 95.83 & 0.00 \\
ToT & 39.77 & 3.76 & 0.10 & 0.00 & 21.19 & 27.64 & 7.94 & 0.46 & 0.08 & 0.94 & 12.89 & 0.00 & 0.41 & 39.30 & 55.58 & 7.23 & 40.21 & 0.01 & 0.55 & 98.51 & 0.00 \\
GToT & 20.94 & 0.79 & 0.02 & 0.00 & 37.43 & 15.68 & 11.60 & 0.38 & 0.00 & 0.71 & 3.59 & 0.00 & 0.06 & 22.43 & 26.44 & 1.46 & 50.94 & 0.14 & 0.42 & 91.82 & 0.00 \\
ZSL & 54.77 & 4.60 & 0.09 & 0.00 & 10.48 & 17.82 & 22.77 & 0.74 & 0.29 & 0.85 & 14.38 & 0.00 & 0.51 & 43.15 & 48.75 & 8.78 & 20.32 & 0.06 & 0.73 & 99.72 & 0.00 \\
\midrule
\textbf{Average} & 38.09 & 3.45 & 0.06 & 0.00 & 20.84 & 19.13 & 12.99 & 0.52 & 0.17 & 0.96 & 11.09 & 0.00 & 0.33 & 35.81 & 44.12 & 6.83 & 37.72 & 0.05 & 0.56 & 96.67 & 0.00 \\
\bottomrule
\end{tabular}
}
}

\vspace{2mm} 

\subfloat[Detected by JNose\label{tab:rq2-1-jnose}]{
\centering
\scalebox{0.55}{
\begin{tabular}{lcccccccccccccccccccccc}
\toprule
\textbf{Prompt} & \textbf{AR} & \textbf{CLT} & \textbf{CI} & \textbf{DfT} & \textbf{EmT} & \textbf{EH} & \textbf{GF} & \textbf{MG} & \textbf{RP} & \textbf{RA} & \textbf{SE} & \textbf{VT} & \textbf{ST} & \textbf{EaT} & \textbf{LT} & \textbf{DA} & \textbf{UT} & \textbf{IT} & \textbf{RO} & \textbf{MT} & \textbf{DpT} \\
\midrule
CoT  & 66.25 & 22.50 & 0.00 & 0.00 & 30.00 & 0.00 & 2.50 & 0.00 & 0.00 & 0.00 & 26.25 & 0.00 & 1.25 & 42.50 & 62.50 & 43.75 & 3.75 & 0.00 & 0.00 & 35.00 & 0.00 \\
FSL  & 92.13 & 0.00  & 0.00 & 0.00 & 7.87  & 0.00 & 8.33 & 0.00 & 0.00 & 0.00 & 32.87 & 0.00 & 0.00 & 59.72 & 59.72 & 15.74 & 11.11 & 0.00 & 0.00 & 35.65 & 0.00 \\
ToT  & 100.00& 0.00  & 0.00 & 0.00 & 0.00  & 0.00 & 6.50 & 2.50 & 0.00 & 0.00 & 11.50 & 0.00 & 0.00 & 53.67 & 73.50 & 3.33 & 24.00 & 0.00 & 2.50 & 38.83 & 0.00 \\
GToT & 65.83 & 0.00  & 0.00 & 0.00 & 35.83 & 0.00 & 0.00 & 0.00 & 0.83 & 0.00 & 6.67  & 0.00 & 0.00 & 32.50 & 24.17 & 1.67 & 5.00  & 0.00 & 0.00 & 11.67 & 0.00 \\
ZSL  & 75.83 & 2.50  & 0.00 & 0.00 & 13.33 & 0.00 & 2.50 & 5.00 & 0.00 & 0.00 & 7.50  & 0.00 & 0.00 & 20.00 & 30.00 & 12.50 & 24.17 & 0.00 & 5.00 & 27.50 & 0.00 \\
\midrule
\textbf{Average} & 80.01 & 5.00 & 0.00 & 0.00 & 17.41 & 0.00 & 3.97 & 1.50 & 0.17 & 0.00 & 16.96 & 0.00 & 0.25 & 41.68 & 49.98 & 15.40 & 13.61 & 0.00 & 1.50 & 29.73 & 0.00 \\
\bottomrule
\end{tabular}
}
}
\end{table*}
\begin{table*}[ht]
\centering
\caption{Test Smells Distribution by Prompt among all projects from Benchmark 2 Detected by TsDetect and JNose}
\label{tab:test-smells-benchmark2-prompts}
\subfloat[Detected by TsDetect\label{tab:testbench-tsdetect-prompts}]{
\centering
\scalebox{0.58}{
\begin{tabular}{lcccccccccccccccccccccc}
\toprule
\textbf{Prompt} & \textbf{AR} & \textbf{CLT} & \textbf{CI} & \textbf{DfT} & \textbf{EmT} & \textbf{ECT} & \textbf{EH} & \textbf{GF} & \textbf{MG} & \textbf{RP} & \textbf{RA} & \textbf{SE} & \textbf{VT} & \textbf{ST} & \textbf{EaT} & \textbf{LT} & \textbf{DA} & \textbf{UT} & \textbf{IT} & \textbf{RO} & \textbf{MT} & \textbf{DpT} \\
\midrule
SCC    & 94.78 & 29.55 & 7.14 & 0.00 & 0.00 & 0.34 & 0.00 & 6.25 & 0.33 & 0.00 & 0.34 & 0.32 & 1.63 & 0.00 & 0.00 & 28.80 & 3.63 & 11.08 & 4.68 & 0.00 & 0.00 & 95.04 \\
SC & 95.16 & 31.44 & 5.56 & 0.00 & 0.00 & 0.37 & 0.00 & 8.46 & 2.00 & 0.00 & 0.32 & 0.00 & 1.67 & 0.00 & 0.65 & 32.52 & 3.67 & 10.21 & 4.23 & 0.00 & 0.00 & 95.34 \\
FC   & 96.23 & 23.82 & 4.85 & 0.00 & 0.00 & 0.37 & 0.00 & 6.23 & 0.64 & 0.00 & 0.00 & 0.00 & 0.32 & 0.00 & 0.32 & 26.96 & 3.85 & 7.69 & 3.64 & 0.00 & 0.00 & 87.90 \\
\midrule
\textbf{Average}   & 95.39 & 28.27 & 5.85 & 0.00 & 0.00 & 0.36 & 0.00 & 6.98 & 0.99 & 0.00 & 0.22 & 0.11 & 1.21 & 0.00 & 0.32 & 29.43 & 3.71 & 9.66 & 4.18 & 0.00 & 0.00 & 92.76 \\
\bottomrule
\end{tabular}
}
}
\vspace{2mm}

\subfloat[Detected by JNose\label{tab:testbench-jnose-prompts}]{
\centering
\scalebox{0.58}{
\begin{tabular}{lcccccccccccccccccccccc}
\toprule
\textbf{Prompt} & \textbf{AR} & \textbf{CLT} & \textbf{CI} & \textbf{DfT} & \textbf{EmT} & \textbf{ECT} & \textbf{EH} & \textbf{GF} & \textbf{MG} & \textbf{RP} & \textbf{RA} & \textbf{SE} & \textbf{VT} & \textbf{ST} & \textbf{EaT} & \textbf{LT} & \textbf{DA} & \textbf{UT} & \textbf{IT} & \textbf{RO} & \textbf{MT} & \textbf{DpT} \\
\midrule
SCC & 93.14 & 7.77 & 0.00 & 0.00 & 0.43 & 2.77 & 0.00 & 0.39 & 0.00 & 0.43 & 0.37 & 1.94 & 7.83 & 0.00 & 8.84 & 3.96 & 4.19 & 3.33 & 0.00 & 0.00 & 43.62 & 0.00 \\
SC  & 91.74 & 5.68 & 0.00 & 0.00 & 0.47 & 5.18 & 2.01 & 2.01 & 0.00 & 0.41 & 0.00 & 2.10 & 9.28 & 0.40 & 16.66 & 4.42 & 5.68 & 3.57 & 0.00 & 0.00 & 42.86 & 0.00 \\
FC  & 93.06 & 12.63 & 0.00 & 0.00 & 0.45 & 5.49 & 0.00 & 0.44 & 0.00 & 0.00 & 0.00 & 0.00 & 9.21 & 0.44 & 15.35 & 4.58 & 3.83 & 2.79 & 0.00 & 0.00 & 46.32 & 0.00 \\
\midrule
\textbf{Average} & 92.65 & 8.69 & 0.00 & 0.00 & 0.45 & 4.48 & 0.67 & 0.95 & 0.00 & 0.28 & 0.12 & 1.35 & 8.77 & 0.28 & 13.62 & 4.32 & 4.57 & 3.23 & 0.00 & 0.00 & 44.27 & 0.00 \\
\bottomrule
\end{tabular}
}
}
\begin{tablenotes}
\scriptsize
\item[1]$^*$ SCC: Self-Contained Context, SC: Simple Context, FC: Full Context, AR: Assertion Roulette, CLT: Conditional Logic Test, CI: Constructor Initialization, DfT: Default Test, EmT: Empty Test, ECT: Exception Catching Throwing, EH: Exception Handling, GF: General Fixture, MG: Mistery Guest, RP: Redundant Print, RA: Redundant Assertion, SE: Sensitive Equality, VT: Verbose Test, ST: Sleepy Test, EaT: Eager Test, LT: Lazy Test, DA: Duplicate Assert, UT: Unknown Test, IT: Ignored Test, RO: Resource Optimism, MT: Magic Number Test, DpT: Dependent Test.
\end{tablenotes}
\end{table*}

\begin{table}[htbp]
\centering
\caption{Summary of Kruskal-Wallis Test Results on Benchmark 1 (LLM-Generated Test Suites)}
\label{tab:kruskal-benchmark1-summary}
\scalebox{0.55}{
\begin{tabular}{lcccccc}
\toprule
\textbf{Tool} & \textbf{Total Comparisons} & \textbf{\#p $<$ 0.01} & \textbf{0.01 $\leq$ p $<$ 0.05} & \textbf{0.05 $\leq$ p $\leq$ 0.1} & \textbf{0.1 $<$ p $<$ 0.5} & \textbf{0.5 $\leq$ p $\leq$ 1.0} \\
\midrule
TsDetect & 180 & 0 & 0 & 0 & 120 & 60 \\
JNose    & 160 & 0 & 0 & 0 & 108 & 52 \\
\bottomrule
\end{tabular}
}
\end{table}
%
\begin{table*}[!htbp]
\centering
\caption{Mutual Information (MI) Results on Benchmark 1 LLM-Generated Tests.}
\label{tab:mi-benchmark1}

\begin{tabular}{cc}
\subfloat[Detected by TsDetect\label{tab:mi-bench1-tsdetect}]{
\scalebox{0.45}{
\begin{tabular}{lcccccccc}
\toprule
 & AR & CLT & MG & EaT & UT & IT & RO & MT \\
\midrule
\#Params (B) & - & 2.22e-16 & - & - & - & - & - & - \\
Context Length & 2.22e-16 & - & 1.11e-16 & 2.22e-16 & 1.11e-16 & 1.11e-16 & 1.11e-16 & 2.22e-16 \\
Top-p & 2.22e-16 & - & 1.11e-16 & 2.22e-16 & 1.11e-16 & 1.11e-16 & 1.11e-16 & 2.22e-16 \\
\#Classes & 2.22e-16 & - & 1.11e-16 & 2.22e-16 & 1.11e-16 & 1.11e-16 & 1.11e-16 & 2.22e-16 \\
\#Methods & 2.22e-16 & - & 1.11e-16 & 2.22e-16 & 1.11e-16 & 1.11e-16 & 1.11e-16 & 2.22e-16 \\
KLOC & 2.22e-16 & - & 1.11e-16 & 2.22e-16 & 1.11e-16 & 1.11e-16 & 1.11e-16 & 2.22e-16 \\
CBO & 2.22e-16 & - & 1.11e-16 & 2.22e-16 & 1.11e-16 & 1.11e-16 & 1.11e-16 & 2.22e-16 \\
RFC & 2.22e-16 & - & 1.11e-16 & 2.22e-16 & 1.11e-16 & 1.11e-16 & 1.11e-16 & 2.22e-16 \\
DIT & 2.22e-16 & - & 1.11e-16 & 2.22e-16 & 1.11e-16 & 1.11e-16 & 1.11e-16 & 2.22e-16 \\
CyCo & 2.22e-16 & - & 1.11e-16 & 2.22e-16 & 1.11e-16 & 1.11e-16 & 1.11e-16 & 2.22e-16 \\
\bottomrule
\end{tabular}
}
}

&

\subfloat[Detected by JNose\label{tab:mi-bench1-jnose}]{
\scalebox{0.45}{
\begin{tabular}{lccccccc}
\toprule
 & CLT & GF & MG & RA & VT & ST & RO \\
\midrule
\#Params (B) & 2.22e-16 & - & - & - & - & - & - \\
Context Length & - & 1.11e-16 & 1.11e-16 & 1.11e-16 & 1.11e-16 & 1.11e-16 & 1.11e-16 \\
Top-p & - & 1.11e-16 & 1.11e-16 & 1.11e-16 & 1.11e-16 & 1.11e-16 & 1.11e-16 \\
\#Classes & - & 1.11e-16 & 1.11e-16 & 1.11e-16 & 1.11e-16 & 1.11e-16 & 1.11e-16 \\
\#Methods & - & 1.11e-16 & 1.11e-16 & 1.11e-16 & 1.11e-16 & 1.11e-16 & 1.11e-16 \\
KLOC & - & 1.11e-16 & 1.11e-16 & 1.11e-16 & 1.11e-16 & 1.11e-16 & 1.11e-16 \\
CBO & - & 1.11e-16 & 1.11e-16 & 1.11e-16 & 1.11e-16 & 1.11e-16 & 1.11e-16 \\
RFC & - & 1.11e-16 & 1.11e-16 & 1.11e-16 & 1.11e-16 & 1.11e-16 & 1.11e-16 \\
DIT & - & 1.11e-16 & 1.11e-16 & 1.11e-16 & 1.11e-16 & 1.11e-16 & 1.11e-16 \\
CyCo & - & 1.11e-16 & 1.11e-16 & 1.11e-16 & 1.11e-16 & 1.11e-16 & 1.11e-16 \\
\bottomrule
\end{tabular}
}
}

\end{tabular}
\end{table*}

\begin{table*}[ht]
\centering
\caption{Test Smells Distribution in Human-written Test Suites Detected by TsDetect and JNose}
\label{tab:rq4-1-human-written}
\subfloat[Detected by TsDetect\label{tab:tsdetect-tsmells-human}]{
\centering
\scalebox{0.55}{
\begin{tabular}{lccccccccccccccccccccccc}
\toprule
\textbf{Dataset} & \textbf{AR} & \textbf{CLT} & \textbf{CI} & \textbf{DfT} & \textbf{EmT} & \textbf{ECT} & \textbf{EH} & \textbf{GF} & \textbf{MG} & \textbf{RP} & \textbf{RA} & \textbf{SE} & \textbf{VT} & \textbf{ST} & \textbf{EaT} & \textbf{LT} & \textbf{DA} & \textbf{UT} & \textbf{IT} & \textbf{RO} & \textbf{MT} & \textbf{DpT} \\
\midrule
Defects4J  & 99.75 & 60.52 & 31.43 & 26.49 & 0.00 & 3.77 & 0.00 & 54.15 & 8.21 & 1.09 & 1.01 & 5.62 & 14.42 & 0.00 & 0.34 & 58.09 & 52.47 & 36.04 & 28.67 & 6.96 & 1.26 & 95.14 \\
SF110      & 99.55 & 44.89 & 26.78 & 30.72 & 0.00 & 4.85 & 0.00 & 58.34 & 7.96 & 2.72 & 3.36 & 3.43 & 6.40 & 0.00 & 1.75 & 47.80 & 42.30 & 14.29 & 32.99 & 2.91 & 4.14 & 81.31 \\
CAT-LM     & 97.19 & 37.55 & 13.04 & 9.38  & 0.02 & 1.43 & 0.00 & 34.14 & 7.02 & 2.83 & 3.43 & 1.15 & 5.37 & 0.00 & 1.32 & 30.98 & 29.34 & 12.91 & 19.39 & 9.24 & 3.11 & 67.21 \\
\midrule
\textbf{Averages} 
           & 98.83 & 47.65 & 23.75 & 22.20 & 0.01 & 3.35 & 0.00 & 48.88 & 7.73 & 2.21 & 2.60 & 3.40 & 8.73 & 0.00 & 1.14 & 45.62 & 41.37 & 21.08 & 27.02 & 6.37 & 2.84 & 81.22 \\
\bottomrule
\end{tabular}
}
}
\vspace{5mm}

\subfloat[Detected by JNose\label{tab:jnose-tsmells-human}]{
\centering
\scalebox{0.55}{
\begin{tabular}{lccccccccccccccccccccccc}
\toprule
\textbf{Dataset} & \textbf{AR} & \textbf{CLT} & \textbf{CI} & \textbf{DfT} & \textbf{EmT} & \textbf{ECT} & \textbf{EH} & \textbf{GF} & \textbf{MG} & \textbf{RP} & \textbf{RA} & \textbf{SE} & \textbf{VT} & \textbf{ST} & \textbf{EaT} & \textbf{LT} & \textbf{DA} & \textbf{UT} & \textbf{IT} & \textbf{RO} & \textbf{MT} & \textbf{DpT} \\
\midrule
Defects4J & 74.11 & 18.04 & 30.00 & 0.00   & 2.86 & 23.04  & 0.00 & 4.64 & 2.14 & 1.61 & 1.25 & 4.46 & 10.89 & 0.71 & 3.39 & 4.82 & 14.11 & 11.96 & 1.07 & 2.32 & 41.43 & 0.00  \\
SF110     & 59.38 & 15.62 & 40.62 & 0.00    & 3.12 & 21.88  & 0.00 & 6.25 & 3.12 & 0.00 & 6.25 & 6.25 & 15.62 & 3.12 & 0.00 & 6.25 & 9.38  & 21.88 & 0.00 & 3.12 & 28.12 & 0.00  \\
CAT-LM     & 57.85 & 9.77  & 11.79 & 0.07 & 12.58 & 7.19  & 0.00  & 5.33 & 2.26 & 3.61 & 1.10 & 3.96 & 9.36  & 0.76 & 1.35 & 2.74 & 6.89  & 24.42 & 9.20 & 2.52 & 19.17 & 0.00  \\
\midrule
\textbf{Average} & 63.78 & 14.48 & 27.47 & 0.07 & 6.19 & 17.37  & 0.00 & 5.41 & 2.51 & 1.74 & 2.87 & 4.89 & 11.96 & 1.53 & 1.58 & 4.60 & 10.13 & 19.42 & 3.42 & 2.65 & 29.57 & 0.00  \\
\bottomrule
\end{tabular}
}
}
\begin{tablenotes}
\scriptsize
\item[1]$^*$ AR: Assertion Roulette, CLT: Conditional Logic Test, CI: Constructor Initialization, DfT: Default Test, EmT: Empty Test, ECT: Exception Catching Throwing, EH: Exception Handling, GF: General Fixture, MG: Mystery Guest, RP: Redundant Print, RA: Redundant Assertion, SE: Sensitive Equality, VT: Verbose Test, ST: Sleepy Test, EaT: Eager Test, LT: Lazy Test, DA: Duplicate Assert, UT: Unknown Test, IT: Ignored Test, RO: Resource Optimism, MT: Magic Number Test, DpT: Dependent Test.
\end{tablenotes}
\end{table*}

\end{document}